\documentclass[aps,prd,twocolumn,groupedaddress,floatfix,nofootinbib]{revtex4}

\newcommand{\apjl}{Astrophys. J. Lett.}

\usepackage{amsmath,amssymb}
\usepackage{textcomp}
\usepackage{hyperref}
\usepackage{bm}
\usepackage{graphicx}
\usepackage{psfrag}
\usepackage[usenames]{color}
\usepackage{empheq}

\allowdisplaybreaks[4]

\newcommand{\uvec}[1]{\bm{\hat{#1}}}
\newcommand{\dvec}[1]{\dot{\bm{#1}}}

\newcommand{\duvec}[1]{\dot{\bm{\hat{#1}}}}

\newcommand{\pdfrac}[2]{\frac{ \partial #1}{\partial #2}}

\newcommand{\be}{\begin{equation}}
\newcommand{\ee}{\end{equation}}
\newcommand{\orb}{{\mbox{\tiny orb}}}
\newcommand{\obs}{{\mbox{\tiny obs}}}
\newcommand{\precc}{{\mbox{\tiny prec}}}
\newcommand{\rr}{{\mbox{\tiny rr}}}
\newcommand{\V}{{\mbox{\tiny V}}}
\newcommand{\NV}{{\mbox{\tiny NV}}}
\newcommand{\C}{{\mbox{\tiny C}}}
\newcommand{\PP}{{\mbox{\tiny P}}}
\newcommand{\SPA}{{\mbox{\tiny SPA}}}
\newcommand{\dett}{{\mbox{\tiny det}}}
\newcommand{\coal}{{\mbox{\tiny coal}}}
\newcommand{\eLISA}{{\mbox{\tiny eLISA}}}

\newcounter{subsubsubsection}[subsubsection]

\begin{document}

\title{Gravitational Waveforms for Precessing, Quasi-circular
Binaries \\ via Multiple Scale Analysis and Uniform Asymptotics: \\ The Near 
Spin Alignment Case}


\author{Antoine Klein}
\affiliation{Department of Physics, Montana State University, Bozeman,
MT 59717, USA}

\author{Neil Cornish}
\affiliation{Department of Physics, Montana State University, Bozeman,
MT 59717, USA}

\author{Nicol\'as Yunes}
\affiliation{Department of Physics, Montana State University, Bozeman,
MT 59717, USA}

\date{\today}

\begin{abstract}
We calculate analytical gravitational waveforms in the time- and 
frequency-domain for precessing quasi-circular binaries with spins of arbitrary 
magnitude, but nearly aligned with the orbital angular momentum.
We first derive an analytical solution to the precession equations by expanding 
in the misalignment angle and using multiple scale analysis to separate 
timescales.
We then use uniform asymptotic expansions to analytically Fourier transform the 
time-domain waveform, thus extending the stationary-phase approximation, which 
fails when precession is present. 
The resulting frequency-domain waveform family has a high overlap with 
numerical 
waveforms obtained by direct integration of the post-Newtonian equations of 
motion and discrete Fourier transformations. 
Such a waveform family lays the foundations for the accurate inclusion of spin 
precession effects in analytical gravitational waveforms, and thus, it can aid 
in 
the detection and parameter estimation of gravitational wave signals from the 
inspiral phase of precessing binary systems. 
\end{abstract}

\maketitle
\section{Introduction}

Gravitational waves are a promising new tool for astrophysics, capable of 
bringing breakthroughs in our understanding of the 
Universe. Currently, an array of ground-based interferometers are undergoing 
upgrades that should lead to the first direct 
detection of gravitational waves within this decade~\cite{ligo,virgo}. A project 
for a space-based interferometer~\cite{2012CQGra..29l4016A} is 
also under way, and could be operational in the next decade.

The likelihood of detecting gravitational waves improves when reliable and 
efficient
waveform models are available for use in the data analysis~\cite{Finn:1992xs}. 
Systems with spins, 
unless they are exactly aligned or anti-aligned with the orbital angular 
momentum, will
undergo a secular precession of the spins and of the orbital 
plane~\cite{Apostolatos:1994mx}.
This precession will impact the waveforms, and it is important to take it into
account to properly extract the spins of the binary components, as well as to 
break degeneracies 
between different system parameters~\cite{Lang:1900bz,Lang:2011je}. 

The purpose of this paper is to develop accurate analytical 
waveforms for spinning and precessing compact binary, quasi-circular inspirals.
Currently, the most accurate waveforms are numerical, time-domain solutions of 
the PN equations~\cite{Lang:1900bz,Lang:2011je} that are then numerically 
Fourier transformed. 
Such numerical solutions, however, can be rather computationally expensive, 
especially when 
a large number of templates are needed, as is the case for spinning systems. 
We will here employ novel mathematical techniques 
(multiple scale analysis and uniform asymptotic expansions)~\cite{Bender} to 
produce analytic, 
Fourier-domain, waveform families that accurately reproduce their numerical 
counterparts.

\subsection{Previous Waveforms for Spinning Binaries}

Over the years, several groups have developed increasingly accurate waveforms 
for use in gravitational wave astrophysics. Post-Newtonian (PN), quasi-circular 
and eccentric inspiral waveforms for compact binaries have been obtained to 
high-order as an expansion in the orbital velocity~\cite{blanchet-review}. The 
leading-order, spin-orbit and spin-spin contributions to the dynamics appear at 
1.5PN and 2PN relative order 
respectively~\cite{PhysRevD.12.329,PhysRevD.47.R4183,PhysRevD.52.821}. These 
calculations have now been extended through 
2.5PN~\cite{PhysRevD.63.044006,PhysRevD.74.104033,PhysRevD.77.064032} and 3PN 
order~\cite{PhysRevD.77.081501} out to 3.5PN 
order~\cite{Marsat:2012fn,ANDP:ANDP201100094,Bohe:2012mr,Bohe:2013cla}.  The 
spin-orbit contributions to the gravitational wave
phase and amplitude are currently known to 3.5PN 
order~\cite{PhysRevD.81.089901,1475-7516-2012-09-028,Bohe:2013cla}. 

Presently, there are three special cases where purely analytic, inspiral 
waveform models exist for spinning binaries: 
\begin{enumerate}
\item[(i)] {\bf{Aligned}}: Systems where the spins are co-aligned or 
anti-aligned with the orbital angular momentum; 
\item[(ii)] {\bf{Partially Non-Spinning}}: Systems where one of the compact 
bodies is not spinning; 
\item[(iii)] {\bf{Equal-mass systems}}. Systems with both components spinning 
but with a mass ratio of unity.
\end{enumerate}
For case (i), the spins stay aligned (or anti-aligned) with the orbital angular 
momentum throughout the evolution and there is no precession. The waveforms 
here 
bear strong resemblance to non-spinning waveforms, but with a spin-corrected 
chirping (see e.g.~\cite{Poisson:1995ef,Arun:2008kb}). 
For cases (ii) and (iii), the system undergoes {\em simple 
precession}\footnote{For case (iii), the system undergoes simple precession if 
the 2PN spin-spin corrections to the spin dynamics are neglected.}, i.e.~the 
precession of the orbital and spin angular momenta about the total angular 
momentum with a single (evolving) precession frequency (see 
e.g.~\cite{Apostolatos:1994mx}). 

Cases (i) and (ii) are of particular interest since they provide good 
approximations to systems that are expected to
be found in Nature. Case (i) pertains to binaries embedded in a gaseous 
environment, where the gas torque tends to align the spin with the orbital 
angular momentum~\cite{2007ApJ.661L.147B,Barausse:2012fy,2013ApJ.762.68D}. Most 
studies of spin 
alignment have focused on supermassive black hole mergers, but recently, 
similar 
mechanisms have been invoked for stellar-mass black hole
mergers~\cite{2013arXiv1302.4442G}. Case (ii) pertains to the case when the 
spin 
angular momentum of one of the bodies is much smaller than the other one. 
Population synthesis models suggest that typical mass ratios in a binary black 
hole system may be of the order of ten or so, so that the spin of the more 
massive object will 
dominate, and the spin of the smaller body can be neglected.

The data analysis of signals usually requires the Fourier transform of the wave 
signal, and the complexity of the latter varies strongly depending on the 
particular case considered. For the non-precessing case (i), the time-domain 
waveforms are especially simple, and they can easily be analytically recast in 
the frequency-domain using the stationary phase approximation 
(SPA)~\cite{cutlerflanagan,Droz:1999qx}. 
In the SPA, the waveform amplitude is assumed to vary much more slowly than the 
phase, which 
then allows for a Taylor expansion of the generalized Fourier 
integral~\cite{Bender}. For the 
simple precession case (ii) and (iii), the waveforms have far more structure 
because of the precession of the orbital plane. Contrary to earlier 
claims~\cite{Apostolatos:1994mx}, however, 
such simple-precessing waveforms are not amenable to a formal stationary phase 
treatment. This is because the mapping between time and frequency becomes 
multi-valued and the SPA amplitude diverges. 

Precessing waveforms have to therefore be modeled differently in the 
frequency-domain.  One approach is to take the discrete Fourier transform (DFT) 
of the time-domain waveform, which requires the fine sampling of the latter at 
the orbital timescale, which can lead to a large computational cost. Another 
approach is to take a spin-aligned (non-precessing) waveform and transform it to 
a frame that tracks some characteristic quantity (like the total angular 
momentum vector) at some precession 
rate~\cite{Schmidt:2010it, O'Shaughnessy:2011fx, Boyle:2011gg, 
O'Shaughnessy:2012vm, Ochsner:2012dj, Schmidt:2012rh, Ossokine:2013zga, 
Lundgren:2013jla}. This approach has been validated to some degree with 
numerical simulations but the matches are not perfect, as these waveforms miss 
some precession effects. Without an analytical treatment, one is left with 
educated guesses as to how to improve on these waveforms. 
Yet another approach is to use an effective SPA, where precession corrections to 
the 
amplitude are completely neglected and phase corrections are only partially 
modeled using insight from the spin-aligned 
case~\cite{Lang:2011je,Klein:2009gza}. Both of these 
approaches are unattractive due to either increased computational cost or 
increased systematic mismodeling error. 

\pagebreak
\subsection{Executive Summary}

The goals of the present work are the following: 
\begin{enumerate}
\item[(a)] To develop a general formalism to perturbatively solve the 
time-domain PN evolution equations and obtain analytic time-domain waveforms 
for 
precessing quasi-circular inspirals;
\item[(b)] To develop a general formalism to perturbatively Fourier transform a 
precessing quasi-circular inspiral, time-domain waveform. 
\end{enumerate}
Goal (a) will be achieved through the technique of {\emph{multiple scale 
expansions}}~\cite{Bender}, where one solves the evolution equations using 
the fact that 
$t_{\orb} \ll t_{\precc} \ll t_{\rr}$, where $t_{\orb}$, $t_{\precc}$ and 
$t_{\rr}$ are the orbital, precession and radiation-reaction timescales 
respectively. Goal (b) is reached through the technique of {\emph{uniform 
asymptotic expansions}}~\cite{Bender,PhysRevE.64.026215}, where one recasts the 
phase modulation induced by precession as a sum of Bessel functions that are 
then amenable to a formal SPA treatment. Both of these techniques have proven 
very successful in various areas, from quantum field theory to aerospace 
engineering. 

Although these formalisms are general, we exemplify them here by generalizing 
case (i) to accommodate systems where the spins are only partially aligned with 
the orbital angular momentum. The accretion torques that drive the spin-orbit 
alignment are not expected to produce perfect alignment, so it is useful to 
have 
analytic waveforms that cover the more realistic, partial alignment case. 
Expanding the spin precession equations in the misalignment angle leads to a 
system of coupled harmonic oscillators that diagonalizes to yield two 
precession 
frequencies $\omega_+$ and $\omega_-$. Multiple scale analysis is then used to 
derive an analytic expression for the evolution of these precession frequencies 
as the black holes spiral together~\cite{Bender}. 

Spin precession alters the phasing of the waveform, and causes the mapping 
between gravitational wave frequency and time to become multi-valued, rendering 
the standard SPA inapplicable. The SPA returns singular results at turning 
points in the time-frequency mapping, resulting in what are known as 
{\emph{fold}} and {\emph{cusp catastrophes}} in the optical literature~
\cite{Berry1980257}. We show that the singularities can be cured using uniform 
asymptotic expansions of the phase 
in terms of Bessel functions~\cite{Bender,PhysRevE.64.026215}. 

The final result is a family of fully analytic, approximate, time and frequency 
domain waveforms for spinning, precessing, 
quasi circular binaries with moderately misaligned spins. The frequency-domain 
waveform family can be constructed from the following recipe: 
\begin{enumerate}
\item The waveform is mode-decomposed as 
\be
\tilde{h}(f) = \sum_{n\geq0} \;\; \sum_{k \in \mathbb{Z}} \;\; \sum_{m = 
\{-2,2\}} \tilde{h}_{n,k,m}(f)\,,
\label{htildef}
\ee 
\item Each Fourier mode is given by Eq.~\eqref{htildef-nkm} in terms of the 
carrier phase $\phi_{\C}$, the precession phases $\phi_{\PP,\pm}$, the 
time-frequency mapping $t=t(f)$, the mode-decomposed, time-domain amplitudes 
${\cal{A}}_{n,k,m}$, and the additional constant, amplitude, and phase 
modulation corrections $A_{0,n,k,m}$,
$A_{\pm,n,k,m}$, and $\phi_{\pm,n,k,m}$ respectively. 
\item The carrier phase $\phi_{\C}$ and its second time derivative are given in 
Appendix~\ref{app:freqevol} as a function of the orbital frequency.
\item The precession phases and their second time derivatives $\phi_{\PP,\pm}$ 
are given in Appendix~\ref{app:precphases} as a function of the orbital 
frequency. 
\item The time-frequency mapping $t=t(f)$ is given as a function of the orbital 
frequency in Appendix~\ref{app:freqevol}.
\item The mode-decomposed, time-domain amplitudes ${\cal{A}}_{n,k,m}$ are given 
as a function of the orbital frequency in Appendix~\ref{app:amplitudes}.
\item The constant, amplitude, and phase modulation corrections 
$A_{0,n,k,m}$, $A_{\pm,n,k,m}$, and 
$\phi_{\pm,n,k,m}$ are given as a function of the orbital frequency in 
Appendix~\ref{app:amplitudeandphasemodulations}.
\item The orbital frequency is given in terms of the Fourier frequency in 
Eqs.~\eqref{xi-eq}-\eqref{u-eq}.
\end{enumerate}
We prove that these new waveforms are accurate (i.e.~faithful) by 
comparing them to the waveforms obtained by numerically solving for the orbital 
evolution and discretely Fourier transforming. 
We find typical matches of $0.99$-$0.999$, maximized only over time and phase 
of 
coalescence, when the misalignment angles 
do not exceed $25^\circ$.

The benefit of our approach is two-fold. On the one hand, we provide 
ready-to-use, analytic waveforms that are computationally inexpensive to 
produce. The 
computational cost of numerically solving for the Fourier transform of spinning 
and precessing systems is currently a roadblock in the data analysis of signals 
for advanced ground detectors. On the other hand, an analytical treatment 
provides insight into the physics of the problem. Our results analytically 
explain why the waveforms of spinning and precessing binaries are essentially 
simple harmonic oscillators, with a carrier band and side-bands induced by 
evolving precession frequencies~\cite{Lundgren:2013jla}. Moreover, our results 
extend the recently-proposed kludge waveforms~\cite{Lundgren:2013jla} to 
account 
for amplitude and additional phase corrections induced by precession, which 
cannot be captured by educated guesses from numerical relativity waveforms. 

The general formalism presented here opens the door to several other 
applications. For instance, one can extend case (ii) to include the first-order 
correction in the ratio of the spins of the two bodies. In this case, our 
formalism can be viewed as a systematic extension of the simple-precessing 
treatment of Apostolatos, et al~\cite{Apostolatos:1994mx}, which allows us to 
compute the time-domain waveforms to higher order in the ratio of the 
timescales 
of the problem. Moreover, our method allows for the correct analytical 
calculation of the Fourier-domain waveforms, which cannot be obtained via a 
standard SPA treatment, contrary to older claims~\cite{Apostolatos:1994mx}. Our 
formalism can also be applied to other systems, such as inspiraling binary 
neutron stars, where the magnitude of both spin angular momenta are much 
smaller 
than the orbital one. 

\subsection{Organization and Conventions}

The remainder of this paper is organized as follows:
Section~\ref{sec:msa} reviews multiple scale analysis through selected 
examples, 
which we use later to solve the equations of precession analytically;
Section~\ref{sec:uaa} discusses the SPA and uniform asymptotic expansions 
applied to the Fourier transform of oscillatory functions, where the mapping 
between time and frequency is multi-valued and the standard SPA fails; 
Section~\ref{sec:near-alignment} derives an analytic formula for the evolution 
of the angular momenta in the case of nearly aligned spins; 
Section~\ref{sec:spa} uses the results of the previous sections to derive an 
analytical gravitational waveform valid for nearly aligned spins; 
Section~\ref{sec:comp} compares our waveform to the results obtained by taking 
a 
discrete Fourier transform of the time series.

Throughout this article we use geometric units with $G = c = 1$.  We also 
employ 
the following conventions:
\begin{itemize}
 \item Three-dimensional vectors are written in boldface and unit vectors carry
a hat over them, e.g.~$\bm{A} = (A_x, A_y, A_z)$, with norm $|\bm{A}| = A$, and
unit vector $\uvec{A} = \bm{A}/A$.
\item Three-dimensional matrices are written in mathematical boldface, 
e.g.~${\mathbb{M}}$
and ${\mathbb{A}}$.
 \item Total time derivatives are denoted with an overhead dot: $\dot{f} =
df/dt$.
 \item $\omega$ is the angular frequency in a frame fixed to the orbital plane.
 \item $\bm{L}$ is the Newtonian orbital angular momentum $3$-vector.
 \item $\bm{S}_A$ is the spin angular momentum  $3$-vector for component $A$. 
 \item $m_A$ is the mass of component $A$, and we assume $m_1 \geq
m_2$.
 \item $\mu = m_1 m_2/M$ is the reduced mass.
 \item $\nu = m_1 m_2/M^2$ is the symmetric mass ratio.
 \item $\chi_{A} \equiv |\bm{S}_{A}|/m_{A}^{2}$ is the dimensionless spin
parameter for component $A$.
 \item $\uvec{N}$ is a unit vector pointing from the observer to the
source.
\end{itemize}

\section{Techniques from Asymptotic Analysis}

\subsection{A Primer on Multiple Scale Analysis}
\label{sec:msa}

Multiple scale analysis is a powerful mathematical formalism that serves as the
theoretical foundations of boundary-layer theory and the 
Wentzel-Kramers-Brillouin
(WKB) approximation. In this section, we review 
some important features of this formalism, as they will be essential  in the
solution to the precession 
equations. We will mostly follow and summarize the treatment in Bender and
Orszag~\cite{Bender}. 

Consider the non-linear oscillator ordinary differential equation $\ddot{y}
+ y + \epsilon y^{3} = 0$, 
where $y$ is a function of time $t$, with initial conditions
$(y(0),\dot{y}(0)) = (1,0)$. If we attempted
the series solution 
\be
y(t) = \sum_{n=0}^{\infty} \epsilon^{n} y_{n}(t)\,,
\ee
assuming $\epsilon \ll 1$, and matched coefficients of the same order in
$\epsilon$, 
we would find the solution
\be
y(t) = \cos{t} + \epsilon \left[ \frac{1}{32} \cos{3 t} - \frac{1}{32} \cos{t} -
\frac{3}{8} t \sin{t} \right] + {\cal{O}}(\epsilon^{2})\,.  
\label{series-exp}
\ee
Clearly, this series approximation diverges as $t \to \infty$, but in fact, it
becomes invalid much sooner, 
when $3\epsilon t /8\sim 1$. As we will show below, however, the exact solution
to this differential equation
remains perfectly bounded in the $t \to \infty$ limit; a multiple-scale
expansion treatment will allow us to find 
such a solution.  

Let us then introduce a new variable $\tau = \epsilon t$ that defines a long
time scale, as $\tau$ does not become
negligible when $t \sim 1/\epsilon$. In multiple scale analysis, we search for
solutions that are functions of all timescales in the problem, in this case  $t$
and $\tau$, treated as if they were independent
variables. This, of course, is just a mathematical trick, since
at the end of the day, we can replace $\tau$ in favor of $\epsilon t$ to obtain
a solution that is only $t$-dependent. 
We then assume a perturbative ansatz of the form
\be
y(t) = \sum_{n=0}^{\infty} \epsilon^{n} Y_{n}(t,\tau)\,.
\label{multiple-scale-expansion}
\ee
Taking the sum to $n=1$, the non-linear oscillator equation leads to the
following two evolution equations
\begin{align}
\frac{\partial^{2} Y_{0}}{\partial t^{2}} + Y_{0} &= 0\,,
\label{0th-order-eq}
\\
\frac{\partial^{2} Y_{1}}{\partial t^{2}} + Y_{1} &= -Y_{0}^{3} - 2
\frac{\partial^{2} Y_{0}}{\partial \tau \partial t}\,.
\label{1st-order-eq}
\end{align}
Notice that the differential operator (the terms on the left-hand side of both 
equations)
is always the same, an expected result in perturbation theory. 
The most general solution to Eq.~\eqref{0th-order-eq} is $Y_{0} = A(\tau) e^{it}
+ A^{*}(\tau) e^{-it}$, where the star
stands for complex conjugation. Inserting this solution into
Eq.~\eqref{1st-order-eq}, we find
\be
\frac{\partial^{2} Y_{1}}{\partial t^{2}} + Y_{1} = e^{it} \left[-3 A^{2}
A^{*} - 2 i \frac{\partial A}{\partial \tau}\right] - e^{3 i t} A^{3} +
{\rm{c.c.}}\,,
\label{1st-order-eq-subed}
\ee
where ${\rm{c.c.}}$ stands for the complex conjugate. The first term of the
right-hand side in Eq.~\eqref{1st-order-eq-subed}
is a solution to Eq.~\eqref{0th-order-eq}, and thus, it is it which induces a
secular growth. We can eliminate this secular growth
by requiring that the term inside square brackets on the right-hand side of
Eq.~\eqref{1st-order-eq-subed} vanishes, which then leads
to a differential equation for $A(\tau)$, whose solution is
\be
A(\tau) = R(0) e^{i \theta(0) + 3 i R^{2}(0) \tau/2}\,,
\ee
where $R(0)$ and $\theta(0)$ are constants of integration. Using the initial
conditions stipulated above and reassembling
the full solution, we find
\be
y(t) = \cos{\left[ t \left(1 + \frac{3}{8} \epsilon t \right) \right]} +
{\cal{O}}(\epsilon)\,.
\ee
Notice that this solution is bounded for all $t$ and it is much more accurate
than the series expansion in Eq.~\eqref{series-exp}
for large $t$. 

An extra degree of complication arises when we consider differential equations
with implicit functional dependence in the source. 
For example, let us consider the oscillator ordinary differential equation
\be
\epsilon^{2} \frac{d^{2}y}{dt^{2}}  + \omega^{2}(\tau) y = 0\,,
\label{oscillator-WKB-eq}
\ee
where again $\tau = \epsilon t$ and try to solve it with multiple-scale 
analysis.
Imposing the same expansion of the solution 
as in Eq.~\eqref{multiple-scale-expansion}, Eq.~\eqref{oscillator-WKB-eq}
becomes
\begin{align}
\frac{\partial^{2} Y_{0}}{\partial t^{2}} + \omega^{2}(\tau) Y_{0} &= 0\,,
\\
\frac{\partial^{2} Y_{1}}{\partial t^{2}} + \omega^{2}(\tau) Y_{1} &= -2
\frac{\partial^{2} Y_{0}}{\partial t \partial \tau}\,.
\label{1st-order-eq-WKB}
\end{align}
The solution to the zeroth-order equation is now $Y_{0} = A(\tau) e^{i
\omega(\tau) t} + A^{*}(\tau) e^{-i \omega(\tau) t}$, 
which when inserted into Eq.~\eqref{1st-order-eq-WKB} leads to
\be
\frac{\partial^{2} Y_{1}}{\partial t^{2}} + \omega^{2}(\tau) Y_{1} = -2 i e^{i
\omega(\tau) t} \left[ \frac{\partial(A \omega)}{\partial \tau} + i t A \omega
\frac{\partial \omega}{\partial \tau} \right] + {\rm{c.c.}}\,.
\label{1st-order-eq-WKB-subed}
\ee
To eliminate secularity, we would want to set the term inside the square 
brackets on the
right-hand side of Eq.~\eqref{1st-order-eq-WKB-subed} 
to zero, but due to the explicit appearance of $t$, this would force $A = 0$.
Multiple scale analysis fails if the long time scale
is proportional to the short time scale {\emph{and}} the frequency of
oscillation is not a constant. 

We can force the frequency to be constant by changing variables to $T=f(t) =
f(\tau/\epsilon)$, which then transforms Eq.~\eqref{oscillator-WKB-eq} into
\be
\frac{d^{2}y}{dT^{2}}  + \frac{f''(t)}{[f'(t)]^{2}} \frac{dy}{dT} +
\frac{\omega^{2}(\epsilon t)}{[f'(t)]^{2}} y = 0\,,
\ee
We can force the frequency oscillation to be constant by choosing
\be
T = f(t) = \int^{t} \omega(\epsilon s) ds = \frac{1}{\epsilon} \int^{\tau}
\omega(s) ds\,, 
\ee
which then leads to 
\be
\frac{d^{2}y}{dT^{2}}  + y + \epsilon \frac{\omega'(\tau)}{\omega^{2}(\tau)}
\frac{dy}{dT} =0\,. 
\ee
Now, this equation can be solved via multiple-scale analysis. Using the 
expansion in
Eq.~\eqref{multiple-scale-expansion}, 
the above equation leads to
\begin{align}
\frac{\partial^{2} Y_{0}}{\partial T^{2}} + Y_{0} &= 0\,,
\\
\frac{\partial^{2} Y_{1}}{\partial T^{2}} + Y_{1} &= -
\frac{\omega'(\tau)}{\omega^{2}(\tau)} \frac{\partial Y_{0}}{\partial T} -
\frac{2}{\omega} \frac{\partial^{2} Y_{0}}{\partial \tau \partial T}\,. 
\label{1st-order-eq-WKB-new}
\end{align}
The solution to the zeroth-order in $\epsilon$ is the same as that of the
non-linear oscillator, and with this, 
the first-order in $\epsilon$ equation becomes
\be
\frac{\partial^{2} Y_{1}}{\partial T^{2}} + Y_{1} = -i e^{iT} \left[
\frac{2}{\omega} \frac{\partial A}{\partial \tau} +
\frac{\omega'(\tau)}{\omega^{2}(\tau)} A \right] + {\rm{c.c.}}\,.
\label{1st-order-eq-WKB-new-subed}
\ee
This time we can eliminate the secularly growing terms by requiring that the
terms inside square brackets in Eq.~\eqref{1st-order-eq-WKB-new-subed} vanish, 
which leads to a partial
differential equation for $A(\tau)$, whose solution is
$A(\tau) = A_{0}/\sqrt{\omega(\tau)}$. The full solution is then 
\be
y(t) = \frac{A_0}{\sqrt{\omega(\epsilon t)}} e^{\frac{i}{\epsilon}
\int^{\epsilon t} \omega(s) ds} + \rm{c.c.}\,,
\ee
which is the same as what one would have obtained through the WKB 
physical-optics
approximation. We see then that multiple-scale analysis is a generic and 
powerful technique that, in certain cases,
allows us to recover the WKB approximation, among others. Of course, The above 
examples employed only 2
scales, but multiple scale analysis is in principle valid given an arbitrary 
number of scales provided they satisfy a
certain scale hierarchy.

\subsection{The Stationary Phase Approximation and Uniform Asymptotic 
Expansions}
\label{sec:uaa}

The leading-order gravitational wave signal from a quasi-circular binary 
inspiral can be expressed in the form
\begin{equation}\label{ht}
h(t) = A(t) e^{-i\Phi(t)}
\end{equation}
where the amplitude $A(t)$ and the phase $\Phi(t)$ are slowly evolving 
functions 
of time. The full signal is given by
a sum of such terms that form a harmonic series in the orbital frequency. The 
function $h(t)$ oscillates on the orbital timescale,
with an amplitude and frequency that evolve on the slower spin-precession and 
radiation-reaction timescales. 

In gravitational wave data analysis, quantities of interest (such as the 
likelihood function) are usually calculated in the frequency domain, where the 
noise-autocorrelation function is assumed to take a simple form. Thus, we are 
faced with the task of Fourier transforming the waveform in Eq.~(\ref{ht}): 
\begin{equation}\label{hf}
\tilde{h}(f) = \int \, A(t) e^{-i\Phi(t)} e^{2\pi i f t} \, dt =  \int \, A(t) 
e^{i\phi(f,t)} \, dt\, .
\end{equation}
 A direct numerical implementation using a fast Fourier transform algorithm is 
possible, 
but the computational cost
can be high since the waveform needs to be sampled at a cadence set by the 
orbital period. 

The quadratic SPA is the standard analytic approach to solve Eq.~\eqref{hf}. At 
a given frequency $f$, the integral is dominated by the contributions where the 
phase $\phi(f,t)$ is a slowly-varying function of time. Away from this region, 
the integrand oscillates rapidly and contributes little. Defining the 
stationary 
phase points implicitly as the times $t_\SPA$ where $\dot\phi(f,t_\SPA) = 0$, 
or 
equivalently, 
\begin{equation}\label{spp}
\dot\Phi(t_\SPA) = 2 \pi f \,,
\end{equation}
the Fourier phase can be expanded as
\begin{align}\label{spx}
\phi(f,t) = \phi(f,t_\SPA) + \frac{1}{2}\ddot\phi(f,t_\SPA) (t-t_\SPA)^2 + 
\ldots 
\end{align}
Given such an expansion, one can then analytically solve the generalized 
Fourier 
integral in Eq.~\eqref{hf} through a change of variables~\cite{Bender}
\begin{align}
\tilde{h}_{{\rm SPA}_2}(f) &=  \left[ \frac{2}{\vert \ddot \Phi(t_\SPA) 
\vert}\right]^{1/2} A(t_\SPA) \; \Gamma(1/2) 
\nonumber \\
& e^{i[2\pi f t_\SPA -  \Phi(t_\SPA)-\sigma \pi/4]} \,,
\label{sspa}
\end{align}
where $\sigma = {\rm sign} (\ddot \Phi(t_\SPA))$, $\Gamma(\cdot)$ is the Gamma 
function and $t_\SPA(f)$ is understood as a function of frequency.  

Several assumptions go into the solution of Eq.~\eqref{sspa}, which have been 
implicitly taken for granted in gravitational wave modeling. First, one assumes 
that there is a unique stationary phase time $t_\SPA$ for a given frequency 
$f$, 
so that the time-frequency mapping $t_\SPA(f)$ is single valued for each 
harmonic. Second, one assumes that the expansion for the phase about the 
stationary point in Eq.~(\ref{spx}) can be truncated at quadratic order, and 
that the amplitude $A(t)$ can be replaced by the constant value $A(t_\SPA)$. 
When the mapping in Eq.~(\ref{spp}) between frequency and time yields multiple 
stationary points, the full solution is given by summing up the contributions 
of 
the form (\ref{sspa}) for all the stationary points. But when this mapping is 
not single valued, the SPA can lead to divergent results, i.e.~$\ddot 
\Phi(t_\SPA)$ can vanish and the amplitude can diverge.

The goal of {\em uniform asymptotic expansions} is to replace non-uniform 
expansions, like that of Eq.~\eqref{sspa}, by a new expansion that remains 
valid 
in a domain containing the singular point. A standard example is the Airy 
function uniformization of a fold catastrophe~\cite{Berry1980257}, which occurs 
when two stationary points coalesce and $\ddot\phi(f,t_\SPA)=\ddot \Phi(t_\SPA) 
=0$. At these catastrophe points, the stationary point is defined by the 
last equation and the Taylor expansion of the phase in Eq.~(\ref{spx}) has to 
be 
continued to higher order. At cubic order, the integral in Eq.~(\ref{hf}) 
yields 
an Airy function, and the cubic SPA is
\begin{align}
\tilde{h}_{{\rm SPA}_3}(f) &= \left[\frac{2}{\vert \dddot \Phi(t_\SPA)\vert} 
\right]^{1/3} A(t_\SPA)
\nonumber \\
&2 \pi  {\rm Ai}\left\{-\sigma [2 \pi f - \dot{\Phi}(t_\SPA)] 
\left[\frac{2}{\vert \dddot 
\Phi(t_\SPA)\vert} \right]^{1/3} \right\} 
\nonumber \\
& e^{i[2\pi f t_\SPA - \Phi(t_\SPA)]} \, ,
\label{sspa3}
\end{align}
where $\sigma = {\rm sign} [\dddot \Phi(t_\SPA)]$, and the amplitude and phase 
are evaluated at the singular point $t_\SPA$.
The expression in Eq.~(\ref{sspa3}) matches the numerical Fourier transform for 
frequencies $f$ in the neighborhood of the critical point 
$f=\dot{\Phi}(t_\SPA)/2\pi$ where 
the phase is well approximated by a cubic Taylor expansion. In many instances, 
there is an overlap region where both approximations [Eqs.~(\ref{sspa}) 
and~(\ref{sspa3})] are valid simultaneously. In such cases, it is possible to 
construct a complete SPA waveform from a piecewise collection of the quadratic 
and cubic SPAs.

A completely different uniformization is required, however, for situations 
where 
the singular points become so dense that the Airy function and related 
techniques breakdown. A good example is when the phase has an oscillatory 
component, which is exactly the situation for precessing black hole binaries. 
One solution is to re-express the original waveform as the sum of simpler 
waveforms that each have a well-behaved SPA~\cite{PhysRevE.64.026215}. For 
example, if the GW phase can be written as the sum of a carrier phase and an 
oscillatory component:
\begin{equation}
\Phi(t) = \Phi_\C(t) + \alpha(t) \cos \beta(t) \label{eq:correctionsimple}
\end{equation}
where $\Phi_\C(t)$, $\alpha(t)$ and $\beta(t)$ are monotonic functions of time, 
then 
\begin{equation}\label{htx}
h(t) = A(t) e^{-i\Phi_\C(t)} \sum_{n=-\infty}^{\infty} (-i)^n J_n(\alpha(t)) 
e^{-i n \beta(t)}
\end{equation}
and
\begin{eqnarray}\label{SPAx}
&& \tilde{h}_{\rm SPA}(f) = \sum_{n=-\infty}^{\infty} \left[ \frac{2 \pi}{\vert
 \ddot\Phi_C(t_n)+ n \ddot \beta(t_n) \vert}\right]^{1/2}A(t_n)  \nonumber \\
&& \times (-i)^n J_n[\alpha(t_n)] e^{i[2 \pi f t_n - \Phi_C(t_n)- n 
\beta(t_n)-\sigma  \pi/4]} \,,
\end{eqnarray}
where $\sigma = {\rm sign} [\ddot \Phi_C(t_n)+ n \ddot \beta(t_n)]$.  Notice 
that there are now $n$ different stationary points $t_n$, defined by the 
stationary phase condition $2 \pi f = \dot\Phi_C(t_n)+ n \dot \beta(t_n)$. In 
Eq.~(\ref{SPAx}), we have assumed that the individual contributions are 
non-singular: $\ddot\Phi_C(t_n)+ n \ddot \beta(t_n)\neq 0$. If a singularity 
does occur in any of the terms, then the standard SPA for this term can be 
replaced by the  Airy uniformization of Eq.~(\ref{sspa3}). The rapid decay of 
the Bessel functions with increasing order $\vert n\vert$ means that just a few 
terms are needed in the sum of Eq.~(\ref{SPAx}) to obtain a good approximation 
to the full Fourier transform.

\begin{figure}[ht]
\begin{center}
\includegraphics[width=\columnwidth]{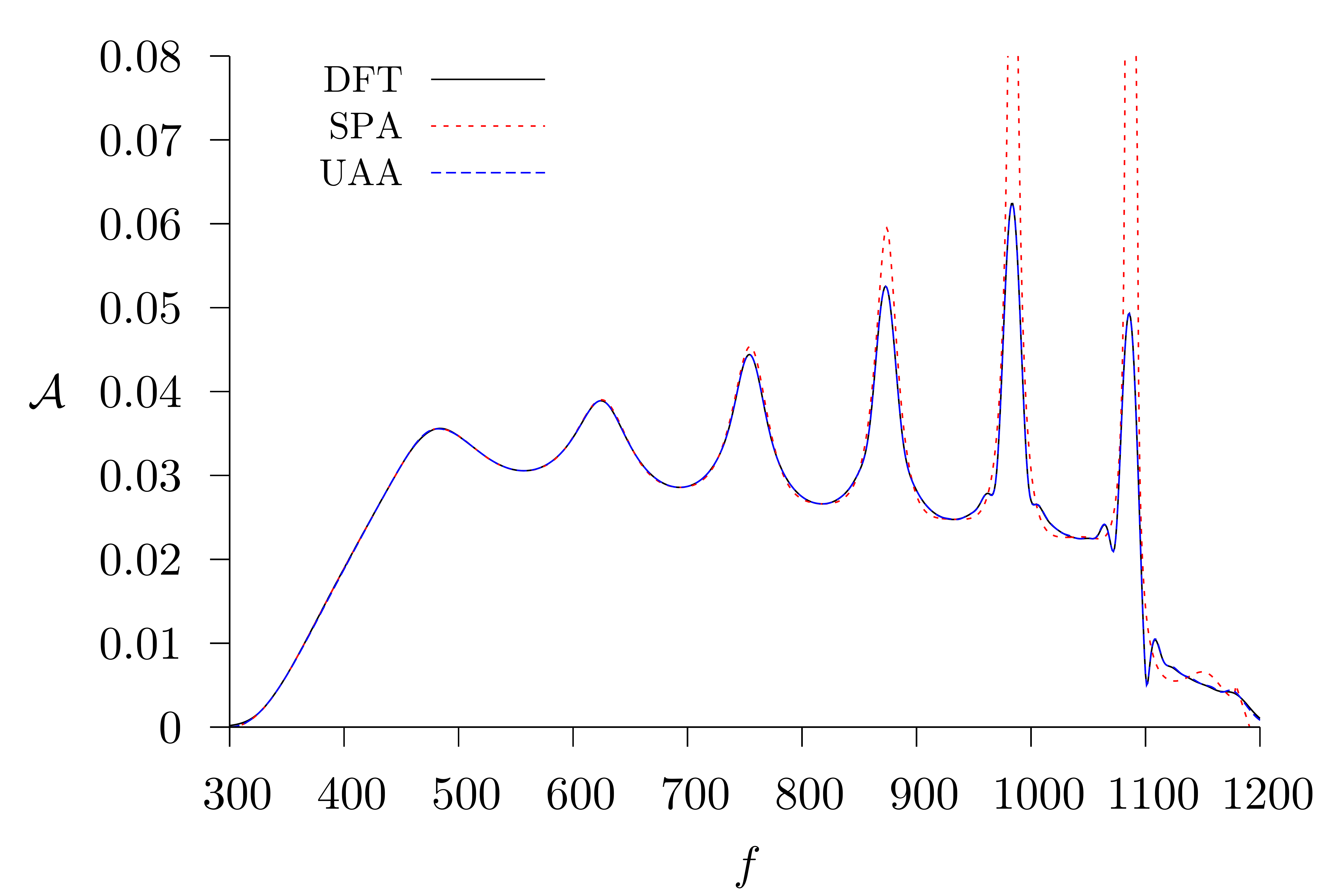}
\caption{\label{fig:SPA} The amplitude of the Fourier transform for the toy 
model waveform computed using a numerical fast Fourier transform (solid, black 
line);
the standard quadratic SPA (dotted, red line); and the Bessel function 
uniformization of the SPA (dashed, blue line).  The numerical transform and the 
uniform asymptotic
expansion are indistinguishable, while the quadratic SPA diverges at turning 
points of the frequency. }
\end{center}
\end{figure}

To illustrate the Bessel uniformization approach, let us consider a simple toy 
model that shares many of the features of the waveforms produced by spinning 
black hole binaries.  Let us then consider the phase given by
\begin{eqnarray}
&&\Phi(t) = 2 \pi \left[ f_0 (t/T) + \frac{1}{2} \dot f_0( t/T)^2 \right. 
\nonumber  \\
&& \quad \left.+ \alpha_0 (t/T) \cos(\omega_0 (t/T)+ \frac{1}{2} \dot\omega_0 
(t/T)^2)\right] \, ,
\end{eqnarray}
and an amplitude given by a Tukey tapered cosine window 
\begin{equation}
A(t) = \begin{cases} 
\frac{1}{2}\left[1+\cos\left(\pi\left(\frac{t}{T\kappa}-1\right)\right)\right] 
, 
& t \leq \kappa T \\ 
 1, &  \kappa T  < t < (1- \kappa)T \\
  
\frac{1}{2}\left[1+\cos\left(\pi\left(\frac{t-T}{T\kappa}+1\right)\right)\right]
 
, & t \geq  (1- \kappa)T \, .
\end{cases}
\end{equation}
The Tukey window helps suppress spectral leakage in the numerical Fourier 
transform. Fig.~\ref{fig:SPA} shows the
amplitude of the Fourier transform computed three different ways: (i) using a 
numerical DFT; (ii)
using the quadratic SPA; and (iii) using the Bessel function uniformization of 
the SPA summing up to $\vert n \vert = 5$. The parameters chosen were $\{T=1, 
f_0 = 300, \dot f_0 = 900, \omega_0 = 30, \dot\omega_0 = 30, \alpha_0 = 0.4,  
\kappa = 0.2\}$. The uniform asymptotic expansion
provides a near perfect match to the numerical Fourier transform, while the 
quadratic SPA diverges at turning points of the frequency.

\section{Spin and Orbital Angular Momentum}
\label{sec:near-alignment}

In this section, we explore the evolution equations of the spin and orbital
angular momentum vectors 
using techniques from multiple-scale analysis. We first develop the
formalism of multiple-scale analysis
as applicable to inspiraling binaries, and then apply it to systems where the
spin angular momentum vectors are nearly aligned 
with the orbital angular momentum vector. Physically, this corresponds to the
inspiral of binary BHs or binary NSs in a gas-rich 
environment, where the latter tends to align the spin and orbital angular
momenta. 


Such a system allows us to make several approximations that enable a 
perturbative analytic solution. First, we expand all quantities in the 
misalignment angle
\begin{align}
\bm{K} = \sum_{n>0} \bm{K}^{(n)}(t) \; \epsilon^{n}\,
\end{align}
for any vector $\bm{K}$, where $\bm{K}^{(n)}(t)$ are undetermined vector 
functions and $\epsilon \ll 1$ is the misalignment order-counting parameter. 
Second, we re-expand all quantities in a separation of timescales
\begin{align}
\bm{K} = \sum_{n,m>0} \bm{K}^{(n,m)}(t) \; \epsilon^{n} \sigma^{m}\,,
\end{align}
where $\bm{K}^{(n,m)}(t) $ are undetermined vector function and we have defined 
the precession order counting parameter
\be
\sigma \equiv \frac{t_{\precc}}{t_{\rr}} \ll 1\,,
\ee
with $t_{\precc}$ and $t_{\rr}$ the precession and radiation-reaction 
timescales. This last expansion is justified for binaries in the PN 
(slow-motion/weak-gravity) regime, where all three characteristic timescales of 
the problem separate. The precession order counting parameter $\sigma$ and the 
PN one $c$ are not independent, but rather ${\cal{O}}(\sigma) = 
{\cal{O}}(c^{-3})$. Henceforth, a term of ${\cal{O}}(c^{-2A})$ will be said to 
be of NPN order.
 
\subsection{Precession Equations}

The spin and orbital angular momentum precession equations for an compact 
binary system 
in a quasi-circular inspiral in the center of mass frame can be written 
as\footnote{These equations correct an error in Eq.~$(2.4)$
of~\cite{Kidder:1995zr}, but they are consistent with Eq.$(4.17)$ in that 
paper, 
as well as with equations in~\cite{Apostolatos:1994mx} 
and~\cite{Buonanno:2002ft}.}
\begin{align}
\dvec{S}_1 &= \Omega_{LS_{1}} \uvec{L} \times \bm{S}_1  + 
\Omega_{S_{1}S_{2}}  \bm{S}_2 \times \bm{S}_1, 
\\
\dvec{S}_2 &= \Omega_{LS_{2}} \uvec{L} \times \bm{S}_2  + 
\Omega_{S_{1}S_{2}}  \bm{S}_1 \times \bm{S}_2, 
\\
\duvec{L} &= \frac{\Omega_{LS_{1}}}{L} \bm{S}_1 \times \uvec{L} +
\frac{\Omega_{LS_{2}}}{L} \bm{S}_2  \times \uvec{L}\,, 
\label{eq:Lhatdot} \\
\dot{L} &= - \frac{1}{3} \frac{a_0}{M} (M\omega)^{8/3} \left[ 1 + \sum_{n=2}^N
a_n
(M\omega)^{n/3}
\right] L , \label{eq:Ldot}
\end{align}
where we have defined
\begin{align}
\Omega_{LS_{1}} &= \frac{\omega^2}{M} \left[\left( 2 + \frac{3m_2}{2m_1} 
\right) 
L -  \frac{3}{2} \left( \uvec{L} \cdot \bm{S}_2 \right) \right]\,,
\\
\Omega_{LS_{2}} &= \frac{\omega^2}{M} \left[\left( 2 + \frac{3m_1}{2m_2} 
\right) 
L -  \frac{3}{2} \left( \uvec{L} \cdot \bm{S}_1 \right) \right]\,,
\\
\Omega_{S_{1}S_{2}} &= \frac{1}{2} \frac{\omega^2}{M}\,,
\end{align}
where $m_{1,2}$ are the component masses, $\omega = M^2(\mu/L)^3$ is the 
orbital 
frequency of the system, with $L$ the magnitude of the {\emph{Newtonian}} 
orbital angular momentum $\bm{L}$.  
The spin angular momentum of the Ath binary component is $\bm{S}_{A}$, while 
$\hat{\bm{L}} = \bm{L}/L$ is an orbital angular momentum unit vector.  All 
cross- 
and dot-products represent the standard (Euclidean) three-dimensional 
operations 
of vector calculus. The quantities $a_i$ are functions of the symmetric mass 
ratio as well as $(\bm{S}_{A} \cdot \hat{\bm{L}})$ and $(\bm{S}_{A} \cdot 
\bm{S}_{B})$, and they are explicitly given in Appendix~\ref{app:freqevol}.

The evolution equations presented above are in principle valid to different PN 
orders. The evolution equation for $L$ (or equivalently for $\omega$) is valid 
to $N/2$ PN order, where here we choose $N = 5$, i.e.~we model the evolution of 
$L$ to $2.5$PN order. The evolution equations for $\bm{S}_{A}$ and
$\hat{\bm{L}}$, however, are only valid to first-subleading PN order, 
i.e.~leading order in spin-orbit (1.5PN)
and spin-spin (2PN) coupling. Since spin enters at $1.5$PN order in the phase, 
the $2.5$PN order corrections that we are leaving out in the evolution of 
$\bm{S}_{A}$ would contribute at $4$PN order. We are then allowed to use 
Newtonian expressions to map between $L$ and $\omega$, and the norm of 
$\bm{S}_{A}$ is conserved. Of course, one could extend the analysis in this 
paper by adding corrections to $\Omega_{LS_{1,2}}$ and $\Omega_{S_{1}S_{2}}$, 
thus including sub-leading PN corrections to the evolution of $\bm{S}_{A}$. But 
to be consistent in PN order counting, one would also have to include 4PN 
corrections to the evolution of $L$, which are currently unknown. 

Let us now pick a frame and implement the near-alignment approximation. We
choose $\uvec{z} =
\uvec{J}(t=0)$, where $\bm{J} = \bm{L} + \bm{S}_1 + \bm{S}_2$ is the total
angular momentum. 
Since $\bm{J}$ is not constant, we have to specify it at a given time for the
frame to be inertial.
In this frame, we can write 
\begin{align}
 \bm{K} &= K_z \uvec{z} +  K_x \uvec{x} + K_y \uvec{y} ,
\end{align}
for any $\bm{K} = \bm{L}$, $\bm{S}_1$, or $\bm{S}_2$. In the near-alignment
approximation, the
components of $\bm{K}$ in this frame can then be expanded as
\begin{align}
 K_z &= \sum_{n \geq 0} K_{z}^{(2n)}(t) \;\epsilon^{2n}, \label{eq:defKz}\\
 K_x &= \sum_{n \geq 0} K_{x}^{(2n+1)}(t) \; \epsilon^{2n+1},
\label{eq:defKx}\\
 K_y &= \sum_{n \geq 0} K_{y}^{(2n+1)}(t) \; \epsilon^{2n+1}.\label{eq:defKy}
\end{align} 
The structure of the equations of motion ensures that odd-powers of $\epsilon$
in $K_z$, as well as the even-powers of $\epsilon$ in $K_j$, $j = x$ or $y$, 
vanish. In
this paper, we will take these sums only up to $\mathcal{O}(\epsilon)$, but 
extending
these results to higher-order is straightforward. 

Before proceeding, let us make an important comment on the near-alignment 
approximation. Consider a binary system at early times, where the misalignment 
angle is $\epsilon_{0}$. As the binary evolves, the norm of the orbital angular 
momentum $L$ shrinks by radiation-reaction. But since the norm of the spin 
angular momentum $S_{A}$ is conserved, the misalignment angle 
will grow. This implies that our perturbation parameter is not a constant, but 
rather an increasing function of time. Thus, just like the PN approximation is 
expected to break in the late stages of inspiral because the orbital velocity 
increases, the mis-alignment approximation is also expected to break as 
$\epsilon$ increases and the series becomes asymptotic.  

\subsection{Analysis to ${\cal{O}}(\epsilon^{0})$}

Let us first focus on the precession equations to leading-order in $\epsilon$. 
The $x$- and $y$-components of these equations are trivially satisfied. 
The $z$-component of the spin angular momentum equation requires that 
$S_{A,z}^{(0)}$ be a constant, which can be obtained by demanding that 
$|\bm{S}_{A}| = m_{A}^{2} \chi_{A}$, and thus
\begin{align}
 |\bm{S}_A| = S_{A,z}^{(0)} + {\cal{O}}(\epsilon^{2}) = m_A^2 \chi_A\,.
 \end{align}
We can use this property to solve for $S_{A,z}^{(2n)}$ at all orders, given the 
lower order solutions for $S_{A,j}^{(2n-1)}$, $j=x$ or $y$.

The $z$-component of the orbital angular momentum evolution equation requires a
bit more work. 
First, let us define the quantity
\be
\xi_0 \equiv \frac{\mu M}{L_z^{(0)}} = {\cal{O}}(c^{-1})\,,
\ee
as a new PN quantity. This parameter is exactly the square-root of the frequency
parameter often used in the literature $x = (M\omega)^{2/3} = \mu^2 M^2 L^{-2}$
only when the spins and orbital
angular momentum are aligned. Using this parameter, we can rewrite the
${\cal{O}}(\epsilon^{0})$ part 
of the evolution equation for the $z$-component of orbital angular momentum as
\begin{align}
 \dot{\xi}_0 =  \frac{1}{3} \frac{a_0}{M} \xi_0^9
\left( 1 + \sum_{n=2}^N a_n \xi_0^n \right),
\label{eq:xi0dot}
\end{align}
where the spin-dependent part of the couplings were evaluated at leading-order
in $\epsilon$:
\begin{align}
 \uvec{L}^{(0)} &= (0,0,1), \\
 \bm{S}_A^{(0)} &= \left(0,0,m_A^2 \chi_A \right) .
\end{align}

Since all coefficients are constants, we can directly integrate
Eq.~\eqref{eq:xi0dot} 
by Taylor expanding $(\dot{\xi}_0)^{-1}$. After inverting the PN
series and integrating, we obtain
\begin{align}
 \xi_0(t) &= \zeta \bigg[ 1 - \frac{a_2}{6} \zeta^2 - \frac{a_3^{(0)}}{5}
 \zeta^3 
\nonumber\\
&+ \frac{5a_2^2 -6a_4^{(0)}}{24} \zeta^4 + \frac{9 a_2 a_3^{(0)} - 5
a_5^{(0)}}{15} \zeta^5 + \mathcal{O}\left(c^{-6}\right) \bigg]\,, 
\label{eq:xi0ofzeta}
\end{align}
with
\begin{align}
\zeta &= \left[ \frac{3M}{8 a_0 (t_\coal - t)} \right]^{1/8}\,.
\label{eq:zeta}
\end{align}
The notation $a_{i}^{(n)}$ means the part of $a_{i}$ at 
${\cal{O}}(\epsilon^{n})$, 
where recall that $a_{2}$ is a 1PN correction that is spin-independent. 
We have here kept terms up to $2.5$PN order, while the spin-dependent
couplings are included to leading-order in $\epsilon$.

For convenience, let us also introduce a new quantity $\xi$, which coincides
with $\xi_0$
at ${\cal{O}}(\epsilon^{0})$ but differs at higher orders. This quantity is
defined via 
\begin{align}
 \xi(t) &= \zeta \bigg[ 1 - \frac{a_2}{6} \zeta^2 - \frac{a_3(t=0)}{5}
\zeta^3 + \frac{5a_2^2 -6a_4(t=0)}{24} \zeta^4\nonumber\\
& + \frac{9 a_2 a_3(t=0) - 5
a_5(t=0)}{15} \zeta^5 + \mathcal{O}\left(c^{-6}\right) \bigg].
\label{eq:xiofzeta}
\end{align}
which extends Eq.~\eqref{eq:xi0ofzeta} by using the full $a_{i}$ coefficients,
evaluated at $t=0$,
instead of their ${\cal{O}}(\epsilon^{0})$ parts $a_{i}^{(0)}$. Therefore, the
difference between $\xi$ and 
$\xi_0$ is of $\mathcal{O}\left(\epsilon^2\right)$, which implies
\begin{equation}
L_z^{(0)} = \frac{\mu M}{\xi_0} = \frac{\mu M}{\xi} +
\mathcal{O}\left(\epsilon^2\right).
\end{equation}
With such a definition, $\xi$ coincides with $x^{1/2} = (M\omega)^{1/3}$ when
the scalar products between 
$\uvec{L}$, $\bm{S}_1$, and $\bm{S}_2$ are time-independent. That is, in the
near-alignment approximation, we can write
\begin{align}
 \xi &= (M\omega)^{1/3} + \mathcal{O}(\epsilon^2).
\end{align}
Henceforth, we will use $\xi$ as our independent variable.

\subsection{Analysis to ${\cal{O}}(\epsilon^{1})$}

Let us now look at the evolution equations to first-order in $\epsilon$.
We can write them in matrix notation as
\begin{align}
 \frac{d \bm{W}_1^{(1)}}{dt} &= 
 -{\mathbb{M}} \, \bm{W}_2^{(1)} - a {\mathbb{A}} \, \bm{W}_1^{(1)}, 
\nonumber \\
 \frac{d \bm{W}_2^{(1)}}{dt} &= 
{\mathbb{M}} \, \bm{W}_1^{(1)} - a {\mathbb{A}} \, \bm{W}_2^{(1)}, 
\label{eq:matrix}
\end{align}
where we have defined the vectors
\begin{align}
 \bm{W}_1^{(1)} &= \left( \begin{array}{c}
L_x^{(1)}\\ S_{1,x}^{(1)}\\ S_{2,x}^{(1)}
                          \end{array} \right),  \qquad
 \bm{W}_2^{(1)} = \left( \begin{array}{c}
L_y^{(1)}\\ S_{1,y}^{(1)}\\ S_{2,y}^{(1)}
                          \end{array} \right),
\end{align}
and the matrices 
\begin{align}   
{\mathbb{M}} &=  \left( \begin{array}{c c c}
         (b+c) & -d & -e \\
         -b & (d+f) & -g \\
         -c & -f & (e+g)
        \end{array}
        \right), \quad
{\mathbb{A}} = \left( \begin{array}{c c c}
         1 & 0 & 0 \\
         0 & 0 & 0 \\
         0 & 0 & 0
        \end{array}
        \right)\,.
\end{align}
with
\begin{align}
 a &= \frac{1}{3} \frac{a_0}{M} \xi(t)^8
\left( 1 + \sum_{n=2}^N a_n \xi(t)^n \right),
\label{eq:a} \\
b &= \frac{\xi(t)^6}{M} \bigg[
\left( 2 + \frac{3m_2}{2m_1} \right) \frac{m_1^2}{M^2} \chi_1 - 
\frac{3}{2} \xi(t) \nu \chi_1 \chi_2
 \bigg] , \\
c &= \frac{\xi(t)^6}{M} \bigg[
\left( 2 + \frac{3m_1}{2m_2} \right) \frac{m_2^2}{M^2} \chi_2 - 
\frac{3}{2} \xi(t) \nu \chi_1 \chi_2
 \bigg] , \\
d &= \frac{\xi(t)^5}{M} \bigg[
\left( 2 + \frac{3m_2}{2m_1} \right) \nu - 
\frac{3}{2} \xi(t) \frac{m_2^2}{M^2} \chi_2
 \bigg] , \\
e &= \frac{\xi(t)^5}{M} \bigg[
\left( 2 + \frac{3m_1}{2m_2} \right) \nu - 
\frac{3}{2} \xi(t) \frac{m_1^2}{M^2} \chi_1
 \bigg] , \\
f &= \frac{1}{2} \frac{\xi(t)^6}{M} \frac{m_2^2}{M^2} \chi_2 , \qquad
g = \frac{1}{2} \frac{\xi(t)^6}{M} \frac{m_1^2}{M^2} \chi_1.
\end{align}

The solution to the system in Eq.~\eqref{eq:matrix} can be obtained via a
standard linear algebra approach. 
First, we diagonalize $\mathbb{M}$ via a similarity transformation in matrix
${\mathbb{R}}$, ${\mathbb{R}}^{-1} {\mathbb{M}} {\mathbb{R}} = {\mathbb{D}}$,
where ${\mathbb{D}}$ is
the diagonal matrix 
\begin{align}
{\mathbb{D}} &= \left(\begin{array}{c c c}
                    0 & 0 & 0\\
                    0 & \omega_{\PP,+} & 0\\
                    0 & 0 & \omega_{\PP,-}
                   \end{array}
\right)\,,
\end{align}
with eigenvalues
\begin{multline}
\omega_{\PP,\pm} = \frac{1}{2} \Big[ b+c+d+e+f+g \\
\pm
\sqrt{(b-c+d-e+f-g)^2 + 4(c-f)(b-g)} \Big].
\end{multline}
The transformation matrix ${\mathbb{R}}$ is given explicitly in
Appendix~\ref{app:RandRm1}. The first few terms of the PN expansion of
$\omega_{\PP,\pm}$ are
\begin{align}
 \omega_{\PP,+} &= \frac{1}{M}\left( 2 \nu + \frac{3}{2} \frac{m_1^2}{M^2} 
\right)
\xi^5 \nonumber\\
 &+ \frac{1}{M}\left[
\left( 2 \frac{m_2^2}{M^2} + \frac{3}{2} \nu \right) \chi_2 - \frac{m_1^2}{M^2}
\chi_1 \right] \xi^6 +
\mathcal{O} \left( c^{-7} \right), \\
 \omega_{\PP,-} &= \frac{1}{M}\left( 2 \nu + \frac{3}{2} \frac{m_2^2}{M^2}
\right) \xi^5 \nonumber\\
 &+ \frac{1}{M}\left[
\left( 2 \frac{m_1^2}{M^2} + \frac{3}{2} \nu \right) \chi_1 - \frac{m_2^2}{M^2}
\chi_2 \right] \xi^6 +
\mathcal{O} \left( c^{-7} \right).
\end{align}

With this at hand, 
Eq.~\eqref{eq:matrix} can be transformed into 
\begin{align}
 \frac{d\bm{Q}_1^{(1)}}{dt} &= - {\mathbb{D}} \, \bm{Q}_2^{(1)} - {\mathbb{E}}
\, \bm{Q}_1^{(1)}, \label{normal-mode-eq-1}\\
 \frac{d\bm{Q}_2^{(1)}}{dt} &= {\mathbb{D}} \, \bm{Q}_1^{(1)} - {\mathbb{E}} \, 
\bm{Q}_2^{(1)}, 
 \label{normal-mode-eq}
\end{align}
where we have defined the transformed $\bm{W}_{i}$, i.e.~the eigenvectors or
quasi-normal modes, via 
\begin{align}
 \bm{Q}_j^{(1)} &\equiv {\mathbb{R}}^{-1} \bm{W}_j^{(1)} =
\left(\begin{array}{c}
                    Q_{0,j}^{(1)}\\
                    Q_{+,j}^{(1)}\\
                    Q_{-,j}^{(1)}
                   \end{array}
\right)\,,
\label{eq:1st-o-system}
\end{align}
with $j = 1$ or $2$, and where the remainder matrix 
\begin{align}
{\mathbb{E}} &= {\mathbb{R}}^{-1} \left(a {\mathbb{A}} {\mathbb{R}} +
\frac{d{\mathbb{R}}}{dt} \right)\,.
\end{align}
The second term in the above equation is necessary because the rotation matrix
${\mathbb{R}}$ is not constant. 
We chose to normalize the eigenvectors such that
\begin{align}
 L^{(1)}_x &= Q^{(1)}_{0,1} + Q^{(1)}_{+,1} + Q^{(1)}_{-,1}, \\
 L^{(1)}_y &= Q^{(1)}_{0,2} + Q^{(1)}_{+,2} + Q^{(1)}_{-,2}\,,
\end{align}
as explained in Appendix~\ref{app:RandRm1}.

We decouple the system in Eq.~\eqref{normal-mode-eq}
by taking an extra time-derivative to obtain
\begin{align}
 \frac{d^2\bm{Q}_1^{(1)}}{dt^2} &= - {\mathbb{D}}^2 \bm{Q}_1^{(1)} \nonumber\\
 &+ \left( \{{\mathbb{D}},{\mathbb{E}}\} -
\frac{d{\mathbb{D}}}{dt}
\right) \bm{Q}_2 + \left( {\mathbb{E}}^2 - \frac{d{\mathbb{E}}}{dt} \right)
\bm{Q}_1^{(1)},
\label{normal-mode-eq-decomp1} \\
 \frac{d^2\bm{Q}_2^{(1)}}{dt^2} &= - {\mathbb{D}}^2 \bm{Q}_2^{(1)} \nonumber\\
 &- \left( \{{\mathbb{D}},{\mathbb{E}}\} -
\frac{d{\mathbb{D}}}{dt}
\right) \bm{Q}_1^{(1)} + \left( {\mathbb{E}}^2 - \frac{d{\mathbb{E}}}{dt}
\right) \bm{Q}_2^{(1)},
\label{normal-mode-eq-decomp2}
\end{align}
where $\{{\mathbb{A}},{\mathbb{B}}\} \equiv {\mathbb{A}} {\mathbb{B}} +
{\mathbb{B}} {\mathbb{A}}$ denotes the matrix anticommutator. In what follows,
we solve this system of equations via multiple scale analysis. 

\subsubsection{Separation of Scales}

Inspection of Eqs.~\eqref{normal-mode-eq-decomp1}
and~\eqref{normal-mode-eq-decomp2} reveals that this is 
a system of perturbed harmonic oscillators with time-dependent frequencies.
Thus, the multiple scale analysis 
methods presented in Sec.~\ref{sec:msa} are well-suited to solve this problem.
As explained in that section, however,
we must first transform to a new independent variable such that the problem
becomes that of a system of perturbed
harmonic oscillators with constant frequencies. 

Because the eigenvalues of the $\mathbb{D}$ matrix are generically not the
same, 
i.e.~$\omega_{\PP,+} \neq \omega_{\PP,-}$, we are forced to introduce a 
different
transformation
for $+$ and $-$ modes. Changing variables $t \to \phi_{\PP,\pm}(t)$ in the
$Q_{\pm,1}^{(1)}$ equation,
Eq.~\eqref{normal-mode-eq-decomp1} becomes
\begin{align}
 \frac{dQ_{0,1}^{(1)}}{dt} &= - \left[ {\mathbb{E}} \, \bm{Q}_1^{(1)}
\right]\cdot \bm{P}_0
, \\
\frac{d^2 Q_{+,1}^{(1)}}{d\phi_{\PP,+}^2} &= -\left( \frac{dt}{d\phi_{\PP,+}}
\right)^2
\omega_{\PP,+}^2
Q_{+,1}^{(1)} \nonumber\\
&- \left( \frac{dt}{d\phi_{\PP,+}} \right)^2
\frac{d^2\phi_{\PP,+}}{dt^2}
\frac{dQ_{+,1}^{(1)}}{d\phi_{\PP,+}} \nonumber\\
&+ \left( \frac{dt}{d\phi_{\PP,+}} \right)^2 \left[
\left( \{{\mathbb{D}},{\mathbb{E}}\} -
\frac{d{\mathbb{D}}}{dt}
\right) \bm{Q}_2^{(1)} \right]\cdot \bm{P}_+ \nonumber\\
&+ \left( \frac{dt}{d\phi_{\PP,+}}
\right)^2\left[ \left(
{\mathbb{E}}^2 -
\frac{d{\mathbb{E}}}{dt} \right) \bm{Q}_1^{(1)}
\right]\cdot \bm{P}_+, \\
\frac{d^2 Q_{-,1}^{(1)}}{d\phi_{\PP,-}^2} &= -\left( \frac{dt}{d\phi_{\PP,-}}
\right)^2
\omega_{\PP,-}^2
Q_{-,1}^{(1)} \nonumber\\
&- \left( \frac{dt}{d\phi_{\PP,-}} \right)^2
\frac{d^2\phi_{\PP,-}}{dt^2}
\frac{dQ_{-,1}^{(1)}}{d\phi_{\PP,-}} \nonumber\\
&+ \left( \frac{dt}{d\phi_{\PP,-}} \right)^2 \left[
\left( \{{\mathbb{D}},{\mathbb{E}}\} -
\frac{d{\mathbb{D}}}{dt}
\right) \bm{Q}_2^{(1)} \right]\cdot \bm{P}_- \nonumber\\
&+ \left( \frac{dt}{d\phi_{\PP,-}}
\right)^2\left[ \left(
{\mathbb{E}}^2 -
\frac{d{\mathbb{E}}}{dt} \right) \bm{Q}_1^{(1)}
\right]\cdot \bm{P}_-,
\end{align}
where we introduced the projectors
\begin{align}
 \bm{P}_0 = \left(\begin{array}{c}
                   1\\0\\0
                  \end{array}
 \right)\, , \quad
 \bm{P}_+ = \left(\begin{array}{c}
                   0\\1\\0
                  \end{array}
 \right)\, , \quad
 \bm{P}_- = \left(\begin{array}{c}
                   0\\0\\1
                  \end{array}
 \right)\, .
\end{align}
We obtain similar equations for $\dot{Q}_{i,2}^{(1)}$ with $i = 0$, $+$, or $-$.
Notice that we have not transformed the time coordinate for the zero-frequency
mode.  
 For the above equations
to have a constant normal frequency, we must set
\begin{align}
 \frac{d \phi_{\PP,\pm}}{dt} = \omega_{\PP,\pm},
 \label{eq-sep-var}
\end{align}
modulo a proportionality constant, which we choose to be unity so that
$\phi_{\PP,\pm}$ are exactly the precession angles.

With this rescaling of the independent variable, the differential system becomes
\begin{align}
 \frac{dQ_{0,1}^{(1)}}{dt} &= - \left[ {\mathbb{E}} \, \bm{Q}_1^{(1)}
\right]\cdot \bm{P}_0
, \\
\frac{d^2 Q_{+,1}^{(1)}}{d\phi_{\PP,+}^2} &= -
Q_{+,1}^{(1)} - \frac{\dot{\omega}_{\PP,+}}{\omega_{\PP,+}^2}
\frac{dQ_{+,1}^{(1)}}{d\phi_{\PP,+}} \nonumber\\
&+ \frac{1}{\omega_{\PP,+}^2} \left[
\left( \{{\mathbb{D}},{\mathbb{E}}\} -
\frac{d{\mathbb{D}}}{dt}
\right) \bm{Q}_2^{(1)} \right]\cdot \bm{P}_+ \nonumber\\
&+  \frac{1}{\omega_{\PP,+}^2} \left[ \left(
{\mathbb{E}}^2 -
 \frac{d{\mathbb{E}}}{dt} \right) \bm{Q}_1^{(1)}
\right]\cdot \bm{P}_+, \\
\frac{d^2 Q_{-,1}^{(1)}}{d\phi_{\PP,-}^2} &= -
Q_{-,1}^{(1)} - \frac{\dot{\omega}_{\PP,-}}{\omega_{\PP,-}^2}
\frac{dQ_{-,1}^{(1)}}{d\phi_{\PP,-}} \nonumber\\
&+ \frac{1}{\omega_{\PP,-}^2} \left[
\left( \{{\mathbb{D}},{\mathbb{E}}\} -
\frac{d{\mathbb{D}}}{dt}
\right) \bm{Q}_2^{(1)} \right]\cdot \bm{P}_- \nonumber\\
&+  \frac{1}{\omega_{\PP,-}^2} \left[ \left(
{\mathbb{E}}^2 -
 \frac{d{\mathbb{E}}}{dt} \right) \bm{Q}_1^{(1)}
\right]\cdot \bm{P}_-,
\end{align}
Note that the source to these oscillators depends on $t$, which must be in
principle solved for through 
inversion of the solution to Eq.~\eqref{eq-sep-var}. We will here leave these
expressions as implicit
functions of $\phi_{\PP,\pm}$. 

Now, let us perform an expansion of the above differential equations in powers
of $\sigma$, which we recall is a book-keeping parameters of
${\cal{O}}(t_{\precc}/t_{\rr})$. In terms of the $\xi$
variable, $\sigma$ counts the powers in $(\dot{\xi}/\xi) \omega_{\PP,\pm}^{-1} =
a/\omega_{\PP,\pm}$, since
 $\dot{\xi} = a \, \xi$. The differential equations then become
\begin{align}
 &\frac{dQ_{0,1}^{(1)}}{dt} = - \sigma a \left[ \mathbb{R}^{-1} \left(\mathbb{A}
\mathbb{R} +
\xi\pdfrac{\mathbb{R}}{\xi} \right) \bm{Q}_1^{(1)} \right]\cdot \bm{P}_0 ,
\label{eq:ddQ0}\\
&\frac{d^2 Q_{+,1}^{(1)}}{d\phi_{\PP,+}^2} = -
Q_{+,1}^{(1)} + \sigma \frac{a}{\omega_{\PP,+}^2}  \left[
 \mathbb{F} \, \bm{Q}_2^{(1)} \right]\cdot \bm{P}_+ \nonumber\\
&+ \sigma^{2} \frac{a^2}{\omega_{\PP,+}^2} \bigg\{ \bigg[ 
\frac{\xi}{\omega_{\PP,+}}\pdfrac{\omega_{\PP,+}}{\xi}\mathbb{R}^{-1} 
\nonumber\\
&\times
\left(\mathbb{A}
\mathbb{R} +
\xi\pdfrac{\mathbb{R}}{\xi} \right) + \mathbb{G} \bigg] \bm{Q}_1^{(1)}
\bigg\}\cdot \bm{P}_+, \label{eq:ddQp}\\
&\frac{d^2 Q_{-,1}^{(1)}}{d\phi_{\PP,-}^2} = -
Q_{-,1}^{(1)} + \sigma \frac{a}{\omega_{\PP,-}^2}  \left[
 \mathbb{F} \, \bm{Q}_2^{(1)} \right]\cdot \bm{P}_- \nonumber\\
&+
\sigma^{2} \frac{a^2}{\omega_{\PP,-}^2} \bigg\{ \bigg[
\frac{\xi}{\omega_{\PP,-}}\pdfrac{\omega_{\PP,-}}{\xi}\mathbb{R}^{-1}\nonumber\\
&\times
\left(\mathbb{A}
\mathbb{R} +
\xi\pdfrac{\mathbb{R}}{\xi} \right)
+ \mathbb{G} \bigg] \bm{Q}_1^{(1)}
\bigg\}\cdot \bm{P}_- ,  \label{eq:ddQm}
\end{align}
where we have defined the new matrices
\begin{align}
&\mathbb{F} = \left\{\mathbb{D},\mathbb{R}^{-1} \left(\mathbb{A} \mathbb{R} +
\xi\pdfrac{\mathbb{R}}{\xi} \right)\right\} , \\
&\mathbb{G} = \mathbb{R}^{-1} \bigg[ \mathbb{A}\mathbb{R} + \xi \mathbb{A}
\pdfrac{\mathbb{R}}{\xi} + \xi \pdfrac{\mathbb{R}}{\xi}
\mathbb{R}^{-1}\mathbb{A}\mathbb{R}
\nonumber\\
&+
\xi^2 \pdfrac{\mathbb{R}}{\xi} \mathbb{R}^{-1} \pdfrac{\mathbb{R}}{\xi} -
\left(\frac{\xi}{a}
\pdfrac{a}{\xi} + \xi \mathbb{R} \pdfrac{\mathbb{R}^{-1}}{\xi} \right)
\nonumber\\
&\times \left(\mathbb{A} \mathbb{R} + \xi
\pdfrac{\mathbb{R}}{\xi}\right) - \xi \mathbb{A} \pdfrac{\mathbb{R}}{\xi} - \xi 
\pdfrac{\mathbb{R}}{\xi} - \xi^2
\pdfrac{^2\mathbb{R}}{\xi^2} \bigg].
\end{align}
 
We can now proceed to the separation of timescales by introducing 
a new time variable $\tau$ such that $\dot{\tau}/\dot{\phi}_{\PP,\pm} =
{\cal{O}}(\sigma)$. In Sec.~\ref{sec:msa}, we used a
linear relation between
$\tau$ and $t$, i.e. $\tau = \sigma t$. Although we could do the same here, we
find it more convenient to use
the non-linear relation $d\tau/dt = \sigma a$, or in angle-variables
$d\tau/d\phi_{\PP,\pm} = \sigma a/\omega_{\PP,\pm}$. 
Such a non-linear mapping between $\tau$, $t$, and $\phi_{\PP\pm}$ leads to a
better match with
numerical solutions because
it allows for the ratio of timescales $t_{\precc}/t_{\rr}$ to vary as the
inspiral proceeds. 
Of course, we can solve for $\tau$ in terms of $t$  via
\begin{align}
 \tau &= \sigma \int a \, dt =
\sigma \int \frac{1}{\xi}
d\xi = \sigma \log\xi \,. 
 \label{eq:tau}
\end{align}

We then postulate the series ansatz
\begin{align}
 \bm{Q}_j^{(1)}(t) = \sum_{n \geq 0} \sigma^n 
\bm{Q}_j^{(1,n)}(t,\tau) = \sum_{n \geq 0} \sigma^n 
\bm{Q}_j^{(1,n)}(\phi_{\PP,\pm},\tau),
\end{align}
where $j = 1$ or $2$. Recall here that the first superscript $1$ reminds us that
these are quantities of ${\cal{O}}(\epsilon)$, 
while the second superscript labels the orders in $\sigma$. Thus, the quantity
$Q^{(m,n)}_{j}$ is of bivariate 
${\cal{O}}(\epsilon^{m},\sigma^{n})$. With this ansatz, we convert the system of
ordinary differential equations of 
Eqs.~\eqref{eq:ddQ0}-\eqref{eq:ddQm} into a system of partial differential
equations (PDEs). 
In doing so, we transform the differential operators via
\begin{align}
 \frac{d}{dt} &= \pdfrac{}{t} + \sigma a \pdfrac{}{\tau},
\\
 \frac{d^2}{d\phi_{\PP,\pm}^2} &= \pdfrac{^2}{\phi_{\PP,\pm}^2} + 2 \sigma
\frac{a}{\omega_{\PP,\pm}}
\pdfrac{^2}{\phi_{\PP,\pm}\partial\tau} \nonumber\\
&+
\sigma^2 \frac{a^2}{\omega_{\PP,\pm}^2} \pdfrac{^2}{\tau^2} + \sigma^2
\frac{a^2}{\omega_{\PP,\pm}^2}
\left( \frac{\dot{a}}{a} - \frac{\dot{\omega}_{\PP,\pm}}{\omega_{\PP,\pm}} 
\right)
\pdfrac{}{\tau},
\end{align}
and re-expand all quantities in $\sigma \ll 1$. In what follows, we solve the
resulting system of PDEs
order by order in $\sigma$. 

\subsubsection{Solution to ${\cal{O}}(\epsilon^{1},\sigma^{0})$}

At lowest order in $\sigma$, the system of PDEs becomes
\begin{align}
 \pdfrac{Q_{0,j}^{(1,0)}}{t} &= 0, \:
 \pdfrac{^2Q_{+,j}^{(1,0)}}{\phi_{\PP,+}^2} = - Q_{+,j}^{(1,0)}, \:
 \pdfrac{^2Q_{-,j}^{(1,0)}}{\phi_{\PP,-}^2} = - Q_{-,j}^{(1,0)},
\end{align}
$j = 1$ or $2$. Solving these equations and requiring that
they satisfy the original first-order differential system of
Eqs.~(\ref{normal-mode-eq-1}-\ref{normal-mode-eq}) at leading order in $\sigma$,
we find
\begin{align}
 Q_{0,j}^{(1,0)} &= A_{0,j}^{(1,0)}(\tau), \\
 Q_{\pm,1}^{(1,0)} &= A_{\pm,1}^{(1,0)}(\tau) \cos \phi_{\PP,\pm} -
A_{\pm,2}^{(1,0)}(\tau) \sin
\phi_{\PP,\pm} , \\
 Q_{\pm,2}^{(1,0)} &= A_{\pm,2}^{(1,0)}(\tau) \cos \phi_{\PP,\pm} +
A_{\pm,1}^{(1,0)}(\tau) \sin
\phi_{\PP,\pm}. 
\end{align}
with $j = 1$ or $2$ as usual. 

\subsubsection{Solution to ${\cal{O}}(\epsilon^{1},\sigma^{1})$}

Let us now consider the differential system at ${\cal{O}}(\sigma^{1})$. 
The zero-frequency equations are
\begin{align}
 \pdfrac{Q_{0,j}^{(1,1)}}{t} &= - a \pdfrac{A_{0,j}^{(1,0)}}{\tau} - 
aA_{0,j}^{(1,0)} \nonumber\\
&- a\left(Q_{+,j}^{(1,0)} + Q_{-,j}^{(1,0)} \right)
 \left[1 + \mathcal{O}\left(c^{-1}\right) \right] ,
 \label{eq:O11-zero-freq}
\end{align}
$j = 1$ or $2$. As we saw in Sec.~\ref{sec:msa}, we must require that secular
terms do not appear, 
which then leads to the equation
\begin{align}
 \pdfrac{A_{0,j}^{(1,0)}}{\tau} + 
A_{0,j}^{(1,0)} &= 0,
\end{align}
whose solution is 
\begin{align}
 A_{0,j}^{(1,0)}(\tau) &= B_{0,j}^{(1,0)} e^{-\tau} =
\frac{B_{0,j}^{(1,0)}}{\xi}.
\end{align}
The terms depending on $Q_{\pm,j}^{(1,0)}$ in Eq.~\eqref{eq:O11-zero-freq} will 
induce an oscillatory term in $Q_{0,j}^{(1,1)}$.

The non-zero frequency equations at ${\cal{O}}(\sigma^{1})$ are
\begin{align}
 \pdfrac{^2Q_{\pm,1}^{(1,1)}}{\phi_{\PP,\pm}^2} &+ Q_{\pm,1}^{(1,1)} \nonumber\\
 &=
\frac{a}{\omega_{\PP,\pm}}
\left[2 \pdfrac{Q_{\pm,2}^{(1,0)}}{\tau} +
\frac{1}{\omega_{\PP,\pm}} \left[ \mathbb{F}
\bm{Q}_{2}{^{(1,0)}}
\right]\cdot \bm{P}_\pm
\right], \\
 \pdfrac{^2Q_{\pm,2}^{(1,1)}}{\phi_{\PP,\pm}^2} &+ Q_{\pm,2}^{(1,1)} \nonumber\\
 &=
- \frac{a}{\omega_{\PP,\pm}} 
\left[2 \pdfrac{Q_{\pm,1}^{(1,0)}}{\tau} + \frac{1}{\omega_{\PP,\pm}}
\left[ \mathbb{F}
\bm{Q}_{1}{^{(1,0)}}
\right]\cdot \bm{P}_\pm
\right].
\end{align}
Expanding these equations, we find
\begin{widetext}
\begin{align}
 \pdfrac{^2Q_{+,1}^{(1,1)}}{\phi_{\PP,+}^2} + Q_{+,1}^{(1,1)} &=
 \frac{a}{\omega_{\PP,+}}
\Bigg\{ 2 \left(\pdfrac{A_{+,2}^{(1,0)}}{\tau} \cos\phi_{\PP,+} +
\pdfrac{A_{+,1}^{(1,0)}}{\tau} \sin\phi_{\PP,+} \right) + \omega_{\PP,+}^{-1}
\Big[ \mathbb{F}_{++} \left( A_{+,2}^{(1,0)} \cos
\phi_{\PP,+} +
A_{+,1}^{(1,0)}
\sin
\phi_{\PP,+}  \right)\nonumber\\
& + \mathbb{F}_{+0} A_{0,2}^{(1,0)} + \mathbb{F}_{+-} \left( A_{-,2}^{(1,0)}
\cos \phi_{\PP,-} +
A_{-,1}^{(1,0)}
\sin
\phi_{\PP,-}  \right)
\Big] \Bigg\},  \label{eq:eqQp11} \\
 \pdfrac{^2Q_{+,2}^{(1,1)}}{\phi_{\PP,+}^2} + Q_{+,2}^{(1,1)} &=
- \frac{a}{\omega_{\PP,+}}
\Bigg\{ 2 \left(\pdfrac{A_{+,1}^{(1,0)}}{\tau} \cos\phi_{\PP,+} -
\pdfrac{A_{+,2}^{(1,0)}}{\tau} \sin\phi_{\PP,+} \right) + \omega_{\PP,+}^{-1}
\Big[ \mathbb{F}_{++} \left( A_{+,1}^{(1,0)} \cos
\phi_{\PP,+} -
A_{+,2}^{(1,0)}
\sin
\phi_{\PP,+}  \right)\nonumber\\
& + \mathbb{F}_{+0} A_{0,1}^{(1,0)} + \mathbb{F}_{+-} \left( A_{-,1}^{(1,0)}
\cos \phi_{\PP,-} -
A_{-,2}^{(1,0)}
\sin
\phi_{\PP,-}  \right)
\Big] \Bigg\}, \label{eq:eqQp21} \\
 \pdfrac{^2Q_{-,1}^{(1,1)}}{\phi_{\PP,-}^2} + Q_{-,1}^{(1,1)} &=
 \frac{a}{\omega_{\PP,-}}
\Bigg\{ 2 \left(\pdfrac{A_{-,2}^{(1,0)}}{\tau} \cos\phi_{\PP,-} +
\pdfrac{A_{-,1}^{(1,0)}}{\tau} \sin\phi_{\PP,-} \right) + \omega_{\PP,-}^{-1}
\Big[ \mathbb{F}_{--} \left( A_{-,2}^{(1,0)} \cos
\phi_{\PP,-} +
A_{-,1}^{(1,0)}
\sin
\phi_{\PP,-}  \right)\nonumber\\
& + \mathbb{F}_{-0} A_{0,2}^{(1,0)} + \mathbb{F}_{-+} \left( A_{+,2}^{(1,0)}
\cos \phi_{\PP,+} +
A_{+,1}^{(1,0)}
\sin
\phi_{\PP,+}  \right)
\Big] \Bigg\},  \label{eq:eqQm11} \\
 \pdfrac{^2Q_{-,2}^{(1,1)}}{\phi_{\PP,-}^2} + Q_{-,2}^{(1,1)} &=
- \frac{a}{\omega_{\PP,-}}
\Bigg\{ 2 \left(\pdfrac{A_{-,1}^{(1,0)}}{\tau} \cos\phi_{\PP,-} -
\pdfrac{A_{-,2}^{(1,0)}}{\tau} \sin\phi_{\PP,-} \right) + \omega_{\PP,-}^{-1}
\Big[ \mathbb{F}_{--} \left( A_{-,1}^{(1,0)} \cos
\phi_{\PP,-} -
A_{-,2}^{(1,0)}
\sin
\phi_{\PP,-}  \right)\nonumber\\
& + \mathbb{F}_{-0} A_{0,1}^{(1,0)} + \mathbb{F}_{-+} \left( A_{+,1}^{(1,0)}
\cos \phi_{\PP,+} -
A_{+,2}^{(1,0)}
\sin
\phi_{\PP,+}  \right)
\Big] \Bigg\}, \label{eq:eqQm21}
\end{align}
\end{widetext}
To prevent secular terms, we must require that there are no source terms
proportional to solutions of the homogeneous equation. This then
imposes
\begin{align}
 \pdfrac{A_{+,j}^{(1,0)}}{\tau} +  \frac{1}{2} \omega_{\PP,+}^{-1} 
\mathbb{F}_{++}
A_{+,j}^{(1,0)} &= 0, \\
 \pdfrac{A_{-,j}^{(1,0)}}{\tau} + \frac{1}{2} \omega_{\PP,-}^{-1} 
\mathbb{F}_{--}
A_{-,j}^{(1,0)} &= 0, 
\end{align}
where $j = 1$ or $2$. 
The solutions to these equations are
\begin{align}
 A_{\pm,j}^{(1,0)}(\tau) &= B_{\pm,j}^{(1,0)} \exp\left[ - \frac{1}{2}\int
(\mathbb{F}_{\pm\pm} / \omega_{\PP,\pm}) d\tau \right]
\nonumber\\
&= B_{\pm,j}^{(1,0)} \exp\left\{ - \frac{1}{2}\int
[\mathbb{F}_{\pm\pm} / (\omega_{\PP,\pm} \xi) ] d\xi \right\}
\nonumber\\
&= B_{\pm,j}^{(1,0)} \left[ 1 + \mathcal{O}(c^{-1}) \right] . \label{eq:Apmj10}
\end{align}
where $B_{\pm,j}^{(1,0)}$ are integration constants and 
we have used Eq.~\eqref{eq:tau}. A proof of Eq.~\eqref{eq:Apmj10} is  
given in appendix~\ref{app:RandRm1}.

\subsubsection{Precession phases}
\label{sec:precphases}

The precession angles $\phi_{\PP,\pm}$ can be computed using 
Eq.~\eqref{eq-sep-var}
and~\eqref{eq:xiofzeta}, which leads to
\begin{align}
 \phi_{\PP,\pm} &= \int \omega_{\PP,\pm} dt = \int 
\frac{\omega_{\PP,\pm}}{a\xi} 
d\xi.
\label{eq:phipm}
 \end{align}
Care must be exercised when computing this integral because 
$\delta\omega_p = \omega_{\PP,+} - \omega_{\PP,-}$ satisfies
\begin{align}
 \delta \omega_p^2 &= \mathcal{O}( \delta m^2) \xi^{10} + \mathcal{O}(\delta
m) \xi^{11} \\
&+ \mathcal{O}(\delta m^0)
\xi^{12} + \mathcal{O}\left( \delta m^0,\xi^{13} \right),
\end{align}
where $\delta m = (m_1 - m_2)/M$ is the dimensionless mass difference. Depending
on the magnitude of $\delta m$ relative to the magnitude
of $\xi$, the PN expansion will be somewhat different: 
if $\delta m \gg {\cal{O}}(c^{-1})$, $\delta \omega_p^2 \sim \xi^{10}$; 
if $\delta m \ll {\cal{O}}(c^{-1})$, $\delta \omega_{\PP}^{2} \sim \xi^{12}$.
Notice that this is not
a problem in the PN treatment of non-spinning inspirals, as there $\delta m$
does 
not appear in the controlling factor of the approximation.

In order to address this feature of the solution, let us separate the
precession phases via 
$\phi_{\PP,\pm} = \phi_{\PP,m} \pm \delta \phi_\PP$. The mean precession
phase $\phi_{\PP,m} \equiv (\phi_{\PP,+} + \phi_{\PP,-})/2$ can be computed 
using
standard PN methods:
\begin{align}
 \phi_{\PP,m} &= \frac{1}{2} \int \left( \omega_{\PP,+} + \omega_{\PP,-} \right)
\frac{dt}{d\xi}
d\xi \nonumber\\
&= \frac{1}{2} \int \frac{ b+c+d+e+f+g}{a\xi} d\xi \nonumber\\
&= -\frac{5}{128}  \left[ \frac{8}{3} + \left( \frac{m_1}{m_2} +
\frac{m_2}{m_1} \right) \right] \xi^{-3} \left[1 +
\mathcal{O}(c^{-1}) \right]. \label{phim}
\end{align}
We can expand the integrand of this expression to any relevant PN order, 
and we provide higher-order PN terms in Appendix~\ref{app:precphases}. 

The calculation of the precession phase difference $\delta \phi_\PP \equiv
(\phi_{\PP,+} - \phi_{\PP,-})/2$ 
must be studied in two different cases: $\delta m \ll {\cal{O}}(c^{-1})$ 
and $\delta m \gg {\cal{O}}(c^{-1})$. Since the PN parameter that
controls the expansion in $c^{-1}$ evolves with time, some systems might comply
with the former case at some point in time, and to the latter case at another.
Thus, we find it useful to consider a third case, i.e.~when $\delta m \sim
{\cal{O}}(c^{-1})$. As we will see below, this will allow us to use a uniform 
approximation
depending only on the  value of $\delta m$.

Let us first focus on the $\delta m \gg {\cal{O}}(c^{-1})$ case. In this
case, 
$\delta m$ can be treated as a quantity of order unity, and we can carry out a
standard
PN expansion:
\begin{align}
 \delta \phi_{\PP,1} &= \frac{1}{2} \int \left( \omega_{\PP,+} - \omega_{\PP,-}
\right)
\frac{dt}{d\xi}
d\xi \nonumber\\
&= \frac{1}{2} \int \frac{1}{a\xi} \big[(b-c+d-e+f-g)^2 \nonumber\\
&\qquad \qquad\qquad \qquad +
4(c-f)(b-g) \big]^{1/2} d\xi
\nonumber\\
&= \frac{15}{256} \left\{ - \frac{1}{3} \left( \frac{m_1}{m_2} - \frac{m_2}{m_1}
\right) \xi^{-3}  
\right. 
\nonumber \\
&- \left. \frac{1}{2} \left[ \chi_1 - \chi_2 + 2 \left(\frac{m_1}{m_2} \chi_1 -
\frac{m_2}{m_1} \chi_2 \right) \right] 
\xi^{-2} + \mathcal{O}(c) \right\}.
\label{phipxi}
\end{align}

Let us now concentrate on the case where $\delta m \sim {\cal{O}}(c^{-1})$.
In this case, 
we must treat $\delta m$ as a quantity of the same order as the PN parameter
$\xi$. Doing so
and PN expanding, we find
\begin{align}
 \delta \phi_{\PP,2} &= \frac{1}{2} \int \left( \omega_{\PP,+} - \omega_{\PP,-}
\right)
\frac{dt}{d\xi}
d\xi \nonumber\\
&= \frac{1}{2} \int \frac{1}{a\xi} \big[(b-c+d-e+f-g)^2 \nonumber\\
&+ 4(c-f)(b-g) \big]^{1/2} d\xi
\nonumber\\
&= -\frac{5}{8192 \delta m^2 \xi^3} \bigg\{ T_{0} \left[ 32 \delta m^2 -
12 (\chi_1 - \chi_2 ) \delta m \, \xi +
\right.  \nonumber \\
&+ \left. 
 \left( 9 \chi_1^2 - 50 \chi_1 \chi_2 + 9 \chi_2^2
\right) \xi^2 \right] + 144 \chi_1 \chi_2 (\chi_1 - \chi_2) \xi^3 
\nonumber\\
&\times
\log \left[ 4 \frac{\delta
m}{\xi} - 3 (\chi_1 - \chi_2) + \frac{T_{0}}{\xi} \right] + \mathcal{O}(c^{-4})
\bigg\}, 
\label{phipDelta} 
\end{align}
where we have defined the quantity 
\begin{multline}
T_{0} = \big[16 \delta m^2 - 24 (\chi_1 - \chi_2) \delta
m \xi \\
+ \left( 9 \chi_1^2 - 2 \chi_1 \chi_2 + 9 \chi_2^2 \right)
\xi^2
\big]^{1/2}.
\end{multline}

Finally, when $\delta m \ll {\cal{O}}(c^{-1})$, we expand $\delta
\omega_p$ in $\delta m$, but leave the $\xi$ factors unexpanded:
\begin{align}
 \delta \phi_{\PP,3} &= \frac{1}{2} \int \left( \omega_{\PP,+} - \omega_{\PP,-}
\right)
\frac{dt}{d\xi}
d\xi \nonumber\\
&= \frac{1}{2} \int \frac{1}{a\xi} \big[(b-c+d-e+f-g)^2 \nonumber\\
&+ 4(c-f)(b-g) \big]^{1/2} d\xi
\nonumber\\
&= - \frac{15}{512 T_1
\xi^2} \left\{ T_2 T_3 + \frac{20}{\sqrt{T_1}} \chi_1^2 \chi_2^2 (\chi_1 -
\chi_2)^2 \xi^2 
\right. \nonumber \\
&\times \left.
\log \left[ \frac{1}{\xi}  \left( T_3 + \sqrt{T_1} T_2 \right)
\right] \right\} + \mathcal{O}(\delta m),
\label{phipdeltam}
\end{align}
where we have defined the quantities
\begin{align}
T_1 &= 9 \chi_1^2 - 2 \chi_1 \chi_2 + 9 \chi_2^2, \\
T_2 &= \big[ 9 \chi_1^2 - 2 \chi_1 \chi_2 + 9 \chi_2^2 \nonumber\\
& -8 \chi_1 \chi_2 (\chi_1 + \chi_2) \xi + 4 \chi_1^2 \chi_2^2 \xi^2
\big]^{1/2}, \\
T_3 &= 9 \chi_1^2 - 2 \chi_1 \chi_2 + 9 \chi_2^2 - 4 \chi_1 \chi_2 (\chi_1 +
\chi_2) \xi.
\end{align}

In this way, we can reconstruct $\phi_{\PP,\pm}$ by combining 
$\phi_{\PP,m}$ in Eq.~\eqref{phim} with one of the $\delta \phi_{\PP}$ 
in Eqs.~\eqref{phipxi}, \eqref{phipDelta} and \eqref{phipdeltam}, 
depending on the magnitude of $\delta m$. Later on, we will use
the particular implementation described in Appendix~\ref{app:precphases}, 
where we found useful, in practice, to use 
$\delta \phi_{\PP} = \delta \phi_{\PP,1}$ for $\delta m \geq 0.2$, 
$\delta \phi_{\PP} = \delta \phi_{\PP,2}$ for $10^{-5} \leq \delta m < 0.2$,
and $\delta \phi_{\PP} = \delta \phi_{\PP,3}$ for $\delta m < 10^{-5}$.

Different classes of compact binaries will, of course, have a different natural 
set of $\delta m$. 
Neutron star binaries must have $\delta m \in \left(0,0.375\right)$, with 
typical values in $\delta m \sim 0.08$ for which $\delta \phi_{\PP} = 
\phi_{\PP,2}$. Neutron star/black hole binaries can have $\delta m \in 
\left(0.375,0.96\right)$, with typical values in $\delta m \sim 0.75$ for which 
$\delta \phi_{\PP} = \phi_{\PP,1}$. Black hole binaries detectable by advanced 
ground detectors with total mass less than $50 M_{\odot}$ must have $\delta m 
\in \left(0,0.82 \right)$, with typical values in $\delta m \lesssim 0.4$ for 
the best gravitational wave candidates. In this case, one may have to switch 
between $\delta \phi_{\PP} = \phi_{\PP,1}$ and $\delta \phi_{\PP} = 
\phi_{\PP,2}$. The choice $\delta \phi_{\PP} = \phi_{\PP,3}$ is only relevant 
for almost exactly equal mass, which has a very low probability of happening. 

We argue in Sec.~\ref{subsec:discontinuity} that the discontinuity in the 
solution $\delta \phi_{\PP}$ at $\delta m = 0.2$ is of no concern due to high 
faithfulness between the two approximations at the boundary.

\subsubsection{Solutions to higher order in $\sigma$}

Going to higher order in  $\sigma$ is straightforward, except perhaps for the
treatment of certain timescale mixing that will generically occur; see
e.g.~Eqs.~(\ref{eq:eqQp11}-\ref{eq:eqQm21})
at higher order in $\sigma$. In order to exemplify how a higher-order in
$\sigma$ calculation would proceed, 
let us return to Eq.~\eqref{eq:eqQp11}. This equation is that of a harmonic
oscillator
with frequency $\omega_{\PP,+}$, sourced by an oscillatory term of frequency
$\omega_{\PP,-}$. Let us transform the left-hand side of this equation from
$\phi_{\PP,+}$ to $t$:
\begin{align}
 \pdfrac{^2Q_{+,1}^{(1,1)}}{\phi_{\PP,+}^2} = \frac{1}{\omega_{\PP,+}^2}
\pdfrac{^2Q_{+,1}^{(1,1)}}{t^2} - \frac{1}{\omega_{\PP,+}^{3}} \frac{d
\omega_{\PP,+}}{dt} \pdfrac{Q_{+,1}^{(1,1)}}{t}.
\end{align}
The last term in this equation contains a $d \omega_{\PP,+}/dt$ factor, which
introduces an extra factor of $\sigma$, and must therefore be kept to
${\cal{O}}(\sigma^{2})$.
Equation~\eqref{eq:eqQp11} then becomes
\begin{align}
 \pdfrac{^2Q_{+,1}^{(1,1)}}{t^2} &+ \omega_{\PP,+}^2
Q_{+,1}^{(1,1)} = a \Big[ \mathbb{F}_{+0} A_{0,2}^{(1,0)} \nonumber\\
&+
\mathbb{F}_{+-} \left( A_{-,2}^{(1,0)}
\cos \phi_{\PP,-} +
A_{-,1}^{(1,0)}
\sin
\phi_{\PP,-}  \right) \Big], \label{eq:eqQp111}
\end{align}
and its solution is
\begin{align}
 Q_{+,1}^{(1,1)} &= A_{+,1}^{(1,1)}(\psi_{\PP,+}) \cos \phi_{\PP,+} -
A_{+,2}^{(1,1)}(\psi_{\PP,+}) \sin \phi_{\PP,+} \nonumber\\
&- \frac{a}{\omega_{\PP,+} -
\omega_{\PP,-}} \left[ \frac{m_2^2 \chi_2}{m_1(m_1-m_2)}
\xi +
\mathcal{O}\left(c^{-2}\right) \right] \nonumber\\
&\times \left( A_{-,2}^{(1,0)} \cos \phi_{\PP,-} +
A_{-,1}^{(1,0)}
\sin \phi_{\PP,-}  \right).
\end{align}
Notice that the above solution has a pole at $\omega_{\PP,+} = \omega_{\PP,-}$, 
because the terms oscillating with frequency $\omega_{\PP,-}$ drive a resonance 
in
Eq.~\eqref{eq:eqQp111}. In practice, however, this happens only when
$\chi_2 = 0$ and at a single point in time. Such a limit, therefore, must be
excluded
from higher-order solutions.

We can now use this ${\cal{O}}(\epsilon,\sigma)$ solution in the source to the
${\cal{O}}(\epsilon,\sigma^{2})$ differential equation
and require that terms oscillating at frequency $\omega_{\PP,+}$ vanish so as 
not
to produce secularly growing terms. This would
then lead to differential equations for $A_{+,j}^{(1,1)}(\psi_{\PP,+})$, just as
we obtained for $A_{+,j}^{(1,0)}(\psi_{\PP,+})$ at 
${\cal{O}}(\epsilon,\sigma)$. The solution to these equations would then lead to
a solution to the precession equations at 
${\cal{O}}(\epsilon,\sigma)$. We will not carry out a higher-order development 
here. 

\subsection{Summary}

Let us here collect all the pieces of the solution to
${\cal{O}}(\epsilon,\sigma)$ obtained in the previous subsections. 
Using the initial conditions $\uvec{J}(0) = \uvec{z}$, we can write
\begin{align}
 \bm{W}_j^{(1)}(t=0) &= \left( \begin{array}{c}
-S_{1,k}^{(1)}(t=0) - S_{2,k}^{(1)}(t=0)\\ S_{1,k}^{(1)}(t=0)\\
S_{2,k}^{(1)}(t=0)
                          \end{array} \right),
\end{align}
where $j = 1$ and $k=x$, or $j=2$ and $k=y$.
Furthermore,
\begin{align}
 \bm{Q}_j^{(1)}(t=0) = \left( \begin{array}{c}
B_{0,j}^{(1,0)}\\ B_{+,j}^{(1,0)}\\
B_{-,j}^{(1,0)}
                          \end{array} \right) + \mathcal{O}(\sigma, c^{-1}).
\end{align}
and since
\begin{align}
 \bm{Q}_j^{(1)}(t=0) = \mathbb{R}^{-1}(t=0) \bm{W}_j^{(1)}(t=0),
\end{align}
we find that
\begin{align}
 B_{0,j}^{(1,0)} = 0.
\end{align}
This happens because we chose $\uvec{z} = \uvec{J}(t=0)$. Any equivalent but 
different choice would result in a nonvanishing $B_{0,j}^{(1,0)} $.

The solution for the orbital angular momentum is then
\begin{align}
\label{eq:LzNAS}
 L_z &= \frac{\mu M}{\xi} + \mathcal{O} \left(\epsilon^2\right), \\
 L_x &= \epsilon \Big( B_{+,1}^{(1,0)} \cos \phi_{\PP,+} - B_{+,2}^{(1,0)} \sin
\phi_{\PP,+} \nonumber\\
&+ B_{-,1}^{(1,0)} \cos \phi_{\PP,-} - B_{-,2}^{(1,0)} \sin
\phi_{\PP,-} \Big)  + \mathcal{O} \left(\sigma,\epsilon^3\right),
\label{eq:LxNAS}\\
 L_y &= \epsilon \Big( B_{+,2}^{(1,0)} \cos \phi_{\PP,+} + B_{+,1}^{(1,0)} \sin
\phi_{\PP,+} \nonumber\\
&+ B_{-,2}^{(1,0)} \cos \phi_{\PP,-} + B_{-,1}^{(1,0)} \sin
\phi_{\PP,-} \Big)  + \mathcal{O} \left(\sigma,\epsilon^3\right), 
\label{eq:LyNAS}
\end{align}
where $\xi$ is given by Eq.~\eqref{eq:xiofzeta}, the precession angles 
$\phi_{\PP,\pm}$ 
are shown in Sec.~\ref{sec:precphases} and Appendix~\ref{app:precphases}, 
and $B_{i,j}^{(1,0)}$ are given in Appendix~\ref{app:RandRm1}.

The $z$-components of the spins are simply
\begin{align}
 S_{1,z} &= m_1^2 \chi_1 + \mathcal{O} \left(\epsilon^2\right), \\
 S_{2,z} &= m_2^2 \chi_2 + \mathcal{O} \left(\epsilon^2\right),
\end{align}
and the $x$ and $y$-components are
\begin{widetext}
\begin{align}
 S_{1,x} = \epsilon \bigg[&
\frac{2(g-b)}{b+c - (d-e)-(f+g) + \delta \omega_p}  (B_{+,1}^{(1,0)} \cos
\phi_{\PP,+} - B_{+,2}^{(1,0)}
\sin
\phi_{\PP,+}) \nonumber\\
&+\frac{2(g-b)}{b+c- (d-e)-(f+g)  - \delta \omega_p} (B_{-,1}^{(1,0)} \cos
\phi_{\PP,-} - B_{-,2}^{(1,0)} \sin
\phi_{\PP,-}) \bigg] + \mathcal{O}(\sigma, \epsilon^3), \\
 S_{1,y} = \epsilon \bigg[&
\frac{2(g-b)}{b+c - (d-e)-(f+g) + \delta \omega_p}  (B_{+,2}^{(1,0)} \cos
\phi_{\PP,+} + B_{+,1}^{(1,0)}
\sin
\phi_{\PP,+}) \nonumber\\
&+\frac{2(g-b)}{b+c- (d-e)-(f+g)  - \delta \omega_p} (B_{-,2}^{(1,0)} \cos
\phi_{\PP,-} + B_{-,1}^{(1,0)} \sin
\phi_{\PP,-}) \bigg] + \mathcal{O}(\sigma, \epsilon^3), \\
 S_{2,x} = \epsilon \bigg[&
\frac{2(f-c)}{b+c - (d-e)-(f+g) + \delta \omega_p}  (B_{+,1}^{(1,0)} \cos
\phi_{\PP,+} - B_{+,2}^{(1,0)}
\sin
\phi_{\PP,+}) \nonumber\\
&+\frac{2(f-c)}{b+c- (d-e)-(f+g)  - \delta \omega_p} (B_{-,1}^{(1,0)} \cos
\phi_{\PP,-} - B_{-,2}^{(1,0)} \sin
\phi_{\PP,-}) \bigg] + \mathcal{O}(\sigma, \epsilon^3), \\
 S_{2,y} = \epsilon \bigg[&
\frac{2(f-c)}{b+c - (d-e)-(f+g) + \delta \omega_p}  (B_{+,2}^{(1,0)} \cos
\phi_{\PP,+} + B_{+,1}^{(1,0)}
\sin
\phi_{\PP,+}) \nonumber\\
&+\frac{2(f-c)}{b+c- (d-e)-(f+g)  - \delta \omega_p} (B_{-,2}^{(1,0)} \cos
\phi_{\PP,-} + B_{-,1}^{(1,0)} \sin
\phi_{\PP,-}) \bigg] + \mathcal{O}(\sigma, \epsilon^3),
\end{align}
where $\delta \omega_p = \omega_{\PP,+} - \omega_{\PP,-}$.
\end{widetext}

\subsection{Comparison with Simple Precession}

Another physically relevant case where the equations of precession can
be solved analytically is simple precession. This occurs when
one of the spins vanishes or when the masses are equal, provided we neglect
spin-spin interactions. In simple precession, the orbital and spin angular
momentum vectors precess around the total angular momentum vector 
at exactly the same frequency. 

Let us begin to study simple precession by rewriting the evolution equations
without spin-spin couplings:
\begin{align}
\duvec{S}_A &= \frac{\omega^2}{M} \left( 2 + \frac{3 m_B}{2m_A}
\right) \bm{L} 
\times
\uvec{S}_A,\\
\duvec{L} &= \frac{\omega^2}{M} \left[ \left( 2 + \frac{3 m_2}{2m_1}
\right) \bm{S}_1 + \left( 2 + \frac{3 m_1}{2m_2}
\right)
\bm{S}_2 \right]  \times
\uvec{L} ,
\end{align}
where $\bm{L}$ as before is the Newtonian orbital angular momentum with norm $L
= \mu M^{2/3} \omega^{-1/3}$. 

If either spin vanishes or if the masses are equal, the
derivative of the total spin vector $\bm{S} = \bm{S}_1 + \bm{S}_2$ is
perpendicular to $\bm{S}$. We can then rewrite the equations of precession for
$\bm{S}$ and $\bm{L}$ as
\begin{align}
\dot{S} &= 0, \\
\dot{L} &= - \frac{1}{3} \frac{a_0}{M} \omega^{8/3} \left\{ 1 + \sum_{n \geq
2}^N a_n
\omega^{n/3}
\right\} L, \\
\duvec{S} &= \frac{\omega^2}{M} \left( 2 + \frac{3 m_\V}{2m_\NV}
\right) \bm{J} 
\times
\uvec{S},\\
\duvec{L} &= \frac{\omega^2}{M} \left( 2 + \frac{3 m_\V}{2m_\NV}
\right) \bm{J}  \times
\uvec{L} ,
\end{align}
where $\bm{J} = \bm{L} + \bm{S}$ is the total angular momentum vector, the
vanishing spin, if any, is
labelled by the subscript $\V$, while the non-vanishing one is labeled by the
subscript $\NV$. We then see that in simple precession both $\uvec{S}$ and
$\uvec{L}$ precess around
$\uvec{J}$ at a frequency 
\begin{align}
\omega_{\PP,sp} = \frac{\omega^2}{M} \left( 2 + \frac{3 m_\V}{2m_\NV}
\right) J . 
\end{align}
If the spins are only slightly misaligned with the orbital angular momentum,
we have to leading order in $\epsilon$
\begin{align}
 J &= L + S_1 + S_2 = \frac{\mu M}{\xi} + m_\NV^2 \chi_\NV + m_\V^2 \chi_\V , 
 \end{align}
 which then leads to
 \begin{align}
M \omega_{\PP,sp} =& \left( 2 + \frac{3 m_\V}{2m_\NV}
\right) \bigg[ \nu \xi^5 
+ \frac{1}{M^2} \left(m_\NV^2
\chi_\NV + m_\V^2 \chi_\V \right) \xi^6
\bigg],
\end{align}
where recall that $\omega = \xi^3/M$ and $\xi$ was defined in 
Eq.~\eqref{eq:xiofzeta}.

Let us now return to our results for the near-alignment, multiple-scale analysis
calculation. 
In order to map our results to those of simple precession, we must neglect
spin-spin interactions,
which naturally vanish in the single spin case. This implies using the following
relations:
\begin{align}
b &= \frac{\xi(t)^6}{M^3} \bigg[
\left( 2 + \frac{3m_2}{2m_1} \right) m_1^2  \chi_1
 \bigg] , \\
c &= \frac{\xi(t)^6}{M^3} \bigg[
\left( 2 + \frac{3m_1}{2m_2} \right) m_2^2  \chi_2
 \bigg] , \\
d &= \frac{\xi(t)^5}{M} \bigg[
\left( 2 + \frac{3m_2}{2m_1} \right) \nu
 \bigg] , \\
e &= \frac{\xi(t)^5}{M} \bigg[
\left( 2 + \frac{3m_1}{2m_2} \right) \nu
 \bigg] , \\
f &= 0 , \qquad
g = 0.
\end{align}
Thus, in the equal-mass case we have
\begin{align}
 M\omega_{\PP,+} &= \frac{7}{8} \left[ \xi^5 + 
\xi^6 (\chi_1 + \chi_2 ) 
  \right] , \\
 M\omega_{\PP,-} &= \frac{7}{8}  \xi^5  , \\
  B_{+,j}^{(1,0)} &=  - S_{1,j} - S_{2,j} , \\
  B_{-,j}^{(1,0)} &=  0 .
\end{align}
On the other hand, in the case where one of the spins vanishes, $\chi_{\V} = 
0$, we get
\begin{align}
 M\omega_{\PP,+} &= \left( 2 + \frac{3 m_\V}{2 m_\NV} \right) \left( \nu \xi^5 +
\frac{m_\NV^2}{M^{2}}
\chi_\NV \xi^6 \right) , \\
  M\omega_{\PP,-} &= \left( 2 + \frac{3 m_\NV}{2 m_\V} \right) \nu \xi^5 ,\\
  B_{+,j}^{(1,0)} &= - S_{\NV,j} , \\
  B_{-,j}^{(1,0)} &=  0 ,
\end{align}
provided
\begin{align}
 \xi &\geq \xi_{c} \equiv \frac{3 \left( m_\NV^2 - m_\V^2 \right)}{ \left(4 
m_\NV + 3 m_\V
\right) m_\NV \chi_\NV}. 
\label{eq:chi0condition}
\end{align}
In the complementary case, when $\xi < \xi_{c}$, the results are
the same modulo $\omega_{\PP,+} \leftrightarrow \omega_{\PP,-}$ and 
$A_{+,j}^{(1,0)}
\leftrightarrow A_{-,j}^{(1,0)}$. We see then clearly that our results in the
simple precession limit reproduce exactly the results of simple precession in 
the nearly aligned limit. 
That is, both in the equal-mass case or in the vanishing single spin case, 
$\omega_{\PP,+}$ becomes equal to $\omega_{\PP,sp}$, while the $\omega_{\PP,-}$ 
mode is irrelevant because its amplitude
vanishes.

An interesting transition occurs if $\xi_{c} > 0$: the only evolution frequency
continuously switches 
between $\omega_{\PP,+}$ and $\omega_{\PP,-}$. This transition only occurs if 
the
vanishing spin is
$\chi_{2}$, because then $m_{\V} = m_{2}$ and by the conventions used in this
paper, 
the numerator of Eq.~\eqref{eq:chi0condition} is positive, 
i.e.~$m_{\NV}^{2} - m_{\V}^{2} = m_{1}^{2} - m_{2}^{2} > 0$. The transition
occurs at a particular
value in time, given by $\xi = \xi_{c}$. At this time, however, $\omega_{\PP,+} 
=
\omega_{\PP,-}$,
and thus, the transition is continuous. 


\section{Gravitational Waves}
\label{sec:spa}

The results of the previous sections can be used to derive a purely analytic
time-domain waveform for precessing nearly aligned binaries. In the rest of 
this 
section,
we will derive such a waveform.

\subsection{Time-Domain Waveforms: \\ Standard Representation}

An impinging GW will induce the following response in an interferometer with 
perpendicular arms in the long wavelength approximation:
\begin{align}
 h(t) &= \sum_{n\geq0} \left[ F_+ h_{n,+} + F_\times h_{n,\times} \right], 
\label{h-time-domain}\\
 h_{n,+} &= \mathcal{A}_{n,+}(i_L) \cos n\phi + \mathcal{B}_{n,+}(i_L) \sin n 
\phi , \label{h-time-domain-plus}\\
 h_{n,\times} &= \mathcal{A}_{n,\times}(i_L) \cos n\phi + 
\mathcal{B}_{n,\times}(i_L) \sin n \phi,
\label{h-time-domain-cross}
\end{align}
where $n \in \mathbb{N}$ is the harmonic number, $\cos i_L = \uvec{L} 
\cdot \uvec{N}$, and the antenna pattern functions are
\begin{align}
 F_+(\theta_N, \phi_N, \psi_N) &= \frac{1}{2} \left( 1 + \cos^2 \theta_N
\right) \cos 2\phi_N \cos
2\psi_N \nonumber\\
&- \cos \theta_N \sin 2\phi_N \sin 2\psi_N, \label{eq:Fplus}\\
 F_\times(\theta_N, \phi_N, \psi_N) &= F_+(\theta_N,\phi_N,\psi_N-\pi/4),
\label{eq:Fcross}
\end{align}
with $(\theta_N,\phi_N)$ the spherical angles that label the position of the
binary in the detector frame, and $\psi_N$ the polarization angle defined 
through
\begin{equation}
 \tan \psi_N = \frac{\uvec{L}\cdot\uvec{z} - (\uvec{L}\cdot\uvec{N})
(\uvec{z}\cdot\uvec{N})}{\uvec{N}\cdot( \uvec{L}\times\uvec{z} )}\,,
\label{eq:psiN}
\end{equation}
where $\uvec{z}$ is the unit normal vector to the detector plane. 

The time-domain GW phase can be decomposed into a carrier phase and a
precession perturbation $\phi = \phi_{\C} + 
\delta\phi$~\cite{Apostolatos:1994mx}. 
Defining the reference of the orbital phase in the orbital plane
as $\uvec{L} \times \uvec{N}$, the equation of motion for the orbital phase is
\begin{align}
\label{eq:phieqmot}
\dot{\phi} &= \dot{\phi}_{\C} + \delta\dot{\phi}, \qquad
\dot{\phi}_{\C} = \omega , \\
\delta\dot{\phi} &= \frac{1}{L} \frac{\bm{L} \cdot \uvec{N}}{\bm{L}^2 -
\left(
\bm{L} \cdot \uvec{N} \right)^2} \left( \bm{L} \times \uvec{N} \right) \cdot
 \dvec{L}\,.
 \label{eq:phieqmot2}
 \end{align}
The carrier $\phi_{\C}$ is a secular, non-precessing phase, while the 
perturbation
$\delta \phi$ models the precession of the orbital plane.

\subsection{Time-Domain Waveforms: Fourier Representation}

Before we can Fourier transform the GW response via uniform asymptotics, we 
need 
to first
figure out the relative scale and variability of all relevant quantities. This 
is important as it will
tell us which quantities can be safely left in the slowly-varying signal 
amplitude, and which 
ones have to be promoted to the rapidly-varying phase. The part of the 
amplitude 
that depends
only on the sky location $(\theta_{N},\phi_{N})$ varies on the timescale of 
variation of the 
normal to the detector. For ground-based instruments, this is roughly $t_{\dett}
\sim \mathcal{O}(1\ \mathrm{day})$, much larger than the typical observation
time of $t_{\obs} \sim \mathcal{O}(100\ \mathrm{s})$. For space-based 
instruments, 
$t_{\dett} \sim \mathcal{O}(1\ \mathrm{year})$, which is of the same order as 
the typical 
observation time, but bigger than the typical precession timescale 
$t_{\precc} \sim \mathcal{O}(1\ \mathrm{month})$. This implies that it is safe 
to leave such
terms in the slowly-varying signal amplitude.

The different phases, however, can vary on a much shorter timescale. 
Using the equations of motion for $\phi$, 
one can show that
\begin{align}
 \dot{\phi}_{\C} &\sim \mathcal{O}(c^{-3}) , & \ddot{\phi}_{\C} & \sim
\mathcal{O}(c^{-11}), \nonumber\\
 \delta\dot{\phi} &\sim \mathcal{O}(c^{-6}) , & \delta\ddot{\phi} & \sim
\mathcal{O}(c^{-11})\,,
\end{align}
while from Eq.~\eqref{eq:psiN} and $\cos i_L = \uvec{L} \cdot \uvec{N}$, one 
finds
\begin{align}
 \dot{\psi}_N &\sim \mathcal{O}(c^{-6}) , & \ddot{\psi}_N & \sim
\mathcal{O}(c^{-11}),  \nonumber\\
 \dot{i}_L &\sim \mathcal{O}(c^{-6}) , & \ddot{i}_L & \sim
\mathcal{O}(c^{-11}).
\end{align}

Clearly then, $\ddot{\phi}_{\C}$, $\delta\ddot{\phi}$, $\ddot{\psi}_N$ and 
$\ddot{i}_L$ 
are all of the same order, and thus, they must all be promoted to the rapidly 
varying signal phase. 
The phase $\phi$ in $h_{n,+,\times}$ can be put into an exponential via 
Euler's formula. Similarly, 
the polarization angle $\psi_N$ can be included in the phase by rewriting the 
antenna pattern functions
and the harmonic polarizations in Eqs.~\eqref{h-time-domain-plus} 
and~\eqref{h-time-domain-cross} via Euler's formula 
as 
\begin{align}
 F_+ &= \frac{1}{2} \left( \mathcal{A}_F  + i \mathcal{B}_F  \right)
e^{2i\psi_N} + {\rm{c.c.}} \, ,\\
 F_\times &= \frac{1}{2} \left( \mathcal{B}_F  - i \mathcal{A}_F  \right)
e^{2i\psi_N} + {\rm{c.c.}} \, , \\
h_{n,+} &= \frac{1}{2} \left(\mathcal{A}_{n,+} - i \mathcal{B}_{n,+} \right) 
e^{in(\phi_{\C} + \delta\phi)} + {\rm{c.c.}} \, , \\
h_{n,\times} &= \frac{1}{2} \left(\mathcal{A}_{n,\times} - i 
\mathcal{B}_{n,\times} \right) 
e^{in(\phi_{\C} + \delta\phi)} + {\rm{c.c.}} \,,
\end{align}
where we have defined the slowly-varying amplitudes
\begin{align}
\mathcal{A}_F &= \frac{1}{2} \left( 1 + \cos^2 \theta_N
\right) \cos 2\phi_N, \\
\mathcal{B}_F &= \cos \theta_N \sin 2\phi_N\,.
\end{align}
Finally, the inclination angle $i_{L}$ can also be included in the signal phase
if the amplitudes $\mathcal{A}_{n,+}$, $\mathcal{B}_{n,+}$, 
$\mathcal{A}_{n,\times}$, and 
$\mathcal{B}_{n,\times}$ are rewritten as Fourier
series. 

Combining all of these results, one can rewrite Eq.~\eqref{h-time-domain} as
\begin{align}
 h(t) &= \sum_{n \geq 0} \ \sum_{k \in \mathbb{Z}} \  \sum_{m = -2,2} \
h_{n,k,m}(t) \\
h_{n,k,m} &= \mathcal{A}_{n,k,m}(\theta_N, \phi_N) e^{i ( n\phi_{\C} + 
n\delta\phi + k
i_L + m \psi_N)} + \rm{c.c.}, \label{eq:hoft}
\end{align}
where the slowly-varying amplitudes $\mathcal{A}_{n,k,m}$ can be computed 
from~\cite{abiq} and
are also given explicitly at 2PN order in Appendix~\ref{app:amplitudes}. 

\subsection{Preparing for a Uniform Asymptotic Expansion}
\label{sec:timedomainphases}

The time-domain waveform in Eq.~\eqref{eq:hoft} is almost ready for a uniform 
asymptotic treatment; 
the last step is to convert $\phi_{\C} + \delta \phi + \psi_{N} + i_{N}$ into a 
phase of the form $\Phi_{\C} + \alpha(t) \cos(\beta(t))$ as in 
Eq.~\eqref{eq:correctionsimple}. Recall that here $\alpha(t)$ varies on the 
radiation reaction timescale, while $\beta(t)$ varies on the precession 
timescale.

The carrier phase can be solved for using standard techniques as a function of 
the orbital frequency:
\begin{align}
 \phi_{\C} &= \int \frac{\xi^3}{M} dt = \int \frac{\xi^2}{Ma} d\xi \nonumber\\
 &= \phi_\coal - \frac{3}{5 a_0} \xi^{-5} \bigg[ 1 - \frac{5 a_2}{3} \xi^2 -
\frac{5}{2} a_3 \xi^3 
\nonumber\\
&- 5 \left( a_4 - a_2^2 \right) \xi^4 + 5 \left( a_5 - 2 a_2 a_3 \right) \xi^5
\log\xi \bigg] + \mathcal{O}\left( c^{-1}\right), \label{eq:phi0}
\end{align}
where we used $d\xi/dt = a\xi$ with $\xi = (M\omega)^{1/3} +
\mathcal{O}(\epsilon^2)$ as in the previous section; recall that the $a_{i}$
coefficients are given in Appendix~\ref{app:freqevol}, and should be evaluated
at $t=0$. In our implementation, in order to
isolate the effects of spin precession, we artificially increase the order of
the above equation to 6PN; that is, we keep terms in
$\dot{\omega}$ up to relative $\mathcal{O}(c^{-5})$, but we also keep
all induced terms up to
relative 
$\mathcal{O}(c^{-12})$. The resulting PN series contains terms of relative
$\mathcal{O}(c^{-6})$ 
and higher that will not match the expected result from full GR; yet, they
provide a more accurate 
result for the integration of the truncated $\dot{\omega}$ equation relative to
a numerical solution. 
The exact form of Eq.~\eqref{eq:phi0} that we used in
our comparisons
is given in Appendix~\ref{app:freqevol}.

In principle, the remaining phase terms can be rewritten in the desired form 
only
if one first distinguishes between two complementary cases: 
\begin{align}
 &\mbox{case 1:} & \hat{N}_x^2 + \hat{N}_y^2 \sim \mathcal{O}(\epsilon^0) ,
\nonumber\\
 &\mbox{case 2:} & \hat{N}_x \lesssim \mathcal{O}(\epsilon), \quad
\hat{N}_y
\lesssim \mathcal{O}(\epsilon) . \nonumber
\end{align}
Assuming case (i), the equation of motion for the correction to the orbital 
phase is
\begin{align}
\delta \dot{\phi}  &= 
\epsilon \frac{\hat{N}_z}{1 -
\hat{N}_z^2}\Bigg[ \hat{N}_x \left( \frac{\dot{L}_y^{(1)}}{L_z^{(0)}} - 
\frac{\dot{L}_z^{(0)} L_y^{(1)}}{\left.L_z^{(0)}\right.^2} \right) \nonumber\\
&-
\hat{N}_y \left( \frac{\dot{L}_x^{(1)}}{L_z^{(0)}} - 
\frac{\dot{L}_z^{(0)} L_x^{(1)}}{\left.L_z^{(0)}\right.^2} \right) \Bigg]
+ \mathcal{O}\left(\epsilon^2\right) \nonumber\\
&= 
\epsilon \frac{\hat{N}_z}{1 -
\hat{N}_z^2} \frac{d}{dt} \left[ \hat{N}_x \frac{L_y^{(1)}}{L_z^{(0)}} -
\hat{N}_y \frac{L_x^{(1)}}{L_z^{(0)}} \right] +
\mathcal{O}\left(\epsilon^2\right)\,,
\end{align}
and therefore
\begin{align}
 \delta\phi = \epsilon \frac{\hat{N}_z}{1 -
\hat{N}_z^2} \left[ \hat{N}_x \frac{L_y^{(1)}}{L_z^{(0)}} -
\hat{N}_y \frac{L_x^{(1)}}{L_z^{(0)}} \right] +
\mathcal{O}\left(\epsilon^2\right).
\label{eq:dphi1}
\end{align}
Similarly, in case (i) the inclination phase becomes 
\begin{align}
i_L &= \arccos \left( \hat{N}_z \right) - \epsilon \frac{(L_x^{(1)} \hat{N}_x +
L_y^{(1)} \hat{N}_y)}{L_z^{(0)} \sqrt{\hat{N}_x^2 + \hat{N}_y^2}} + 
\mathcal{O}\left( \epsilon^2 \right).
\end{align}
and the polarization angle is 
\begin{widetext}
\begin{align}
 \psi_N &= \arctan \left( \frac{\hat{z}_z - \hat{N}_z \uvec{N} \cdot 
\uvec{z}}{\hat{N}_y \hat{z}_x - \hat{N}_x \hat{z}_y} \right) + 
\epsilon \bigg\{ \frac{(\hat{N}_y \hat{z}_x - \hat{N}_x 
\hat{z}_y)[\hat{L}_x^{(1)} \hat{z}_x + \hat{L}_y^{(1)} \hat{z}_y - 
(\hat{L}_x^{(1)} \hat{N}_x + \hat{L}_y^{(1)} \hat{N}_y) \uvec{N} \cdot 
\uvec{z}]}{L_z^{(0)} [(\hat{N}_y 
\hat{z}_x - \hat{N}_x \hat{z}_y)^2 + (\hat{z}_z - \hat{N}_z \uvec{N} \cdot 
\uvec{z})^2]} \nonumber\\
&+ \frac{(\hat{z}_z - \hat{N}_z 
\uvec{N} \cdot 
\uvec{z} )[L_x^{(1)}(\hat{N}_y \hat{z}_z - \hat{N}_z \hat{z}_y) + 
L_y^{(1)}(\hat{N}_z \hat{z}_x - \hat{N}_x \hat{z}_z)] }{L_z^{(0)} [(\hat{N}_y 
\hat{z}_x - \hat{N}_x \hat{z}_y)^2 + (\hat{z}_z - \hat{N}_z \uvec{N} \cdot 
\uvec{z})^2]} \bigg\} .
\end{align}
\end{widetext}
Case (ii) leads at first to different expressions, but when these are 
re-expanded
in the PN approximation, assuming that $L_{x,y}^{(1)}/L_{z}^{(0)} \ll 1$, one 
recovers
the above expressions. Furthermore, the expansions for $\psi_N$ also depend on 
whether
$\uvec{z}$ is nearly aligned with $\uvec{N}$ or not. But as before, the results 
obtained when 
they are not aligned is recovered by re-expanding the nearly aligned result in 
a 
PN expansion. 
However, we expect our result not to yield a match to the numerical 
solutions as good when $\uvec{N}$ or $\uvec{z}$ are nearly aligned.

Using the results from Eqs.~(\ref{eq:LzNAS}-\ref{eq:LyNAS}), we can express 
$i_{N}$, $\psi_{N}$
and $\delta \phi$ in a Fourier series of the precession phases $\phi_{\PP,+}$ 
and 
$\phi_{\PP,-}$. That is, we can rewrite the phase modulation in 
Eq.~\eqref{eq:hoft} as
\begin{widetext}
\begin{align}
n\delta\phi + k i_L + m \psi_N &= A_{0,n,k,m} + n A_{\delta\phi,+} \cos 
(\phi_{\PP,+} + 
\phi_{\delta\phi,+}^{(0)}) 
+ n A_{\delta\phi,-} \cos (\phi_{\PP,-} + \phi_{\delta\phi,-}^{(0)}) + k 
A_{i_L,+} \cos (\phi_{\PP,+} + \phi_{i_L,+}^{(0)}) 
\nonumber\\
&+  k A_{i_L,-} \cos (\phi_{\PP,-} + \phi_{i_L,-}^{(0)}) + m 
A_{\psi_N,+} \cos (\phi_{\PP,+} + \phi_{\psi_N,+}^{(0)}) 
+ m A_{\psi_N,-} \cos (\phi_{\PP,-} + \phi_{\psi_N,-}^{(0)})  + 
{\cal{O}}(\epsilon,c^{-1})
\nonumber\\
&=A_{0,n,k,m} +  A_{+,n,k,m} \cos(\phi_{\PP,+} + \phi_{+,n,k,m}) + A_{-,n,k,m} 
\cos(\phi_{\PP,-} 
+ \phi_{-,n,k,m}) + {\cal{O}}(\epsilon,c^{-1}),
\label{conversion}
\end{align}
\end{widetext}
where the amplitudes $A_{0,n,k,m}$ and $A_{\alpha,\pm}$, and the phases 
$\phi_{\alpha,\pm}^{(0)}$ 
are given in Appendix~\ref{app:amplitudeandphasemodulations}, while the harmonic 
amplitudes are given 
by
\begin{align}
 A_{\pm,n,k,m} &= \mbox{sign}(A_{c,\pm}) \sqrt{A_{c,\pm}^2 + A_{s,\pm}^2}, \\
 \phi_{\pm,n,k,m} &= \arctan \left( \frac{A_{s,\pm}}{A_{c,\pm}} \right), \\
 A_{c,\pm} &= n A_{\delta\phi,\pm} \cos (\phi_{\delta\phi,\pm}^{(0)}) + k 
A_{i_L,\pm} \cos (\phi_{i_L,\pm}^{(0)}) \nonumber\\
&+ m 
A_{\psi_N,\pm} \cos ( \phi_{\psi_N,\pm}^{(0)}), \\
A_{s,\pm} &= n A_{\delta\phi,\pm} \sin 
(\phi_{\delta\phi,\pm}^{(0)}) + k 
A_{i_L,\pm} \sin (\phi_{i_L,\pm}^{(0)}) \nonumber\\
&+ m 
A_{\psi_N,\pm} \sin ( \phi_{\psi_N,\pm}^{(0)}) ,
\end{align}
This then puts the time-domain waveforms in the desired form to carry out a
uniform asymptotic expansion.

Before proceeding, let us comment on the remainders of Eq.~\eqref{conversion}. 
In going from the left-hand side of this equation to the right-hand side, we 
have neglected
terms of ${\cal{O}}(\epsilon)$ and terms of ${\cal{O}}(c^{-1})$, when 
$\uvec{N}$ 
and $\uvec{J}$
are aligned. When these vectors are misaligned, the remainders are actually 
smaller, namely
of ${\cal{O}}(\epsilon^{2})$. We will see later on that the neglect of 
higher-order terms in $c^{-1}$
is the dominant source of discrepancy between our analytic frequency-domain 
waveforms
and the DFT of numerical time-series waveforms.

\subsection{Frequency-Domain Gravitational Waveform via Uniform Asymptotic 
Expansions}
\label{sec:GW}

We are interested in the Fourier transform of the GW signal. 
Taking advantage of the linearity of the Fourier transform, Eq.~\eqref{hf} can 
be rewritten as
\begin{align}
  \tilde{h}(f) &= \sum_{n \geq 0} \ \sum_{k \in \mathbb{Z}} \  \sum_{m = -2,2} \
\tilde{h}_{n,k,m}(t),
\end{align}
where the Fourier harmonic components are
\begin{align}
 \tilde{h}_{n,k,m}(f) &= \int \mathcal{A}_{n,k,m}^* e^{i(2 \pi ft - n \phi_{\C} 
- n
\delta\phi - k i_L - m \psi_N)} dt \nonumber\\
&+ \int \mathcal{A}_{n,k,m} e^{i(2 \pi ft + n
\phi_{\C} + n
\delta\phi + k i_L + m \psi_N)} dt, \label{eq:htildedef}
\end{align}
and recall that the star denotes complex conjugation.

Our particular asymptotic uniformization requires that we transform the above 
integrands via
\begin{align}
&e^{-i(n
\delta\phi + k i_L + m \psi_N)} \nonumber\\
&= e^{-i A_{0,n,k,m}} \sum_{\{k_+,k_-\} \in \mathbb{Z}^2}  J_{k_+} ( 
A_{+,n,k,m} 
) J_{k_-} ( 
A_{-,n,k,m} ) \nonumber\\
&\times e^{-i [k_+ 
(\phi_{\PP,+} + \phi_{+,n,k,m} + 
\pi/2) + k_- (\phi_{\PP,-} + \phi_{-,n,k,m} + 
\pi/2)]}, 
\end{align}
and similarly for the second term.

After this transformation, we can apply the SPA to 
compute the integrals in Eq.~\eqref{eq:htildedef}.
Since $\dot{\phi}_{\C} \gg \dot{\phi}_{\PP,\pm}$, we can safely neglect the 
second term as it will only contribute to negative frequencies.
We then obtain
\begin{widetext}
\begin{align}
 \tilde{h}_{n,k,m}(f) &= \sum_{\{k_+,k_-\} 
\in \mathbb{Z}^2} \sqrt{\frac{2\pi}{\ddot{\phi}_{\C} + k_+ \ddot{\phi}_{\PP,+} 
+ k_- \ddot{\phi}_{\PP,-}}} 
\mathcal{A}_{n,k,m}^* J_{k_+} ( A_{+,n,k,m} )  J_{k_-} (A_{-,n,k,m}  ) 
\nonumber \\
& \qquad \times\exp[ i (2\pi f t - n \phi_{\C} - A_{0,n,k,m} - k_+  
(\phi_{\PP,+} + 
\phi_{+,n,k,m} +  \pi/2) 
- k_- (\phi_{\PP,-} + \phi_{-,n,k,m} + \pi/2) - \pi/4],
\label{htildef-nkm}
\end{align}
\end{widetext}
where all time dependent functions are evaluated at $t=t_\SPA$, defined via
\begin{align}
 2 \pi f = n\dot{\phi}_{\C}(t_\SPA) + k_+ \dot{\phi}_{\PP,+}(t_\SPA) + k_- 
\dot{\phi}_{\PP,-}(t_\SPA).
\end{align}
We can invert the above equation to find
\begin{align}
 \xi &= u - \frac{1}{24 n} [ k_+(7 + 6 \delta m - \delta m^2) + k_- (7 - 6 
\delta m - \delta m^2)  ] u^3 \nonumber\\
 &+ \frac{1}{24 n} \{ k_+ [ 2( 1 - \delta m)^2 \chi_2 - (7 + 8 \delta m + 
\delta m^2)\chi_1] \nonumber\\
&+ k_- [ 2( 1 + \delta m)^2 \chi_1 - (7 - 8 \delta 
m + 
\delta m^2)\chi_2] \} u^4 + \mathcal{O}(u^5), 
\label{xi-eq} 
\end{align}
where the dimensionless mass difference $\delta m = (m1 - m_2)/M$, and we have 
defined the reduced frequency parameter
\begin{align}
 u &= \left(\frac{2\pi M f}{n}\right)^{1/3},
 \label{u-eq}
\end{align}
and integrate $dt = \int (d\xi/dt)^{-1} d\xi$ to find
\begin{align}
 t_\SPA &= t_\coal - \frac{3M}{8a_0} \xi^{-8} \bigg[ 1 - \frac{4 a_2}{3} \xi^2 -
\frac{8 a_3}{5} \xi^3 \nonumber\\
&+ 2 \left( a_2^2 - a_4 \right) \xi^4 +
\frac{8}{3} \left(
2 a_2 a_3 - a_5 \right) \xi^5  + \mathcal{O}(c^{-6}) \bigg] .\label{eq:toff}
\end{align}
In our implementation, similar to Eq.~\eqref{eq:phi0}, we chose to artificially 
increase the above equation to 6PN. The exact result can be found in 
Appendix~\ref{app:freqevol}.

By inspecting the results of Sec.~\ref{sec:timedomainphases}, we can see that 
$A_{\pm,n,k,m} \sim \mathcal{O} (c^{-1})$, and therefore the Bessel functions 
$J_{k_\pm}(A_{\pm,n,k,m})$ will be rapidly suppressed for high values of 
$k_{\pm}$. This suggests that only a few terms may be needed in the Bessel
expansion to accurately approximate the Fourier transform of the time-domain 
signal.  

\section{Waveform Comparisons}
\label{sec:comp}

In this section, we study how well the analytic frequency-domain waveform 
calculated in the previous section compares to others presented in the 
literature. First, we compare the phase and amplitude of the waveforms against 
each other. Then, we use a {\emph{faithfulness}} measure to carry out 
integrated 
comparisons, without maximizing over intrinsic parameters. We perform a Monte 
Carlo study over a variety of systems with different spin misalignments, 
positions in the sky and relative orientation with respect to the detector 
plane.    

\subsection{Comparison Preliminaries}
\subsubsection{Waveform Models}

The waveforms we compare against each other are the following:
\begin{itemize}
\item {\bf{DFT:}} The discrete Fourier transform of the numerically-calculated, 
time-domain response function, tapered by a Tukey window to remove spectral 
leakage. The time-domain response is constructed from 
Eq.~\eqref{h-time-domain}, 
with all angular momenta and phases obtained numerically by solving the 
evolution 
equations in the time-domain.  
\item {\bf{UAA:}} The fully-analytic, frequency-domain, uniform asymptotic 
approximate waveform of Sec.~\ref{sec:GW}. 
\item {\bf{HSPA~\cite{Lang:1900bz}:}} A hybrid, semi-analytic, frequency-domain 
template, given by the non-precessing, spin-aligned SPA waveform with higher 
harmonics (un-restricted PN), where the spin couplings are promoted {\emph{a 
posteriori}} to functions of the frequency and the phase is enhanced by the 
precession correction $\delta \phi$ obtained by numerically integrating 
Eq.~\eqref{eq:phieqmot2}. All angular momenta are obtained by solving all 
evolution equations numerically in the time-domain, and then numerically 
inverting them to find $\bm{S}_{1,2}$ and $\bm{L}$ as a function of orbital 
frequency. 
\item {\bf{Aligned SPA:}} A non-precessing, spin-aligned, frequency-domain 
waveform, computed in the SPA with higher harmonics (un-restricted PN).
\end{itemize} 

The different waveforms described above have different advantages and 
disadvantages. Perhaps the most accurate one is the DFT family, where the only 
mis-modeling systematic is induced by numerical error, from the numerical 
solutions to the evolution equations and the DFT. Unfortunately, however, this 
is also the most computationally expensive family to evaluate and the one that 
provides the least analytical insight. The aligned SPA family contains the 
largest 
mis-modeling systematics, since it attempts to model the system as 
non-precessing, but it is also the cheapest to evaluate. The HSPA family is 
somewhere in between the DFT and aligned SPA families, being computationally 
less 
expensive than DFT, but containing some systematics due to the improper use of 
the stationary phase approximation. Moreover, although less expensive than DFT, 
the HSPA family is more 
expensive to evaluate than the analytical waveforms, since each template 
requires the numerical solution and inversion of the evolution equations.  

Let us make an important clarification regarding the DFT family. In multiple
scale analysis, one usually compares approximations to some exact answer to
determine, for example, the region of validity and accuracy of the former. Here, 
however,
we lack such an exact solution. The DFT is perhaps the closest quantity to an 
exact Fourier
transform that we possess, but of course, it is not an exact solution, as 
numerical error is
non-negligible and filtering has been employed to prevent spectral leakage. We 
have checked, however, that the DFT is robust upon changes to the Tukey filter 
and numerical resolution. 
Thus, we here adopt the DFT as ``exact'' and compare the different 
approximations to it.

Care must be exercised, however, when comparing analytical and numerical 
spinning waveforms. Even when spins are exactly aligned with the orbital 
angular 
momentum, the analytical expansion of the carrier phase does not match the 
numerical integration of Eq.~\eqref{eq:phi0} to sufficiently high accuracy. 
Similarly, the analytic, perturbative inversion of the time-frequency relation, 
Eq.~\eqref{eq:toff}, is not sufficiently accurate relative to the numerical 
inversion. Therefore, to isolate spin precession effects, we will keep terms in 
Eqs.~\eqref{eq:phi0} and~\eqref{eq:toff} up to 6PN order. This is sufficient
to guarantee that any discrepancies in the compared waveforms arise due
to spinning effects only. The exact relations we use in our comparisons 
are given in Appendix~\ref{app:freqevol}.

\subsubsection{Detector Models}

The comparisons of response functions are, of course, sensitive to the 
particular 
detector considered. We here consider both a typical ground-based detector and 
a 
typical space-based detector, both in the long-wavelength approximation. 

Different detectors will operate in different frequency bands, for different 
observation times, and they will lead to different relations between the 
detector frame and
a  fixed frame tied to the distant stars. The latter will impact the functional
form of the angles $\theta_N$, $\phi_N$, and $\psi_N$ in 
Eqs.~(\ref{eq:Fplus}-\ref{eq:psiN}): 
for a typical ground-based detector, since the observation time is very short,
we can approximate the angles $\theta_N$ and $\phi_N$ as constant; for a 
typical space-based detector, the
observation time is not short, and thus, one must properly model the 
time-dependence of the
angles. We here use an eLISA configuration~\cite{elisa}, where a LISA-type 
configuration trails
behind Earth at a rate of $7.5^\circ$ per year. 

The relation between the detector frame $(\uvec{x}_\dett,
\uvec{y}_\dett, \uvec{z}_\dett)$ and a frame tied to the distant stars 
$(\uvec{x},
\uvec{y}, \uvec{z})$ for the space-based detector is given by
\begin{align}
 \uvec{x}_\dett &= \left( \frac{3}{4} - \frac{1}{4} \cos 2 \Phi_\eLISA(t) 
\right)
\uvec{x} - \frac{1}{4} \sin 2 \Phi_\eLISA(t)
\uvec{y}\nonumber \\
&+
\frac{\sqrt{3}}{2} \cos \Phi_\eLISA(t) \uvec{z}, \\
 \uvec{y}_\dett &= - \frac{1}{4} \sin 2 \Phi_\eLISA(t) \uvec{x} + \left(
\frac{3}{4} + \frac{1}{4} \cos 2 \Phi_\eLISA(t) \right) \uvec{y}
\nonumber\\
&+
\frac{\sqrt{3}}{2} \sin \Phi_\eLISA(t) \uvec{z}, \\
 \uvec{z}_\dett &= - \frac{\sqrt{3}}{2} \cos \Phi_\eLISA(t) \uvec{x} -
\frac{\sqrt{3}}{2} \sin \Phi_\eLISA(t) \uvec{y} + \frac{1}{2}
\uvec{z},
\end{align}
where the detector barycenter is located at 
$(\cos{\Phi_\eLISA(t)},\sin{\Phi_\eLISA(t)},0)$ in the Solar System frame with 
$\Phi_\eLISA(t) = \omega_{\eLISA} \; t$ and $\omega_{\eLISA} = 2 \pi 
(352.5/360) \; \mathrm{yr}^{-1}$. In addition, we modify the carrier phase by 
adding the so-called Doppler term~\cite{cutler1998}
\begin{align}
 \phi_\C \to \phi_\C + \omega R \sin {\theta}_N \cos
(\Phi_\eLISA(t) -
{\phi}_{N}),
\end{align}
due to the fact that the barycenter of the detector moves in the frame tied to 
the distant 
stars. 
Here, $R = 1$~AU, and $\theta_N$ and $\phi_{N}$ are the spherical 
angles of 
$\uvec{N}$ in the frame tied to the distant stars.

\subsubsection{Comparison Measures}

We use two distinct comparison measures: 
\begin{itemize}
\item {\bf{Waveform Comparison.}} A direct waveform amplitude and  phase 
comparison as a function of GW frequency and PN expansion parameter.
\item {\bf{Match Comparison.}} An integrated overlap waveform comparison, with 
white noise and without maximizing over intrinsic parameters.
\end{itemize}
The waveform comparison consists of comparing the Fourier amplitudes (and 
the Fourier phases) computed with different waveform families against each 
other,
as a function of the dimensionless PN parameter $x$ and the GW frequency in Hz. 
The dimensionless PN parameter $x$ corresponding to the frequency $f$ for 
harmonic 
$n$ is computed using the standard SPA relation $x^{3/2} = 2 \pi M f/n$. 

When comparing amplitudes and phases, we isolate spin precession effects by
normalizing or subtracting by the controlling factors in the non-precessing 
case. 
The Fourier amplitude of the $n$-th harmonic is normalized by the 
amplitude pre-factor of the SPA: 
\be
{\cal{A}}_{0} = \sqrt{5 \pi \nu} \frac{M^{2}}{8 D_{L}} \left( \frac{2 
\pi M f}{n} \right)^{-7/6}\,,
\ee
The Fourier phase is subtracted from the non-precessing SPA phase:
\be
\Psi_{0} = 2 \pi f t(f) - n \phi_{\C}(f) - \frac{\pi}{4}\,,
\ee
where $t(f)$ and $\phi_{\C}[t(f)]$ are obtained from the numerical inversion 
of $n \dot{\phi}_{\C}(t) = 2 \pi f$ and from the numerical solution to the 
evolution equations, respectively.

The match comparison is carried out through the so-called 
{\emph{faithfulness}}:
\begin{align}
 F_{\tilde{h}_1,\tilde{h}_2} &\equiv
\frac{\left(\tilde{h}_{1}\left|\right.\tilde{h}_{2}\right)}{\sqrt{\left(\tilde{h
}_{1}\left|\right.\tilde{h}_{1}\right)
\left(\tilde{h}_{2}\left|\right.\tilde{h}_{2}\right)}}\,,
\label{match}
\end{align}
where $\tilde{h}_{1,2}$ are different Fourier-domain waveforms with the 
{\emph{same}} physical 
parameters. We define the inner-product in the usual way:
\begin{align}
 \left(\tilde{h}_{1}\left|\right.\tilde{h}_{2}\right) &\equiv 
4 \Re \int_{f_{\min}}^{f_{\max}} \frac{\tilde{h}_1 \tilde{h}_2^*}{S_{n}} \; 
df\,,
\end{align}
where $\Re[\cdot]$ the real part operator, $(f_{\min},f_{\max})$ are the
boundaries of the detector's sensitivity band and $S_{n}$ is the detector's 
spectral noise 
density. For ground-based detectors, we choose $f_{\min}=10$ Hz and 
$f_{\max}=10^{3}$ Hz, with a maximum observation time of $1$ hr. For 
space-based detectors, we choose $f_{\min} = 10^{-5}$ Hz and $f_{\max}=1$ 
Hz, with a maximum observation time of $2$ yrs. In all cases, we also terminate
the comparisons if the system reaches a separation of 6 times the total mass, 
i.e.~the innermost stable circular orbit of a point particle in a Schwarzschild 
spacetime, 
prior to reaching the frequency $f_{\max}$. We here employ white noise, so 
$S_{n}$ can be taken out of the integral and it cancels when computing the 
faithfulness parameter. We expect the fitting factor, maximized over physical 
parameters, to be in general higher when computed with colored noise than when 
computed with white noise. This is because colored noise has the effect of 
weighing one part of the frequency spectrum more heavily, and thus, the 
maximization occurs in a reduced frequency window, leading to a higher overlap.

The faithfulness parameter exists in the interval $[-1,1]$ and it indicates how 
well waveforms agree with each other, with unity representing perfect agreement. 
The
integrations required are carried out numerically, with errors of 
${\cal{O}}(10^{-5})$; thus, we consider $F_{\tilde{h}_{1}\tilde{h}_{2}} = 
0.9999$ to be consistent 
with perfect agreement. A $98\%$ fitting factor is sometimes argued to be ``good 
enough'' for detection
purposes,  but this of course depends on the detection tolerance chosen.

Both when using a waveform measure or a match measure, we will compare the 
UAA and aligned SPA models to the DFT model. That is, we will treat the DFT 
model as a reference waveform, say $\tilde{h}_{2}$, and let $\tilde{h}_{1}$ 
be the UAA, the aligned SPA or the HSPA waveform.

The comparisons we show below should be considered {\emph{conservative}} 
because 
the faithfulness parameter is maximized only over the non-physical parameters 
$t_{\coal}$ (appearing in the Fourier phase of the analytical models 
through Eq.~\eqref{eq:toff}) and $\phi_{\coal}$ (appearing in the Fourier phase 
of the analytical models through Eq.~\eqref{eq:phi0}). All other 
parameters, including the physical ones, such as the total mass and 
mass ratio, are kept unchanged. Higher matches would be obtained by allowing 
all 
parameters to vary, as is done in parameter estimation. 

\subsubsection{Systems Considered}

\begin{figure*}[th!]
\begin{center}
\includegraphics[width=\columnwidth]{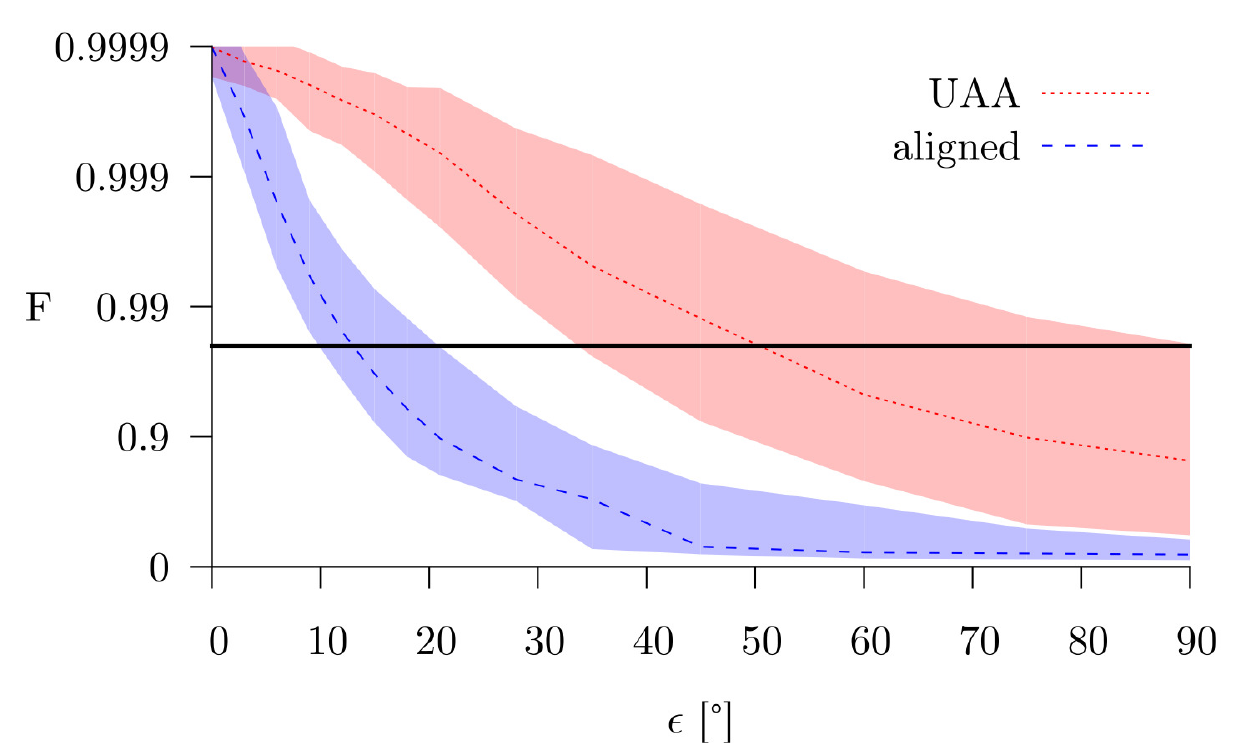}
\includegraphics[width=\columnwidth]{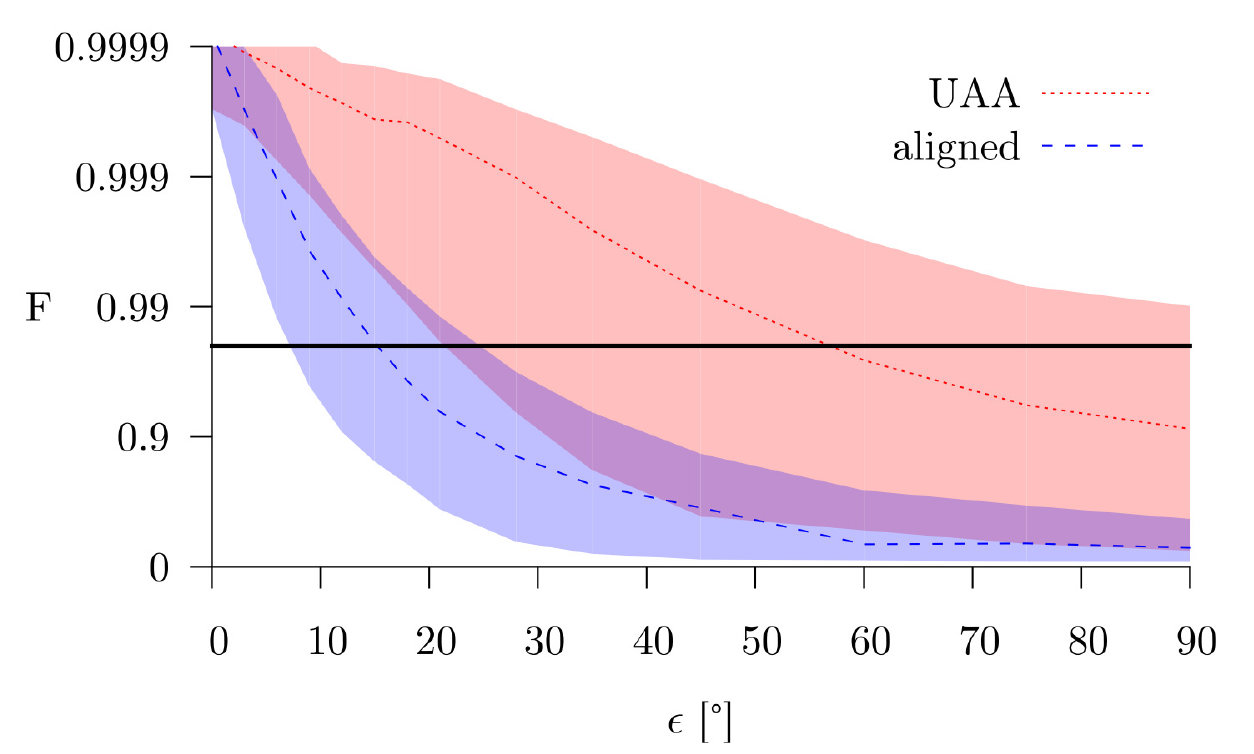}
\caption{\label{fig:misalignment}Median faithfulness and $1$-$\sigma$ 
deviations between DFT-UAA waveforms (red dotted curve)and DFT-aligned SPA 
waveforms (blue dashed curve), as a function of misalignment angle $\epsilon$ 
in degrees for ground-based (left panel) and space-based systems (right 
panels). 
All waveforms satisfy $\cos \epsilon = \uvec{L} \cdot \uvec{S}_1 = \uvec{L} 
\cdot \uvec{S}_2$. 
The solid horizontal line corresponds to a faithfulness of $98\%$. Observe how 
the UAA waveforms
are systematically better than the aligned-SPA ones for misalignments $\epsilon 
\gtrsim 5^{\circ}$.}
\end{center}
\end{figure*}
\begin{figure*}[h!t]
\begin{center}
\includegraphics[width=\columnwidth]{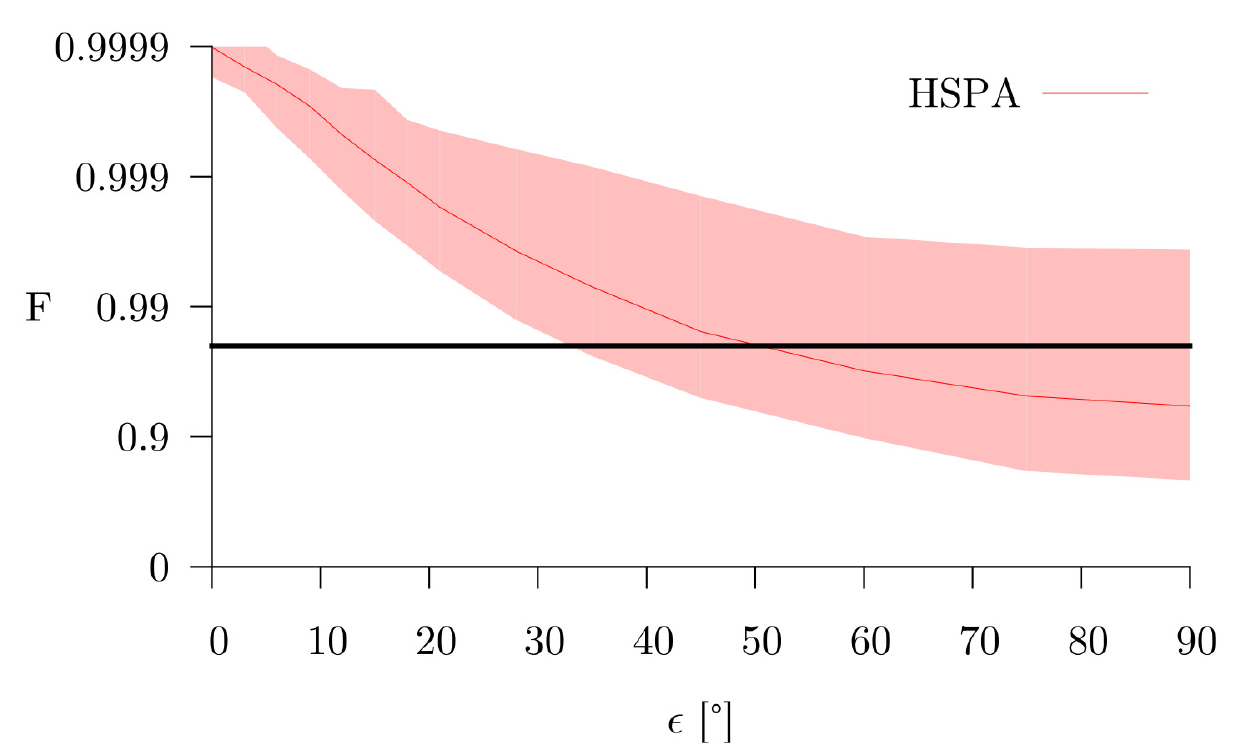}
\includegraphics[width=\columnwidth]{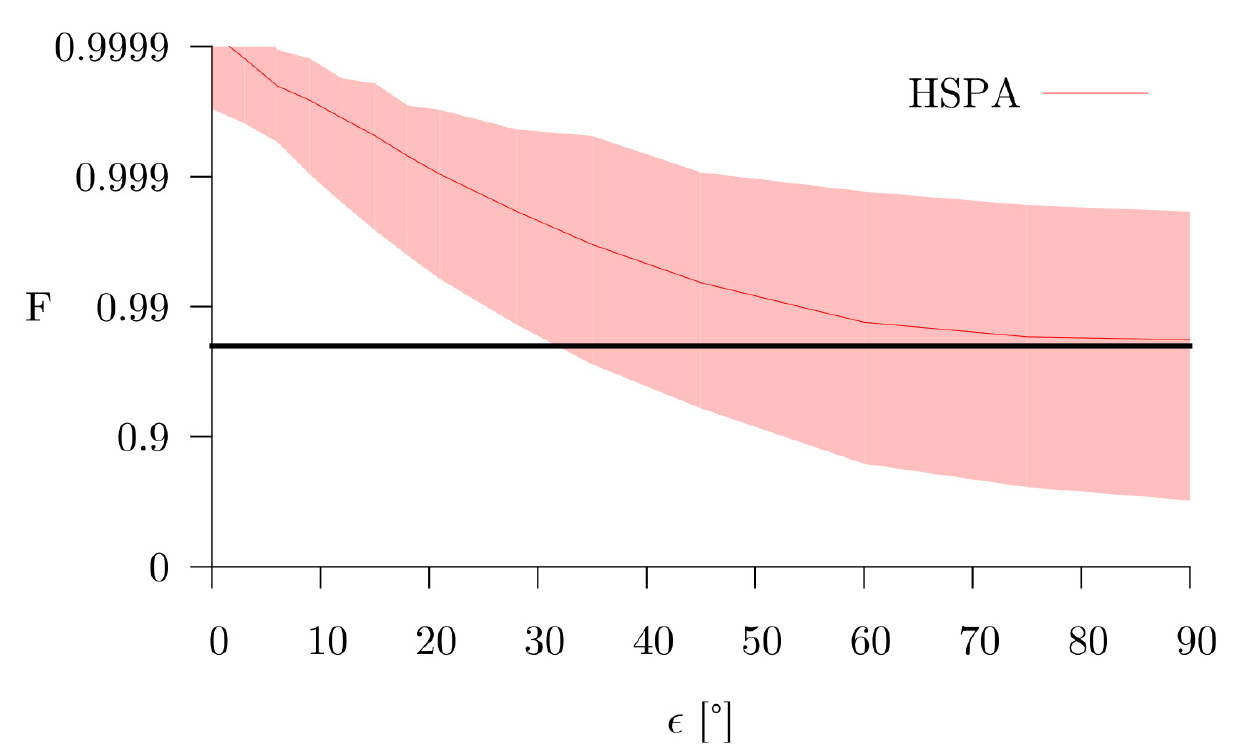}
\caption{\label{fig:misalignment-HSPA}Median faithfulness and $1$-$\sigma$ 
quantiles between DFT-HSPA
waveforms, as a function of the misalignment 
angle
$\epsilon$ in degrees 
for ground-based (left) and space-based systems (right). All 
waveforms used
for this plot satisfy
$\cos \epsilon = \uvec{L} \cdot \uvec{S}_1 = \uvec{L} \cdot \uvec{S}_2$. The
solid horizontal line corresponds to a faithfulness of $98\%$.}
\end{center}
\end{figure*}

Different systems will be studied when using different comparison measures. 
When 
using the waveform measure, we will investigate the following four systems:
\begin{itemize}
\item{\bf{System A}}: $\delta m = 0.5$, $\alpha_{0} = 57^{\circ}$,
\item{\bf{System B}}: $\delta m = 0.1$, $\alpha_{0} = 57^{\circ}$,
\item{\bf{System C}}: $\delta m = 0.5$, $\alpha_{0} = 23^{\circ}$,
\item{\bf{System D}}: $\delta m = 0.1$, $\alpha_{0} = 23^{\circ}$,
\end{itemize}
where $\alpha_{0}$ is the angle between the line of sight vector $\uvec{N}$ and 
the Newtonian orbital angular momentum vector $\bm{L}$ at $t=0$. For these four 
systems, we choose $(\chi_1,\chi_{2}) = (0.89,0.77)$, and initial 
misalignment angles of $22^\circ$ and $25^\circ$. When considering 
space-based detectors, we choose a total redshifted mass of $M = 5 \times 10^6 
M_\odot$ and a total observation time of $T_{\obs} = 2$~yrs. When considering 
ground-based detectors, we choose a total mass of $M = 20 M_{\odot}$ and a 
total 
observation time of $T_{\obs} = 100$~secs. We have also investigated other 
systems, but the results presented will be representative. 
 When computing a UAA waveform,
we use for $\delta\phi_\PP$ Eq.~\eqref{phipxi} for Systems $A$ and $C$, and 
Eq.~\eqref{phipDelta} 
for Systems $B$ and $D$ (recall that for systems with small mass differences,
different
PN expansions are needed). 

When using the match measure, we will perform a Monte-Carlo study over 200 
points
in parameter space for each type of detector, involving systems randomized over 
all 
waveform parameters. The misalignment angles will be set equal to each other, 
but
they will be allowed to vary between $0^{\circ}$ and $90^{\circ}$. 
All throughout, we consider typical systems for ground-based detectors with 
masses in $(5,20) M_\odot$, and systems for space-based detectors with masses 
in $(10^5,10^{8}) 
M_\odot$ and mass ratio $m_1/m_2 \leq 10$. The distribution of the spin 
magnitudes $\chi_1$ and $\chi_2$ is chosen to be flat in $[0,1]$, and the 
distributions of unit vectors are chosen to be flat on the sphere.

\subsection{Match Comparison}
\label{subsec:matchmeasure}

Fig.~\ref{fig:misalignment} shows the median match and $1$-$\sigma$ deviations
between DFT-UAA waveforms (red dotted curve) and DFT-aligned SPA waveforms (blue 
dashed curve), 
as a function of the misalignment angle $\epsilon$ in degrees for ground-based 
systems (left panel) 
and space-based systems (right panel). 

Several observations are due at this time. First, observe that the match for the 
aligned-SPA family is 
significantly worse than that of the UAA family, as soon as the system is even 
slightly misaligned. 
This is mostly due to the fact that the spin couplings in the phase evolution 
equation are 
greatly overestimated in the aligned-SPA model. Second, observe also that even 
for misalignment
angles around $50^\circ$ the UAA family achieves matches around $98\%$ for half 
the systems
considered.This is surprising given that UAA waveforms rely on an expansion in 
misalignment angle.
Third, observe that the lower $1$-$\sigma$ match deviation for space-based 
systems is 
significantly lower than that for ground-based systems. This is in part 
because of the impact 
of the detector's motion on the waveform, and in part because typical 
space-based systems spend more time 
in the detector band than ground-based ones, thus leading to more important 
phase discrepancies. 
Fourth, observe that the median and upper $1$-$\sigma$ match deviations are 
slightly better for 
space-based than for ground-based systems. 

Fig.~\ref{fig:misalignment-HSPA} shows a similar match calculation, but this 
time using
the HSPA family of~\cite{Lang:1900bz}. Observe that, for ground-based 
systems, these waveforms 
fail to provide a high median match for misalignments $\epsilon \gtrsim 
30^{\circ}$. Observe
also that similar poor behavior is observed for space-based systems, which have 
a lower 1-$\sigma$
deviation that dips below $98\%$ at roughly the same value of $\epsilon$. Recall 
that such 
poor behavior is in spite of HSPA waveforms using the same numerical solution to 
the precession
equations used to compute DFT waveforms. The poor behavior is because one of the 
requirements 
of the SPA used to derive HSPA waveforms (that the amplitude of the signal 
varies much more 
slowly than its phase) breaks down for highly misaligned systems. While the 
first time derivative 
of the phase is much larger than that of the amplitude, their second time 
derivatives are of the 
same order. One should keep in mind when comparing HSPA waveforms to UAA ones 
that the former require the numerical integration of the equations of 
precession, while the latter are fully analytic.

\subsection{Discontinuity in the Solution to the Equations of Precession}
\label{subsec:discontinuity}

One concern with the waveform family developed here is that the precession 
phase difference
$\delta \phi_\PP$ is a discontinuous function of the mass difference $\delta 
m$. This quantity
satisfies $\delta \phi_\PP = \delta \phi_{\PP,1}$ as given by 
Eq.~\eqref{phipxi} if $\delta m \geq 0.2$, 
$\delta \phi_\PP = \delta \phi_{\PP,2}$ as given by Eq.~\eqref{phipDelta} if 
$10^{-5} \leq \delta m < 0.2$
and $\delta \phi_\PP = \delta \phi_{\PP,3}$ as given by Eq.~\eqref{phipdeltam} 
if $\delta m < 10^{-5}$. 
Formally then, the waveform derivatives with respect to $\delta m$ are 
ill-defined
at the boundaries of the piecewise function. 

\begin{figure}[th]
\begin{center}
\includegraphics[width=\columnwidth]{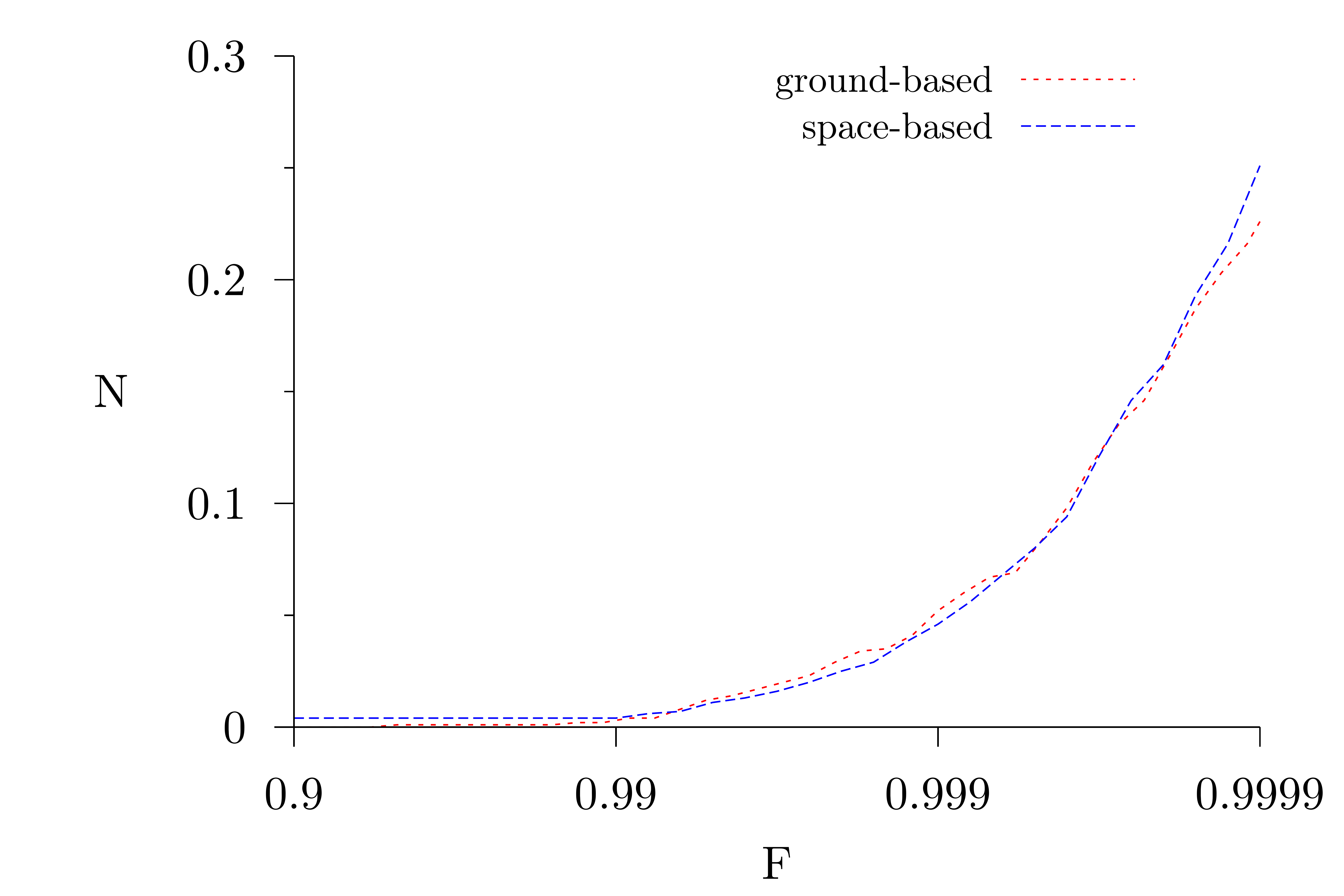}
\caption{\label{fig:misalignment-deltam} Cumulative distributions of 
faithfulnesses between
a waveform that uses $\delta \phi_{\PP,1}$ and one that uses $\delta 
\phi_{\PP,2}$ at $\delta m = 0.2$
for a ground-based (dotted red curve) and space-based (dashed blue curve) set of 
detections. Observe that
for over $95\%$ of the systems investigated, the match is higher than $0.999$, 
implying the
discontinuity would not have a serious effect in parameter estimation.}
\end{center}
\end{figure}

Let us then investigate whether this discontinuity is a problem. To do so, we 
compute the
match at $\delta m = 0.2$ between a waveform that uses $\delta \phi_{\PP,2}$ 
and one that uses
$\delta \phi_{\PP,1}$. Fig.~\ref{fig:misalignment-deltam} shows cumulative 
distributions of faithfulnesses for ground-based (dotted red curve) and 
space-based (dashed blue curve) detections. Observe
that the match is above $0.999$ for over $95\%$ of the systems investigated. 
This then implies
that the formal discontinuity in the waveform derivative with respect to $\delta 
m$ at the boundary
of the piecewise function should not affect parameter estimation.      

%

\subsection{Waveform Measure}
\label{subsec:waveformmeasure}

\begin{figure*}[!htp]
\begin{center}
\includegraphics[width=\columnwidth]{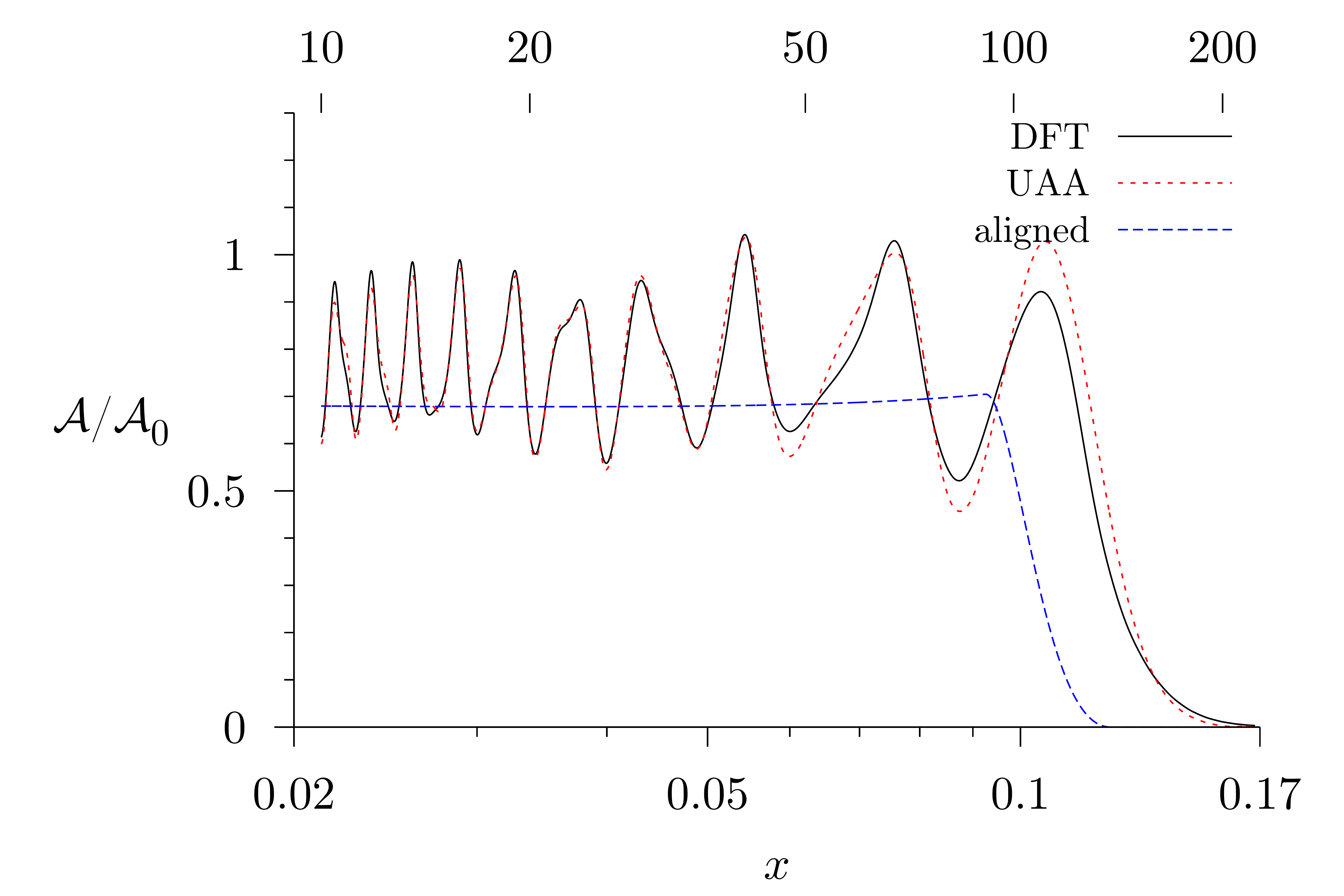}
\includegraphics[width=\columnwidth]{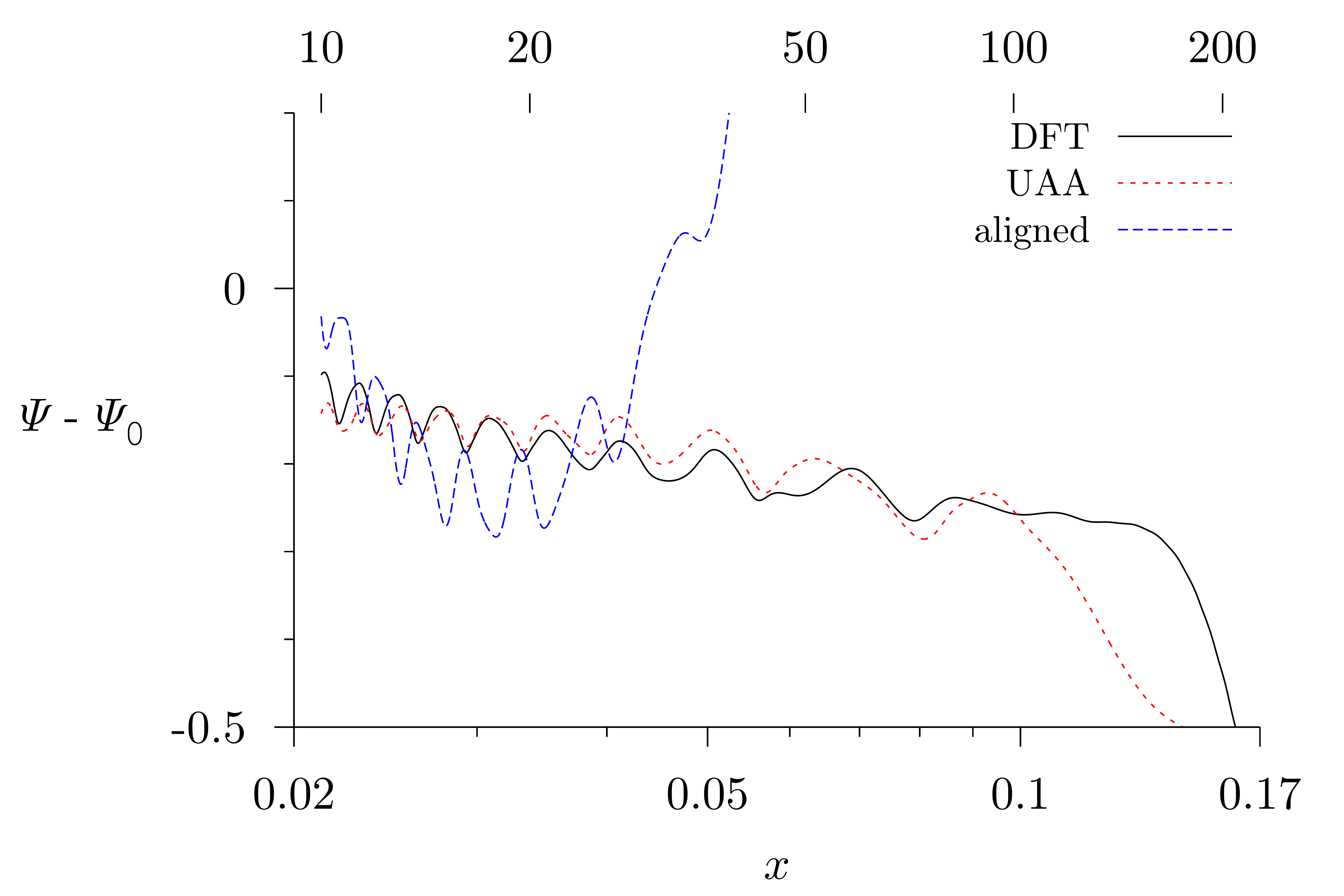} \\
\includegraphics[width=\columnwidth]{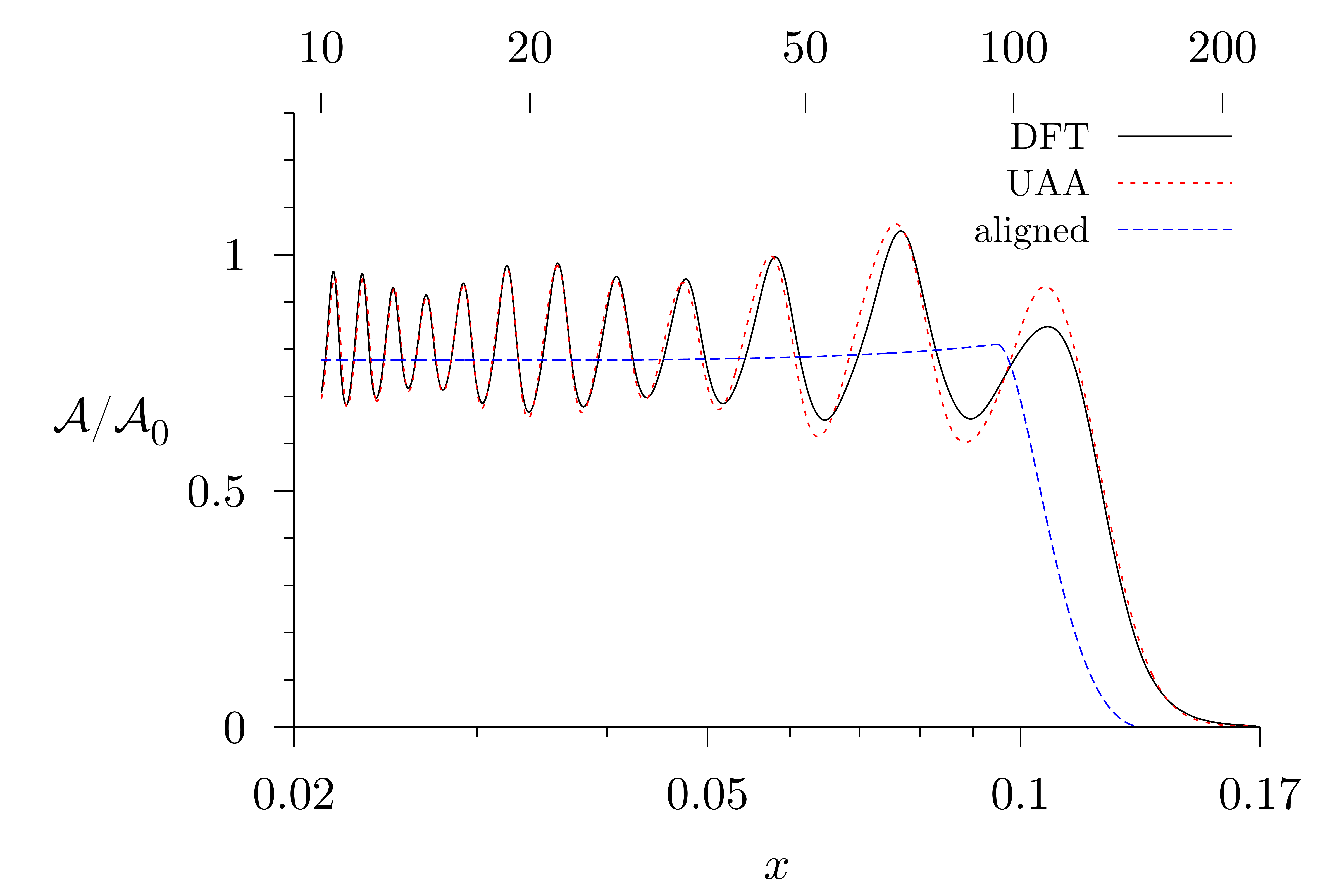}
\includegraphics[width=\columnwidth]{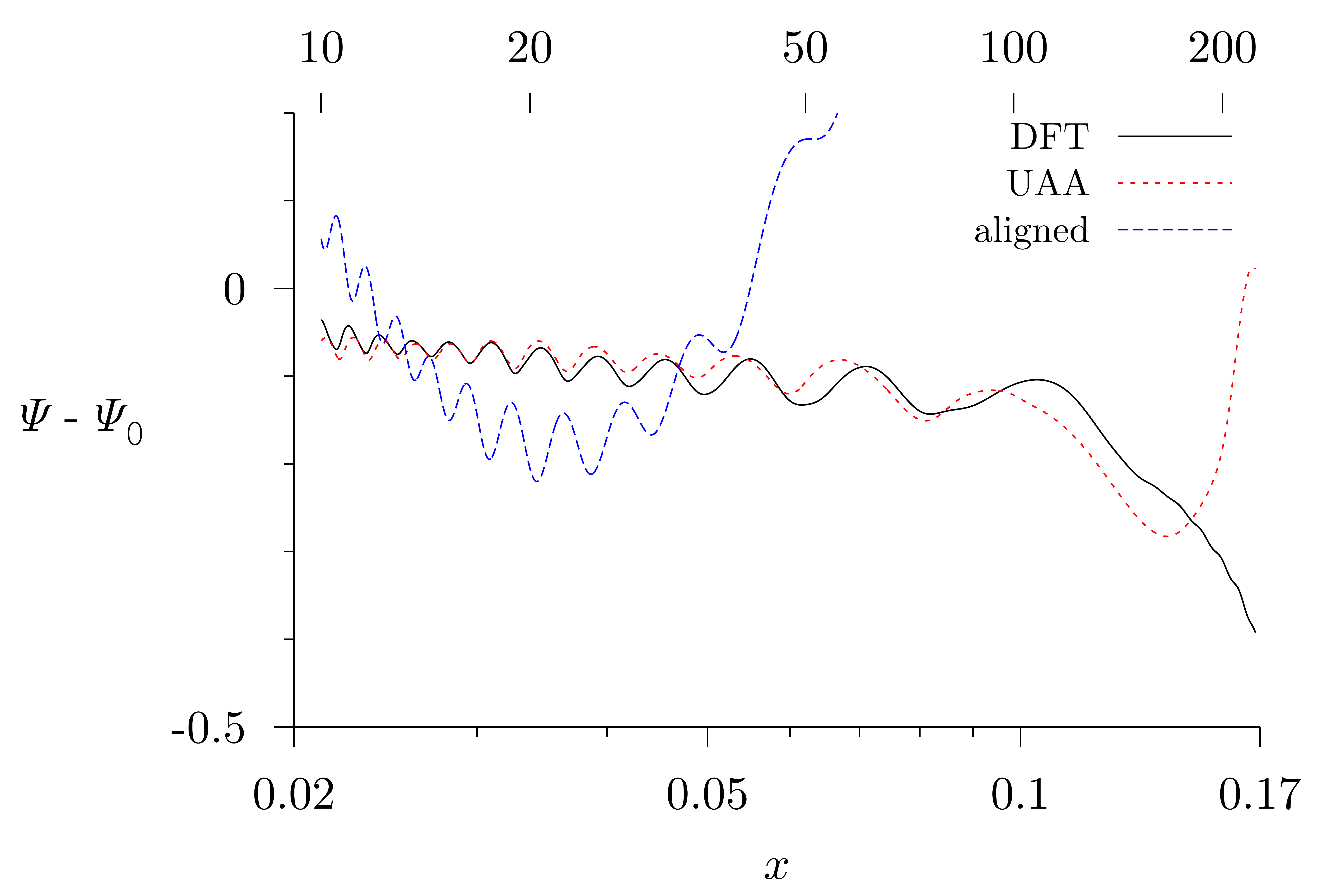} \\
\includegraphics[width=\columnwidth]{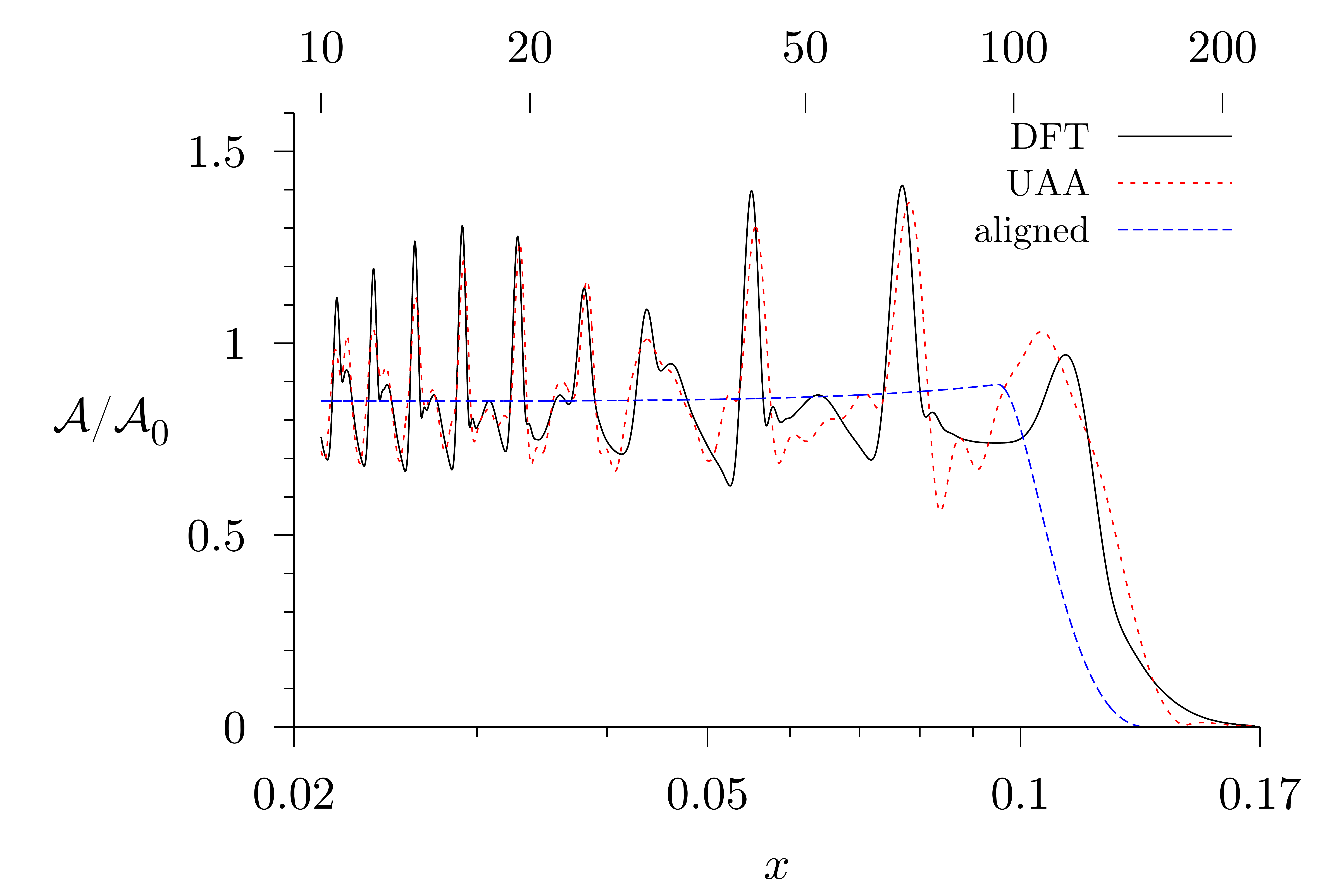}
\includegraphics[width=\columnwidth]{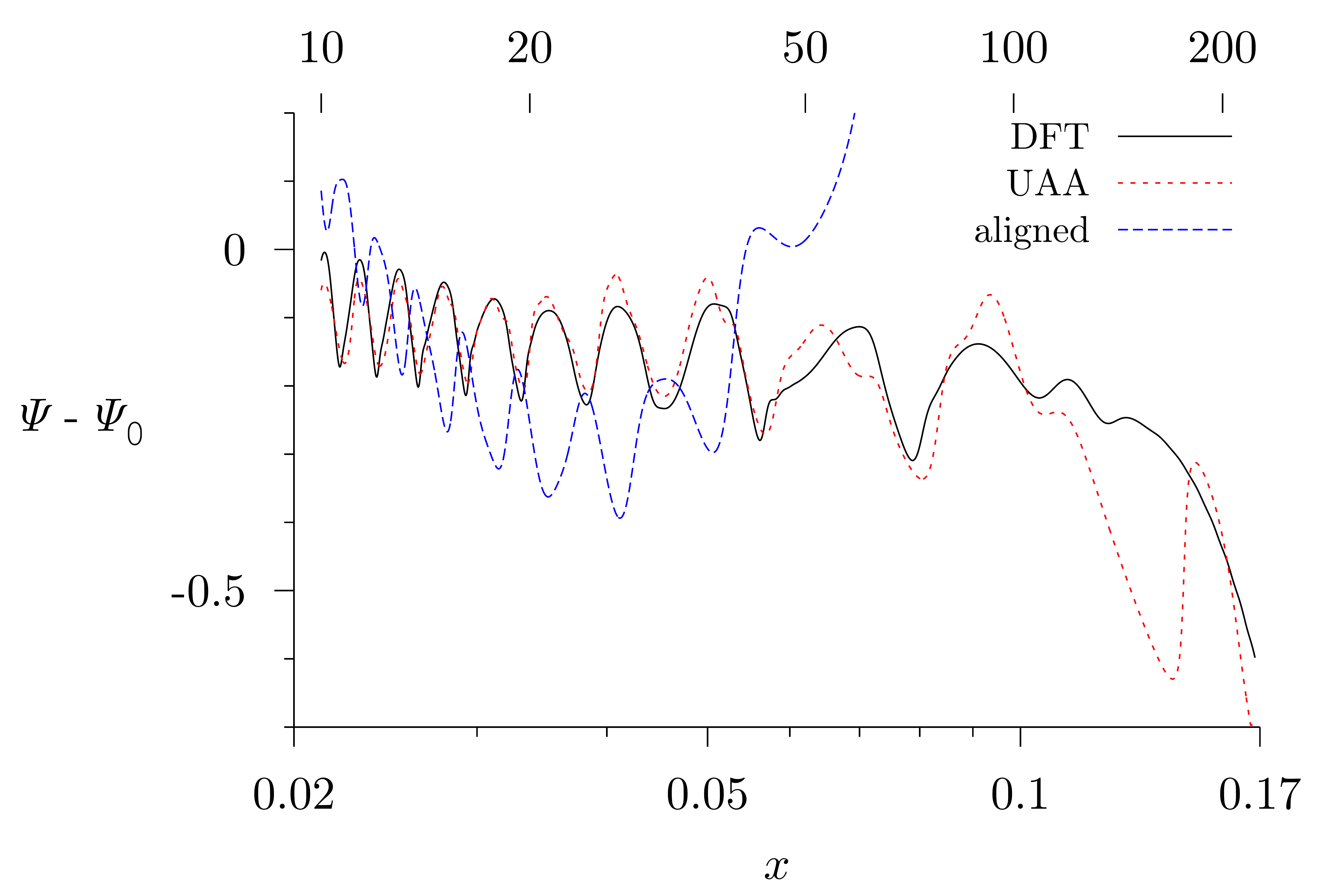} \\
\includegraphics[width=\columnwidth]{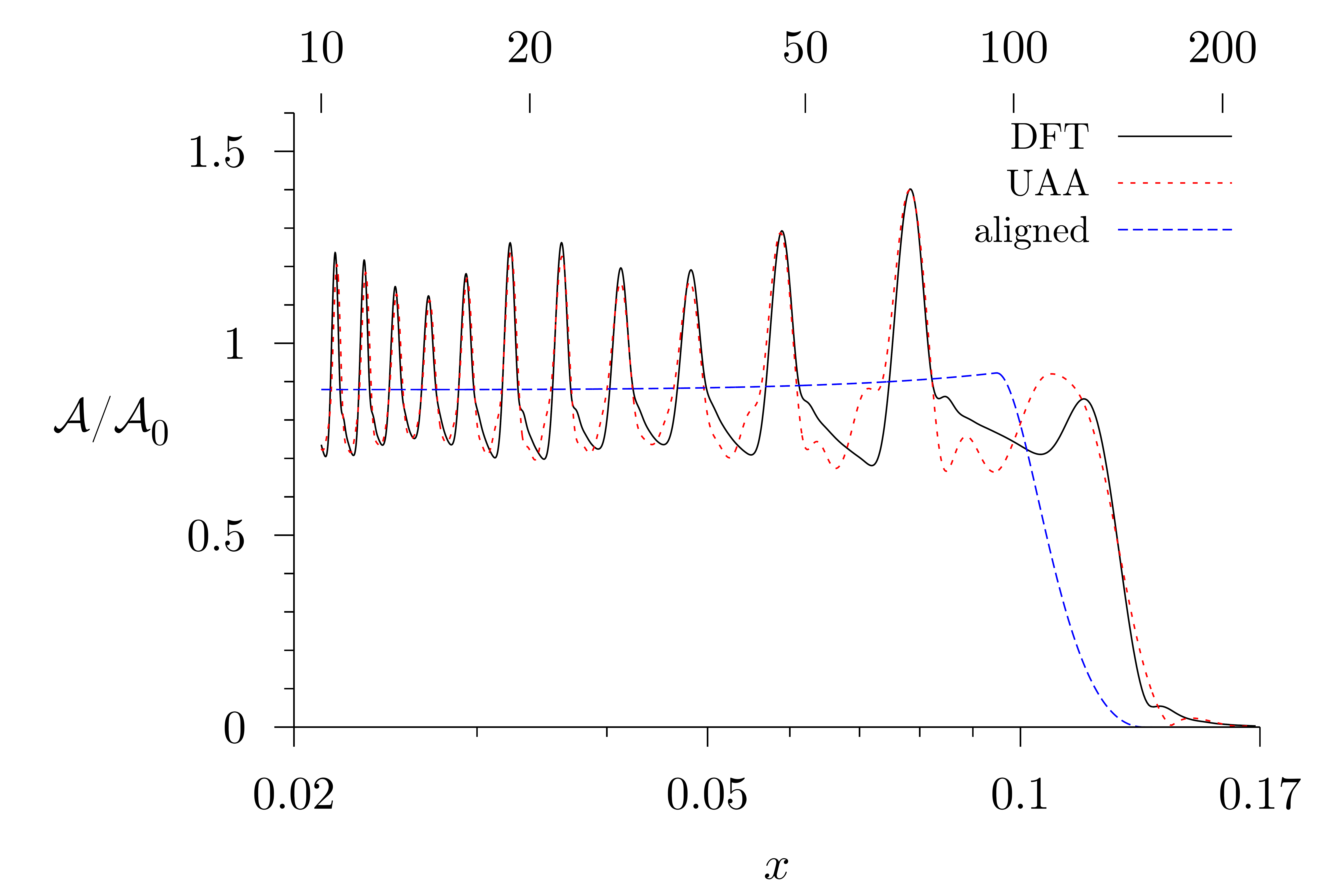}
\includegraphics[width=\columnwidth]{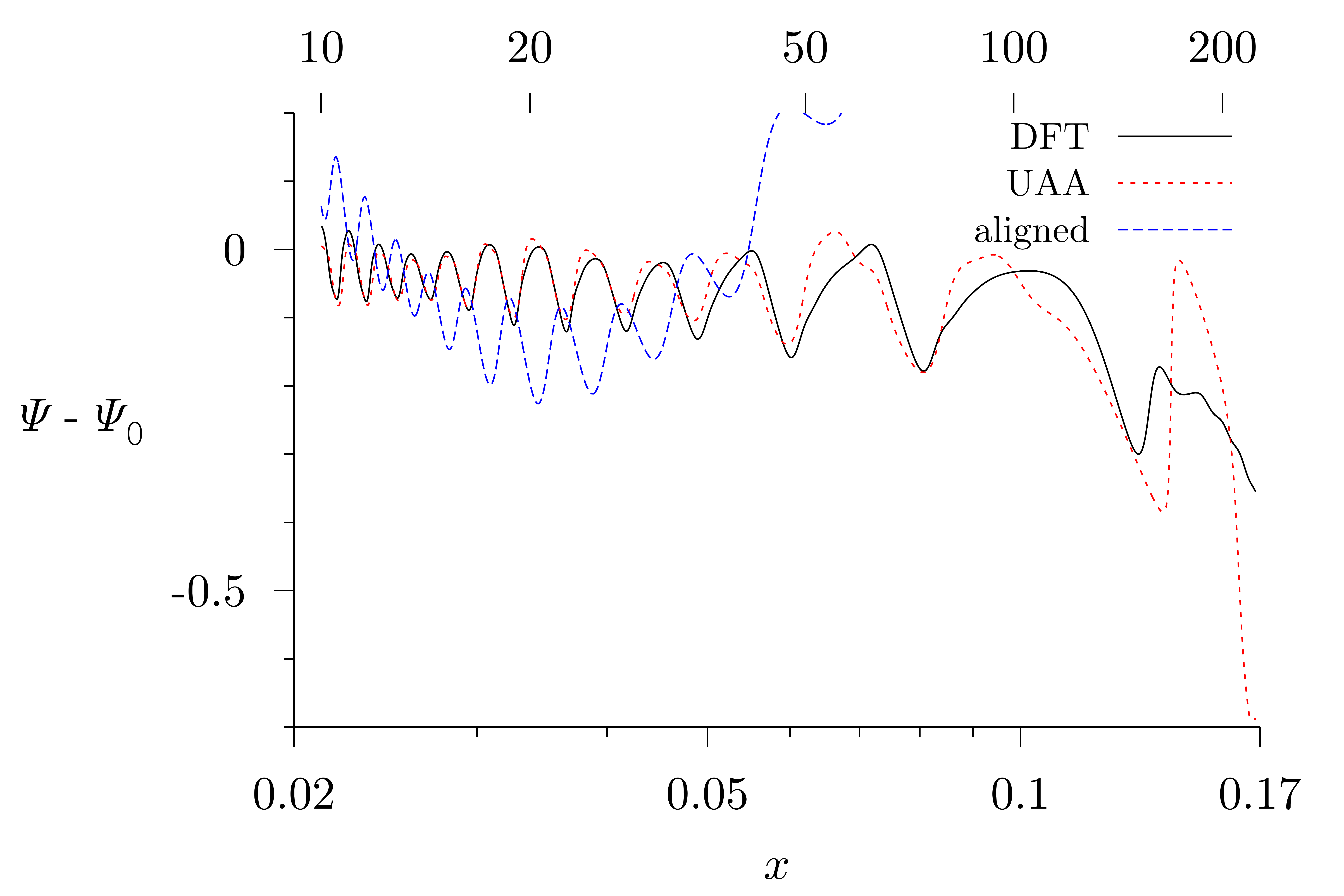} 
\caption{\label{fig:comp_LIGO} Comparison of the Fourier amplitude (left) 
and phase (right) of the $\ell=2$ waveform harmonic for ground-based
systems as a function of the PN parameter $x=(\pi M f)^{2/3}$ (bottom axis) and 
as
a function of the frequency in Hz (the top axis). The solid black curve 
corresponds to the DFT 
result and the dashed red curve to the UAA. The accumulated phase of the
time-domain $\ell=2$ harmonic for each system is $2 \Delta \phi_{orb} \sim
850$~cycles. From top to bottom, we present results for Systems A, B, C and D.}
\end{center}
\end{figure*}

\begin{figure*}[!htp]
\begin{center}
\includegraphics[width=\columnwidth]{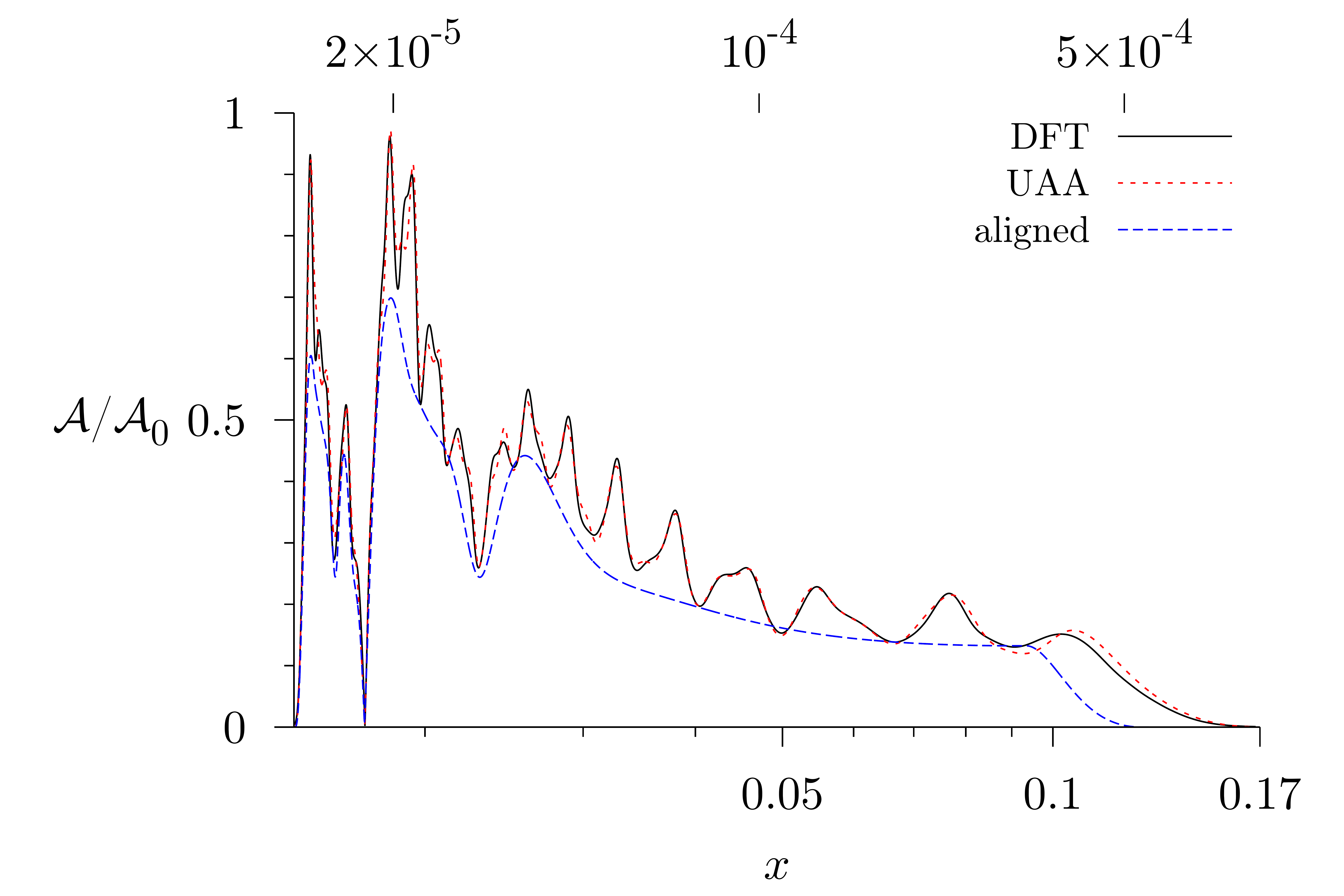}
\includegraphics[width=\columnwidth]{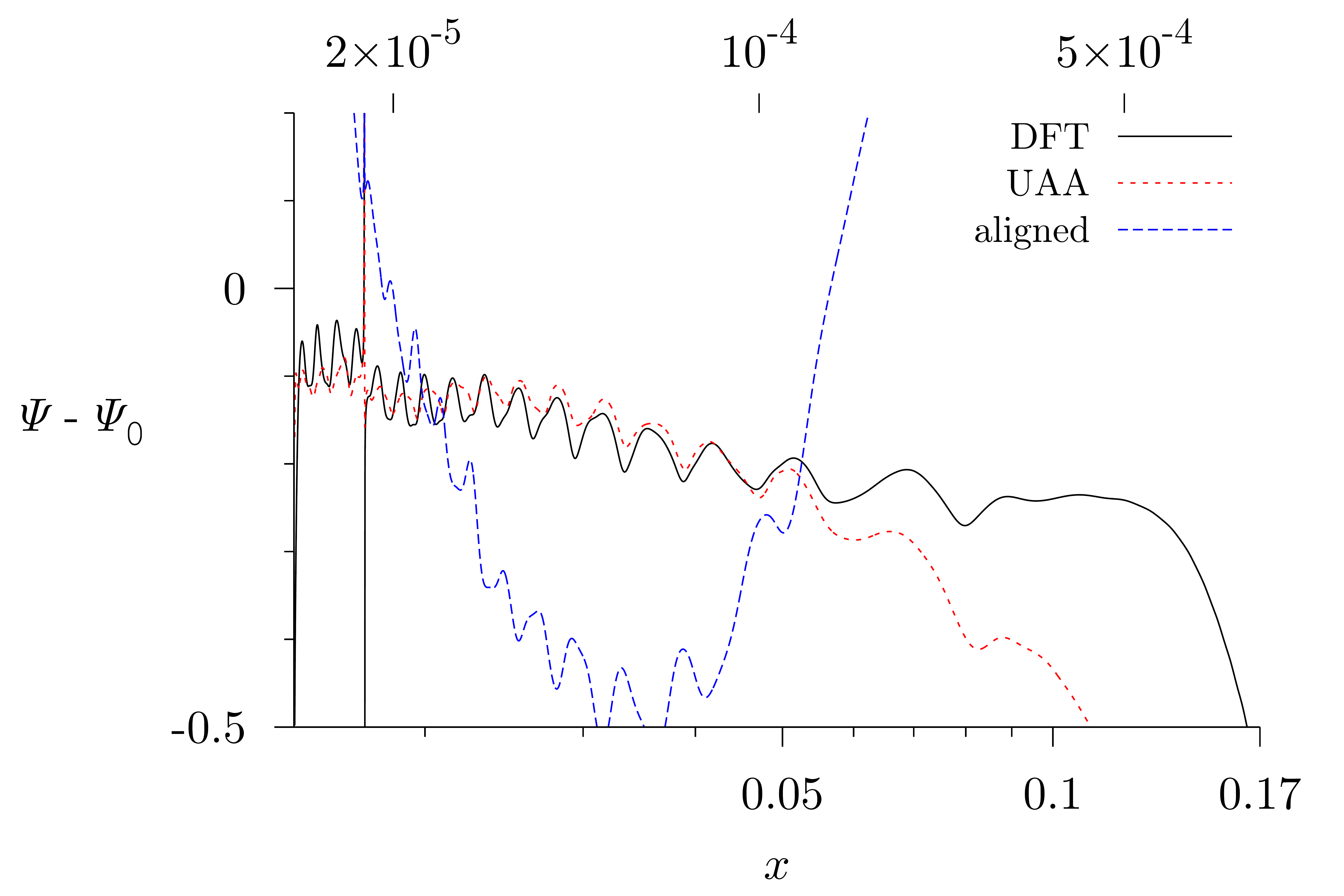} \\
\includegraphics[width=\columnwidth]{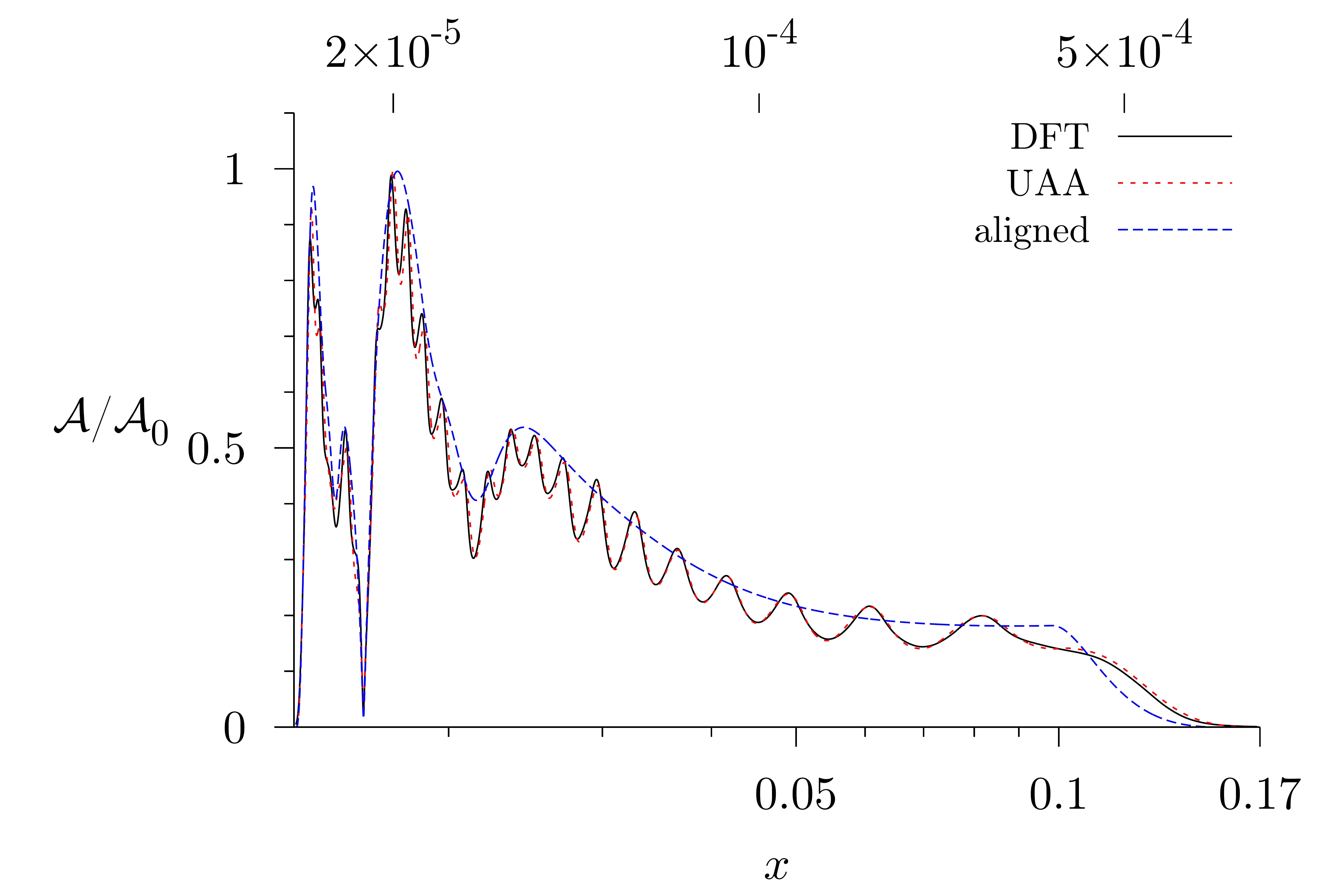}
\includegraphics[width=\columnwidth]{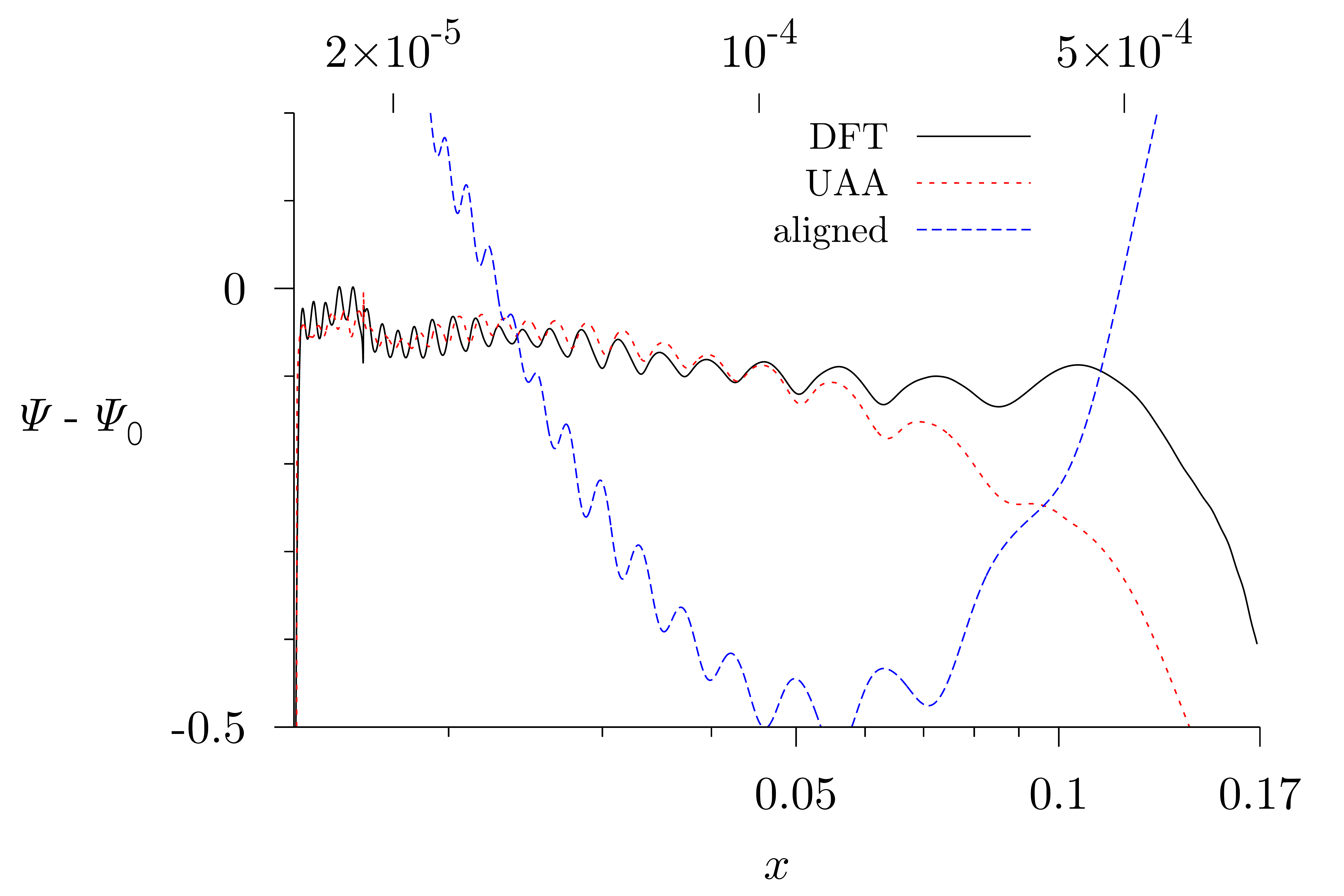} \\
\includegraphics[width=\columnwidth]{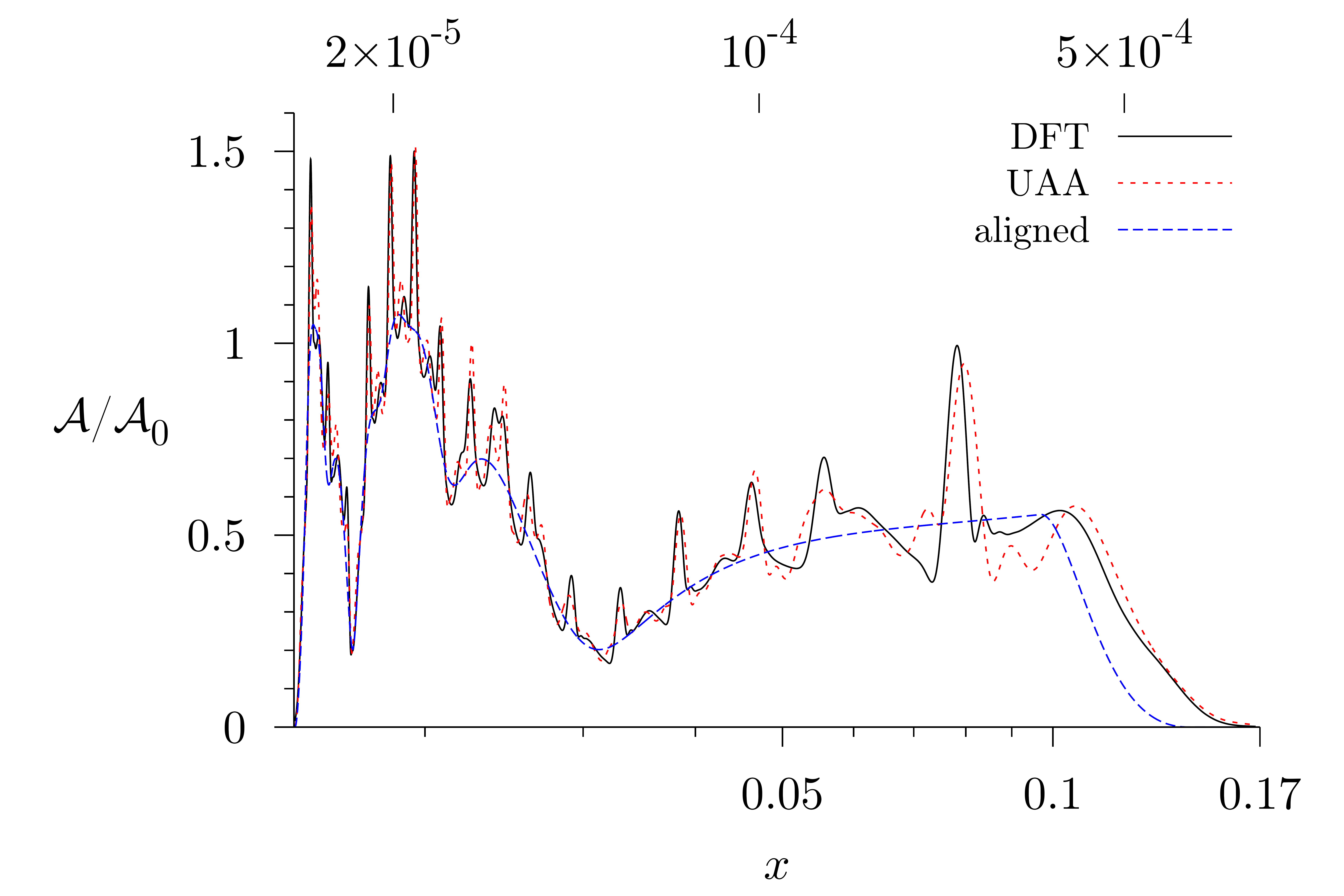}
\includegraphics[width=\columnwidth]{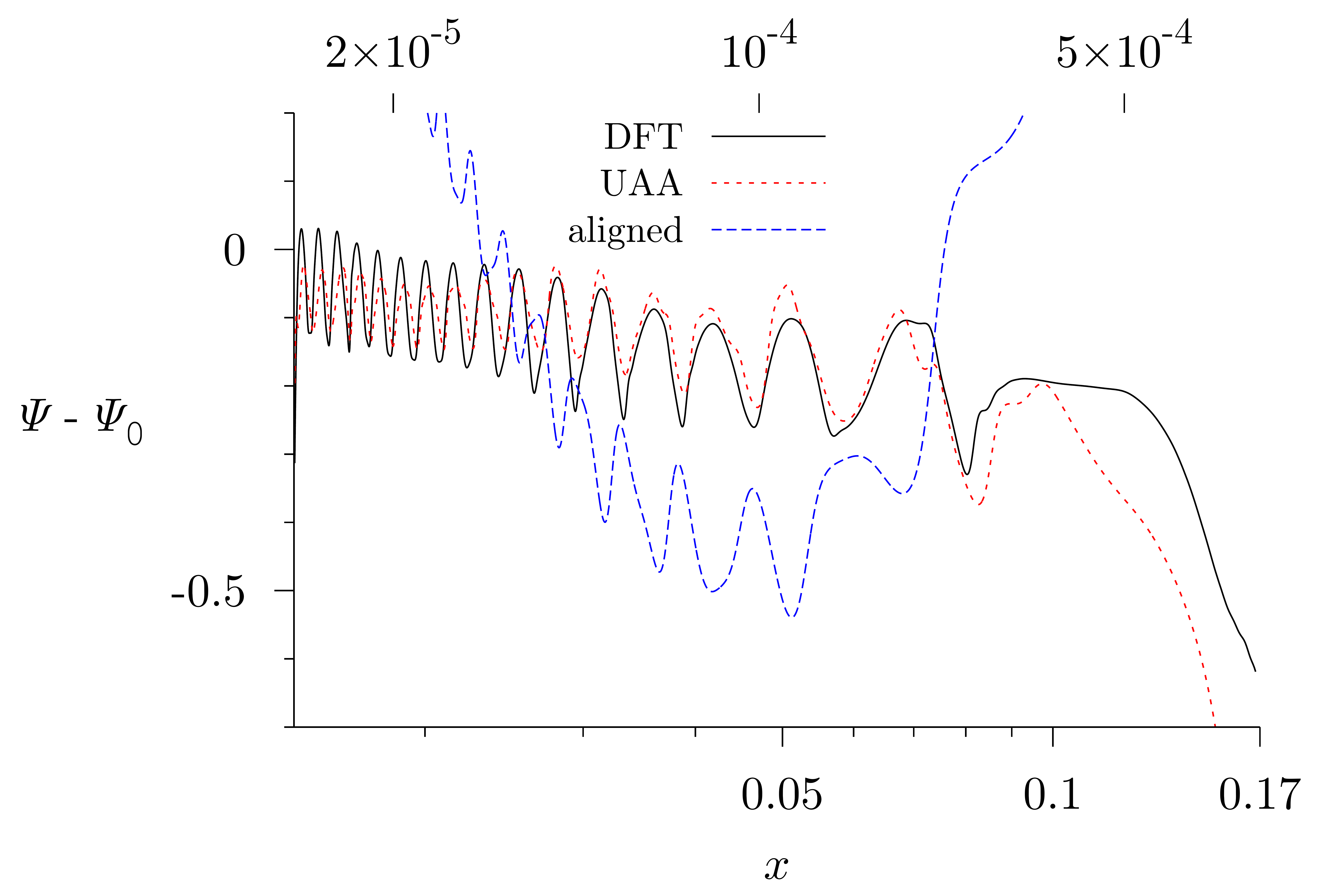} \\
\includegraphics[width=\columnwidth]{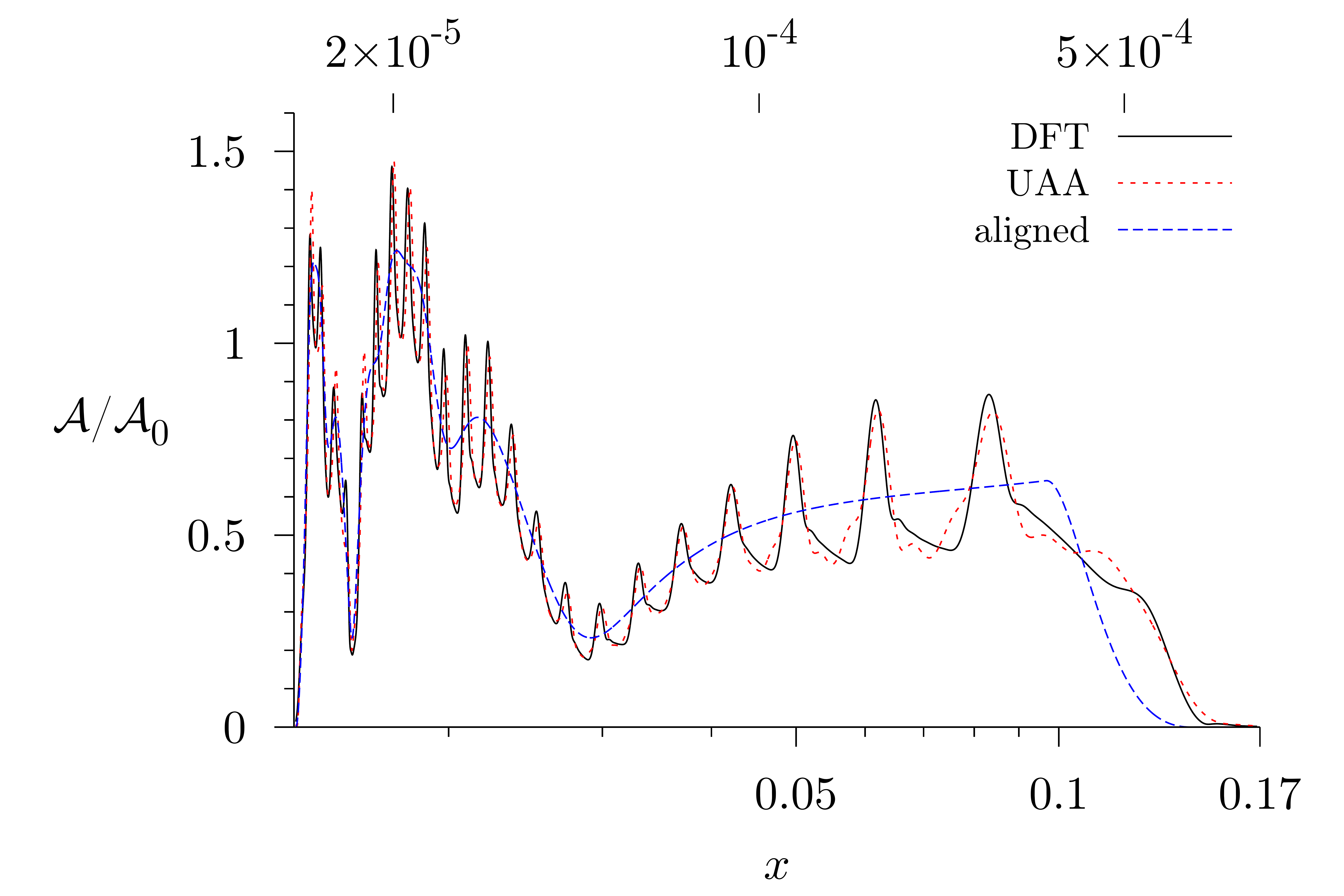}
\includegraphics[width=\columnwidth]{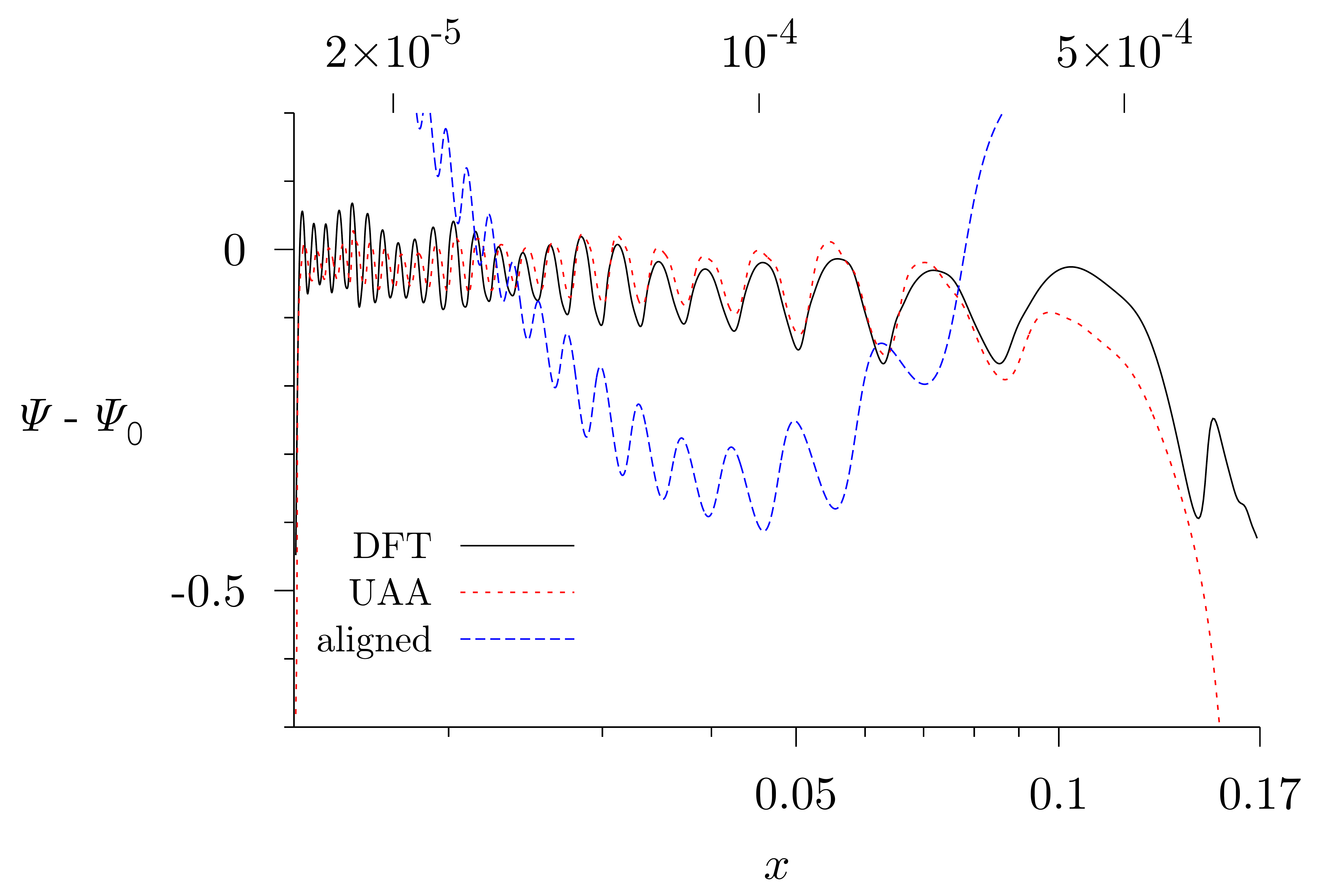} 
\caption{\label{fig:comp_LISA} Same as Fig.~\ref{fig:comp_LIGO} for space-based
systems. The accumulated phase of the $\ell=2$
harmonic is $2 \Delta \phi_{orb} \sim 2000$~cycles.}
\end{center}
\end{figure*}

Fig.~\ref{fig:comp_LIGO} and~\ref{fig:comp_LISA} compare the dominant 
$\ell=2$, Fourier waveform 
amplitude (left panels) and phase (right panels) for Systems A through D as a 
function of the PN 
parameter $x$ (bottom axis) and the GW frequency in Hz (top axis) for 
ground-based and space-based systems respectively. The solid black curves 
correspond to the DFT waveform, the red dotted curves to
the UAA waveform and the blue dashed curves to the aligned-SPA waveforms. For 
reference,
the total accumulated phase of the time-domain $\ell=2$ harmonic is about $850$ 
cycles for all
ground-based systems and $2000$ cycles for space-based systems.

\begin{figure}[!htp]
\begin{center}
\includegraphics[width=\columnwidth]{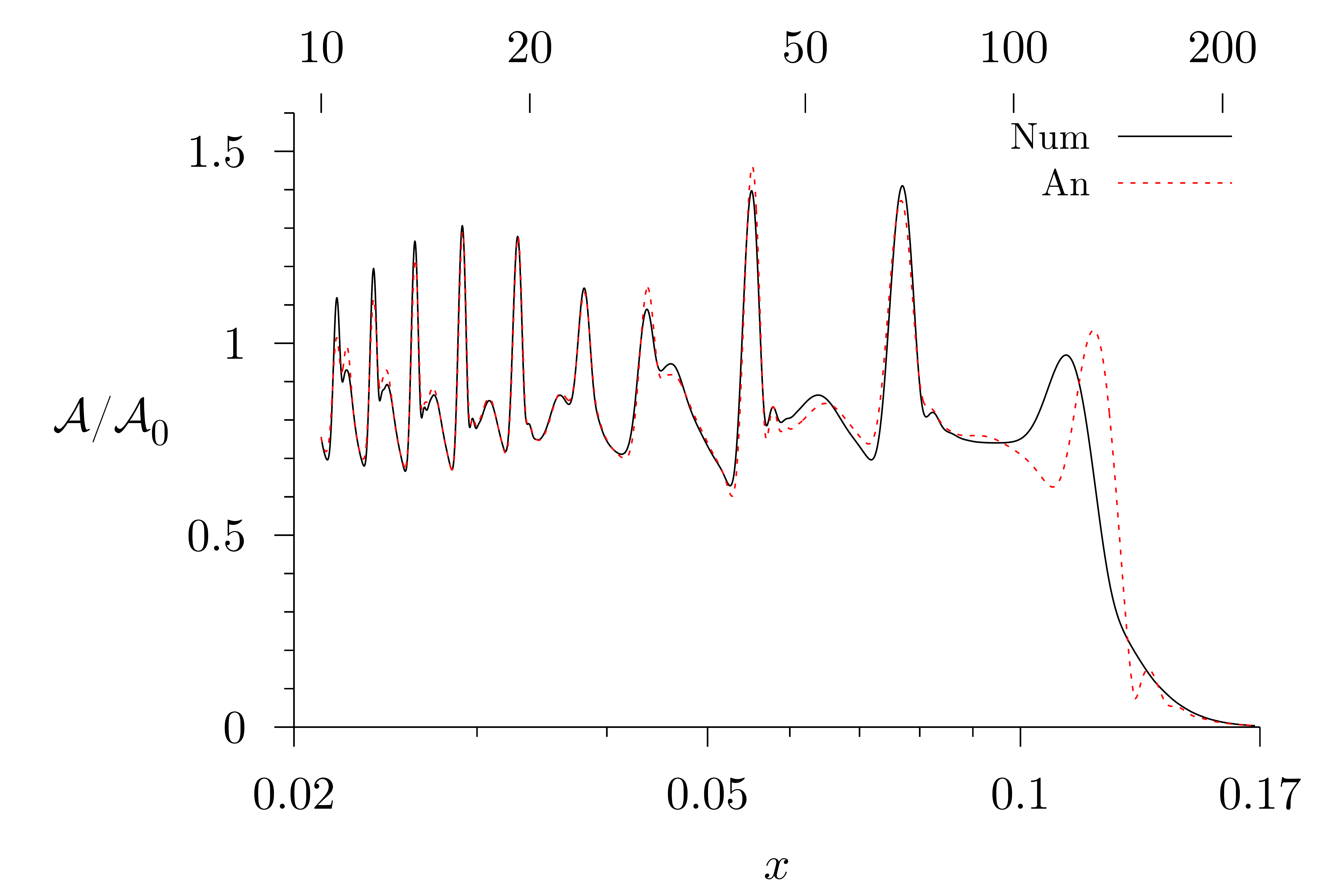} \\
\includegraphics[width=\columnwidth]{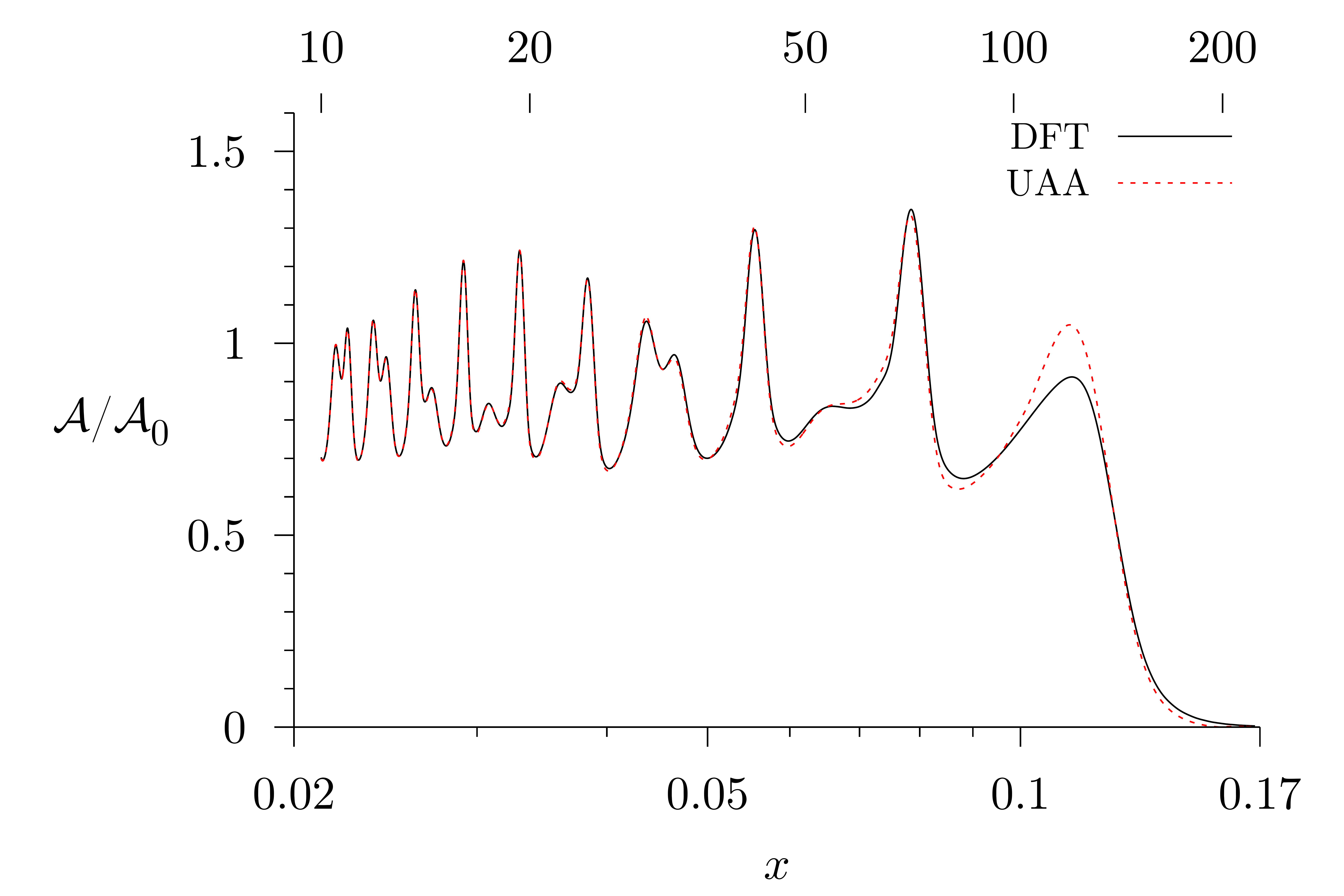}
\caption{\label{fig:DFT-UAAcomp} Comparison between DFT and UAA waveform 
amplitudes, with same parameters as the third row of 
Fig.~\ref{fig:comp_LIGO} (ground-based system C). On the top, 
comparison between the amplitudes of two DFT waveforms, one constructed with 
the fully numerical solution to the equations of motion (black solid line, the 
DFT waveform that we used in the rest of this section), and the other 
constructed with the analytical solution for $\bm{L}(t)$ derived in 
Sec.~\ref{sec:near-alignment}, as well as $\phi_{\C}(t)$ from 
Eq.~\eqref{eq:phi0} (dotted red line). At the bottom, comparison 
between the amplitudes of a DFT waveform constructed with the phases 
$\delta\phi(t)$, $\psi_N(t)$, and $i_L(t)$ derived in 
Sec.\ref{sec:timedomainphases}, as well as $\phi_{\C}(t)$ from 
Eq.~\eqref{eq:phi0}  (solid black line),
and
the UAA waveform (dotted red line).}
\end{center}
\end{figure}

Let us make several observations about these figures. 
First, recall that all phase quantities are here presented relative to the 
carrier, non-spinning phase of
the corresponding system. Therefore, the $\sim \mathcal{O}(10)$ oscillations in 
the phase plots
(right panels) occur on a precession timescale, while in reality $850$ and 
$2000$ total GW cycles
have elapsed for ground-based and space-based systems respectively. 
Second, observe that spin precession clearly induces modulations on the phase 
and amplitude 
that depend sensitively on $\delta m$ and $\alpha_{0}$. These modulations are 
captured much
better by the UAA family than the aligned-SPA one. 
Third, observe that all approximations agree on the frequency of these 
modulations but 
not on the amplitudes or overall trends, i.e.~the troughs and valleys do occur 
roughly at the 
same values of $x$ for all waveforms. 
Fourth, in the ground-based detectors for system C, the Fourier amplitude of 
the DFT shows 
peculiar features (e.g.~at $x \approx 0.06$) that are approximated by the UAA 
waveform. 
Thus, these features are not an artifact of the DFT, and we have checked that 
they are not 
induced by spectral leakage.  
Fifth, we can observe a spike in the DFT and UAA phase difference $\Psi - 
\Psi_0$ for 
space-based system $A$ around $x = 0.017$ that is missed by the aligned SPA 
approximation. 
By inspecting the corresponding amplitude plots, we can see that these spikes 
correspond to 
moments when the amplitudes of the waveforms almost vanish, i.e.~the detector 
is going through a node in 
the waveform power distribution. We can see that the UAA waveform reproduces 
this feature, and we checked that it agrees with the DFT when it is present.
Sixth, the phase discrepancy between the aligned SPA and DFT 
models does not seem to be consistent from system to system. This is because 
the match is too small for the maximization method that we used to yield 
trustworthy results for $\phi_\coal$ and $t_\coal$ in the aligned SPA case.
Seventh, the amplitudes are much better recovered by the UAA for systems A and 
B 
than C and D. This is because the precession modulation angles $\delta\phi$ and 
$\psi_N$ are worse approximations for the latter systems, as discussed in 
Sec.~\ref{sec:timedomainphases}, and shown in Fig.~\ref{fig:DFT-UAAcomp}.

We compare four waveform models in Fig.~\ref{fig:DFT-UAAcomp}, using 
ground-based system C. Three of those models are based on a discrete Fourier 
transform, and the fourth one is the UAA model. The first DFT model, DFT1, is 
the one used in the rest of this section, constructed using the full numerical 
solution to the equations of motion. The second one, DFT2, is computed using the 
carrier orbital phase $\phi_{\C}(t)$ from 
Eq.~\eqref{eq:phi0}, together with precession modulation phases 
$\psi_N(t)$, and $i_L(t)$ computed with the analytical 
solution for $\bm{L}(t)$ derived in 
Sec.~\ref{sec:near-alignment}, and using
\begin{align}
 \delta\phi(t) &= \hat{N}_z \arctan \left( \frac{L_z
\hat{N}_x - L_x}{L_z \hat{N}_y - L_y}\right),
\end{align}
an approximation valid for any degree of misalignment between $\uvec{N}$ and 
$\uvec{L}$.
The third one, DFT3, is identical to 
DFT2 but for the precession modulation phases $\delta\phi(t)$, $\psi_N(t)$, and 
$i_L(t)$, using those used 
to derive the UAA waveform, derived in Sec.\ref{sec:timedomainphases}.
The top panel of Fig.~\ref{fig:DFT-UAAcomp} shows that the amplitude 
discrepancy between the DFT1 and the DFT2 models is much smaller than between 
the DFT1 and UAA models (Fig.\ref{fig:comp_LIGO}, third plot from the top on the 
left panel), meaning that the main source of inaccuracy is not due to the 
inaccuracy in $\bm{L}(t)$. The bottom panel shows that the DFT3 amplitude is 
very well approximated by the UAA amplitude. The main source of amplitude 
discrepancy between the DFT and UAA models for systems C and D that can be 
observed in Figs.\ref{fig:comp_LIGO} and~\ref{fig:comp_LISA} is thus the 
Fourier decomposition derived in Sec.\ref{sec:timedomainphases}, which is less 
accurate when $\uvec{N}$ and $\uvec{L}$ are close to being aligned.

\section{Discussion}

The coming enhancements of ground-based detectors will allow for the first 
direct
detection of gravitational waves. In order to carry out efficient searches, one
needs computationally cheap and accurate waveforms. Systems with spins will
generically undergo precession, unless the spins are perfectly aligned with the 
orbital
angular momentum. Precession will induce a drastic modification to the 
waveform, 
generating corrections in both the phase and amplitude. Such modifications 
cannot
be captured by spin-aligned waveform families, as we demonstrate in this paper. 

Binaries in the presence of gas, however, will tend to have spin vectors almost 
aligned with
the orbital angular momentum 
vector~\cite{2007ApJ.661L.147B,Barausse:2012fy,2013arXiv1302.4442G}, 
i.e. the misalignment angles should be small. Motivated
by this, we have constructed a waveform family that captures faithfully 
the main features of GWs emitted by compact binaries with small 
spin-orbital angular momentum 
misalignment angles. One can think of this waveform family as a perturbation of 
the spin-aligned 
family, with corrections that model precession effects that enter both the 
waveform amplitude 
and phase.   

The waveforms calculated here are purely analytical, constructed both in the 
time- and in the frequency-domain. Such analytical waveforms have several 
advantages. On the one hand, analytical waveforms are usually computationally 
more efficient to evaluate. Given the large number of templates needed for 
parameter estimation of spinning systems, computational efficiency is highly 
desirable. On the other hand, the analytic structure provides important physical 
insight into how each precession effect comes into play. Such insight can then 
be used to construct simpler, phenomenological waveforms, such as those recently 
constructed for binaries where one object is not 
spinning~\cite{Lundgren:2013jla}. 

The mathematical methods used to construct these analytical waveforms had never 
been used in waveform modeling, to our knowledge. These methods, however, are 
very well-known in other fields, such as non-linear optics and aerodynamics. The 
first method is that of multiple scale analysis, amenable to problems with 
several timescales that separate. This method allows us to solve the precession 
equations analytically as an expansion in the ratio of the precession to the 
radiation-reaction timescale. The second method is that of uniform asymptotics, 
which allows us to construct a single asymptotic expansion to the solution of a 
given problem, instead of a series of asymptotic expansions in different 
regimes. This method is essential to cure the stationary phase approximation, 
which fails in the presence of precession due to the coalescence of stationary 
points. 

Many other problems would benefit from the application of the mathematical 
methods implemented here. For example, one could study compact binary inspirals, 
where the spin angular momenta has a small magnitude (relative to the orbital 
angular momentum) but arbitrary orientation. This application would be 
complementary to the example studied here. The resulting analytic waveform can 
be thought of as a perturbation of the non-spinning SPA waveform. Similarly, one 
can study compact binaries where one component has arbitrary angular momentum, 
but the companion has a small spin with arbitrary orientation. This case would 
be intermediate between the one studied here and the one where both binary 
components have small spin. The resulting analytic waveform can be thought of as 
a perturbation of a simple precessing waveform. We are currently investigating 
both of these cases. 

\acknowledgments

We thank Katerina Chatziioannou for her useful comments and suggestions.
A.~K.~and N.~C.~acknowledge support from the NSF Award PHY-1205993  and NASA 
grant
NNX10AH15G. N.~Y.~acknowledges support from NSF grant PHY-1114374 and 
NASA grant NNX11AI49G, under sub-award 00001944.

\appendix

\section{Frequency evolution}
\label{app:freqevol}

The PN evolution equation for the orbital frequency for binaries on
quasicircular orbits is given at 2.5PN order 
by~\cite{racinebuonannokidder,blanchet3PN}
\begin{align}
 \dot{\omega} &= \frac{a_0}{M^2} (M\omega)^{11/3} \left( 1 + \sum_{n \geq 2}^N
a_n (M\omega)^{n/3}
\right), \\
a_0  &= \frac{96\nu}{5}, \\
a_2 &= - \left( \frac{743}{336} + \frac{11\nu}{4} \right), \\
a_3 &= 4\pi -
\beta_{1.5} ,\\
a_4 &= \frac{34103}{18144} + \frac{13661\nu}{2016} + \frac{59\nu^2}{18} -
\sigma , \\
a_5 &= -\left(\frac{4159}{672} + \frac{189}{8}\nu \right)
\pi - \beta_{2.5} , \\
 \beta_{1.5} &= \sum_{A=1}^2 \left( \frac{113}{12} +
\frac{25\mu M}{4m_A^2} \right) \bm{S}_A \cdot \uvec{L}, \\
 \sigma &= \frac{1}{\mu M^3} \left[ \frac{247}{48} \bm{S}_1 \cdot \bm{S}_2
- \frac{721}{48} \left( \bm{S}_1 \cdot \uvec{L} \right) \left( \bm{S}_2 \cdot
\uvec{L} \right) \right] \nonumber\\
&+ \sum_{A=1}^2 \frac{1}{m_A^2M^2} \left[
\frac{233}{96} \bm{S}_A^2 - \frac{719}{96} \left( \bm{S}_A \cdot \uvec{L}
\right)^2 \right], \\
 \beta_{2.5} &= \sum_{A=1}^2 \bigg[ \left( \frac{31319}{1008} - \frac{1159}{24}
\nu \right) \nonumber\\
&+ \frac{\mu M}{m_A^2} \left(
\frac{809}{84} -
\frac{281}{8}\nu \right) \bigg] \bm{S}_A \cdot \uvec{L}.
\end{align}

Using $\xi^3 = M \omega$, we can express the carrier phase $\phi_\C$ at 6PN 
order in terms of these couplings as 
\begin{align}
 \phi_{\C} &= \int \frac{\xi^3}{M} dt = \int \frac{\xi^3}{M \dot{\xi}} d\xi 
\nonumber\\
 &= \phi_\coal - \xi^{-5} [ \phi_0 + \phi_2 \xi^2 + \phi_3 \xi^3 + \phi_4 \xi^4 
+ \phi_5 \xi^5 + \phi_6 \xi^6 \nonumber\\
&+ \phi_7 \xi^7 + 
\phi_8 \xi^8 + \phi_9 \xi^9 + \phi_{10} \xi^{10} + \phi_{11} \xi^{11} + 
\phi_{12} \xi^{12} ], \\
\phi_0 &= \frac{3}{5 a_0}, \\
\phi_2 &= -\frac{a_2}{a_0}, \\
\phi_3 &= -\frac{3 a_3}{2 a_0}, \\
\phi_4 &= -\frac{3 ( a_4 - a_2^2)}{a_0}, \\
\phi_5 &= \frac{3(a_5 - 2 a_2 a_3)}{a_0} \log \xi, \\
\phi_6 &= -\frac{3}{a_0} ( 2 a_2 a_4 + a_3^2 - a_2^3 ), \\
\phi_7 &= -\frac{3}{2a_0} ( 2 a_2 a_5 + 2 a_3 a_4 - 3 a_2^2 a_3 ), \\
\phi_8 &= -\frac{1}{a_0} ( 2 a_3 a_5 + a_4^2 - 3 a_2^2 a_4 - 3 a_2 a_3^2 + 
a_2^4 ) - 6 \log \xi, \\
\phi_9 &= -\frac{3}{4a_0} ( 2 a_4 a_5 - 3 a_2^2 a_5 - 6 a_2 a_3 a_4 - a_3^3 + 4 
a_2^3 a_3 ), \\
\phi_{10} &= -\frac{3}{5 a_0} ( a_5^2 - 6 a_2 a_3 a_5 - 3 a_2 a_4^2 \nonumber\\
&- 
3 a_3^2 a_4 + 4 a_2^3 a_4 + 6 a_2^2 a_3^2 - a_2^5) + 3 \nu \log \xi, \\
\phi_{11} &= \frac{1}{2 a_0} (6 a_2 a_4 a_5 + 3 a_3^2 a_5 - 4 a_2^3 a_5 
\nonumber\\
&+ 3 
a_3 a_4^2 - 12 a_2^2 a_3 a_4 - 4 a_2 a_3^3 + 5 a_2^4 a_3 ), \\
\phi_{12} &= \frac{3}{7a_0} ( 3 a_2 a_5^2 + 6 a_3 a_4 a_5 - 12 a_2^2 a_3 a_5 
+ a_4^3 - 6 a_2^2 a_4^2 \nonumber\\
&- 12 a_2 a_3^2 a_4 + 5 a_2^4 a_4 - a_3^4 + 
10 a_2^3 a_3^2 - a_2^6 ).
\end{align}
The factors proportional to $\log\xi$ in $\phi_8$ and $\phi_{10}$ come 
from the reabsorption of $\log (\omega/\omega_0)$ terms from the amplitude to 
the phase that would lead to an unphysical arbitrary amplitude 
factor~\cite{abiq}.

We can also find a time-orbital frequency relation using $\int dt = \int 
d\xi/\dot{\xi} $:
\begin{align}
 t &= t_\coal - \xi^{-8} [ t_0 + t_2 \xi^2 + t_3 \xi^3 + t_4 \xi^4 
+ t_5 \xi^5 + t_6 \xi^6 \nonumber\\
&+ t_7 \xi^7 + 
t_8 \xi^8 + t_9 \xi^9 + t_{10} \xi^{10} + t_{11} \xi^{11} + 
t_{12} \xi^{12} ], \\
t_0 &= \frac{3}{8a_0}, \\
t_2 &= -\frac{a_2}{2a_0}, \\
t_3 &= -\frac{3 a_3}{5 a_0}, \\
t_4 &= -\frac{3 (a_4 - a_2^2)}{4 a_0}, \\
t_5 &= -\frac{a_5 - 2 a_2 a_3}{a_0}, \\
t_6 &= \frac{3}{2a_0} ( 2 a_2 a_4 + a_3^2 - a_2^3), \\
t_7 &= \frac{3}{a_0} ( 2 a_2 a_5 + 2 a_3 a_4 - 3 a_2^2 a_3 ), \\
t_8 &= -\frac{3}{a_0} ( 2 a_3 a_5 + a_4^2 - 3 a_2^2 a_4 - 3 a_2 a_3^2 + a_2^4 ) 
\log \xi, \\
t_9 &= -\frac{3}{a_0} ( 2 a_4 a_5 - 3 a_2^2 a_5 - 6 a_2 a_3 a_4 - a_3^3 + 4 
a_2^3 a_3 ), \\
t_{10} &= -\frac{3}{2a_0} ( a_5^2 - 6 a_2 a_3 a_5 - 3 a_2 a_4^2 \nonumber\\
&- 3 
a_3^2 a_4 + 4 a_2^3 a_4 + 6 a_2^2 a_3^2 - a_2^5 ), \\
t_{11} &= \frac{1}{a_0} ( 6 a_2 a_4 a_5 + 3 a_3^2 a_5 - 4 a_2^3 a_5 
\nonumber\\
&+ 3 a_3 a_4^2 - 12 a_2^2 a_3 a_4 - 4 a_2 a_3^3 + 5 a_2^4 a_3 ),\\
t_{12} &= \frac{3}{4a_0} ( 3 a_2 a_5^2 + 6 a_3 a_4 a_5 - 12 a_2^2 a_3 a_5 + 
a_4^3 - 6 a_2^2 a_4^2 \nonumber\\
&- 12 a_2 a_3^2 a_4 + 5 a_2^4 a_4 - a_3^4 + 10 a_2^3 a_3^2 - a_2^6 ).
\end{align}

\begin{widetext}
\section{Transformation between angular momenta and normal modes}
\label{app:RandRm1}

The transformation matrices that allow us to change between angular momenta and
normal modes are given below.

$\mathbb{R}^{-1}$ is given through $\bm{Q}_j = \mathbb{R}^{-1} \bm{W}_j$ by:
\begin{align}
 Q_{0,j}^{(1)} &= C_0 \left[ L_k^{(1)} + S_{1,k}^{(1)} + S_{2,k}^{(1)}
\right], \label{eq:Rm1_1}\\
 Q_{+,j}^{(1)} &= C_+ \Big\{ [(c d + c f + b f)(b + d + f - \omega_{\PP,+}) -
(c-f)
(b e + b g + c g)] L_k^{(1)} \nonumber\\
&+ [(c d + c f + b f) (b + d + f - \omega_{\PP,+}) + 
  (c - f) (c d + c f + b f + d e + e f + d g)] S_{1,k}^{(1)} \nonumber\\
  &-[(b e + b g + c g + d e + e f + d g) (b + d + f - \omega_{\PP,+}) + (c - f)
(b e + b g + c g)]  S_{2,k}^{(1)} \Big\}, \\
 Q_{-,j}^{(1)} &= C_- \Big\{ -[(c d + c f + b f)(b + d + f - \omega_{\PP,-}) -
(c-f)
(b e + b g + c g)] L_k^{(1)} \nonumber\\
&- [(c d + c f + b f) (b + d + f - \omega_{\PP,-}) + 
  (c - f) (c d + c f + b f + d e + e f + d g)] S_{1,k}^{(1)} \nonumber\\
  &+[(b e + b g + c g + d e + e f + d g) (b + d + f - \omega_{\PP,-}) + (c - f) 
(b
e + b g + c g)]  S_{2,k}^{(1)} \Big\}, \label{eq:Rm1_3}
\end{align}
where $C_i$ are arbitrary functions
of time, and either $j=1$ and $k=x$, or $j=2$ and $k=y$.
To simplify the inverse transformation, we choose
\begin{align}
 C_0 &= \frac{de+dg+ef}{(b + c) (f + g) + b e + c d + d e + d g + e f }, \\
 C_+ &= \frac{b+c+d-e-f-g + \delta\omega_p}{2\delta\omega_p (f-c) [(b + c) (f +
g) + b e + c d + d e + d g + e f]}, \\
 C_- &= \frac{b+c+d-e-f-g - \delta\omega_p}{2\delta\omega_p (f-c) [(b + c) (f +
g) + b e + c d + d e + d g + e f]}, \\
\delta\omega_p &= \omega_{\PP,+} - \omega_{\PP,-}, \label{eq:deltaomega}
\end{align}
which gives $\mathbb{R}$ through $\bm{W}_j = \mathbb{R} \bm{Q}_j$ as
\begin{align}
 L_k^{(1)} &= Q_{0,j}^{(1)} + Q_{+,j}^{(1)} + Q_{-,j}^{(1)}, \\
 S_{1,k}^{(1)} &= \frac{be + bg + cg}{de + dg + ef} Q_{0,j}^{(1)} + 
\frac{2(g-b)}{b+c - (d-e)-(f+g) + \delta \omega_p}  Q_{+,j}^{(1)}
+\frac{2(g-b)}{b+c- (d-e)-(f+g)  - \delta \omega_p} Q_{-,j}^{(1)}, \\
 S_{2,k}^{(1)} &= \frac{cd + bf + cf}{de + dg + ef} Q_{0,j}^{(1)} + 
\frac{2(f-c)}{b+c + (d-e)-(f+g) + \delta \omega_p}  Q_{+,j}^{(1)}
+\frac{2(f-c)}{b+c + (d-e)-(f+g)  - \delta \omega_p} Q_{-,j}^{(1)}.
\end{align}

The amplitudes $B_{i,j}^{(1,0)}$ with $i= +$ or $-$, and $j = 1$ or $2$ can be
computed using Eqs.~(\ref{eq:Rm1_1}-\ref{eq:Rm1_3}) at $t=0$, with 
$Q_{i,j}^{(1)} \to
B_{i,j}^{(1,0)}$ and $L_k^{(1)} \to - S_{1,k} - S_{2,k}$. Using
$S_{j,k}(t = 0) = m_j^2 \chi_j \hat{S}_{j,k,0}$, we get
\end{widetext}
\begin{align}
 B_{+,1}^{(1,0)} &= T_{+,1} \hat{S}_{1,x,0} + T_{+,2} \hat{S}_{2,x,0}, \\
 B_{+,2}^{(1,0)} &= T_{+,1} \hat{S}_{1,y,0} + T_{+,2} \hat{S}_{2,y,0}, \\
 B_{-,1}^{(1,0)} &= T_{-,1} \hat{S}_{1,x,0} + T_{-,2} \hat{S}_{2,x,0}, \\
 B_{-,2}^{(1,0)} &= T_{-,1} \hat{S}_{1,y,0} + T_{-,2} \hat{S}_{2,y,0}, \\
 T_{+,1} &= - \frac{(T_4 + T_2) T_5}{T_3}, \\
 T_{+,2} &= - \frac{(T_4 + T_2) (T_6 + m_2 T_2)}{T_3}, \\
 T_{-,1} &= \frac{(T_4 - T_2) T_5}{T_3}, \\
 T_{-,2} &= \frac{(T_4 - T_2) (T_6 - m_2 T_2)}{T_3},
 \end{align}
\begin{widetext}
\begin{align}
 T_1 &= m_1 m_2 (\chi_1 - \chi_2) \xi_0 + 
 m_2^2 (1 - 2 \chi_2 \xi_0) - m_1^2 (1 - 2 \chi_1 \xi_0), \\
 T_2 &= \left[ 4 m_1 m_2\chi_1 \chi_2 (M - 
    m_1 \chi_1 \xi_0) (M - m_2 \chi_2 \xi_0) \xi_0^2  + T_1^2
\right]^{1/2}, \\
T_3 &= 4 (M - m_1 \chi_1 \xi_0) T_2  \xi_0, \\
T_4 &= - M (m_1 - m_2) +  \left[ 2 m_1^2 \chi_1 + m_1 m_2 (\chi_1 + 
 \chi_2 ) \right] \xi_0 - 2 m_1 m_2 \chi_1 \chi_2 \xi_0^2, \\
 T_5 &= 2 m_1^2 \chi_1 (M - 
    m_1 \chi_1 \xi_0) \xi_0, \\
 T_6 &= m_2 \left[m_1^2 (1 - 2 \chi_1 \xi_0) - m_1 m_2 (\chi_1 - \chi_2) \xi_0 
- m_2^2 (1 - 2 \chi_2 \xi_0)\right].
\end{align}
where $\xi_0 = \xi(t=0)$.

To describe the time dependence of the amplitudes $A_{i,j}^{(1,0)}$, we solve 
Eq.~\eqref{eq:Apmj10}.
When $\delta m = (m_1-m_2)/M \gg \mathcal{O}(c^{-1})$, we get
\begin{align}
 A_{+,j}^{(1,0)}(t) &= B_{+,j}^{(1,0)} \left[ 1 + \left(\frac{m_1 
\chi_1}{m_1 - m_2} - \frac{m_2 \chi_2}{m_1} \right) \xi(t) \right] + 
\mathcal{O}(c^{-2}) , \\
 A_{-,j}^{(1,0)}(t) &= B_{-,j}^{(1,0)} \left[ 1 - \left(\frac{m_1 
\chi_1}{m_2} - \frac{m_2 \chi_2}{m_1 - m_2} \right) \xi(t) \right] + 
\mathcal{O}(c^{-2}) ;
\end{align}
when $\delta m = (m_1-m_2)/M \sim \mathcal{O}(c^{-1})$, we get
\begin{align}
 A_{+,j}^{(1,0)}(t) &= B_{+,j}^{(1,0)} \left[ 1 + \frac{
\chi_1}{2 \delta m} \xi(t) \right] + 
\mathcal{O}(c^{-2}) , \\
 A_{-,j}^{(1,0)}(t) &= B_{-,j}^{(1,0)} \left[ 1 + \frac{ 
\chi_1 - 3 \chi_2}{4\delta m} \xi(t) 
\right] + 
\mathcal{O}(c^{-2}) ;
\end{align}
when $\delta m = (m_1-m_2)/M \ll \mathcal{O}(c^{-1})$, we get
\begin{align}
 A_{+,j}^{(1,0)}(t) &= B_{+,j}^{(1,0)} \left\{1 - \frac{1}{18} \left[3 T + 9 
\chi_1 + 5 \chi_2 + \frac{42 \chi_1 \chi_2}{T} - \frac{(44\chi_1 - 36 
\chi_2)\chi_2^2}{T^2} \right]\xi(t) \right\} + 
\mathcal{O}(c^{-2}) , \\
 A_{-,j}^{(1,0)}(t) &= B_{-,j}^{(1,0)} \left\{1 + \frac{1}{18} \left[3 T - 9 
\chi_1 - 5 \chi_2 + \frac{42 \chi_1 \chi_2}{T} + \frac{(44\chi_1 - 36 
\chi_2)\chi_2^2}{T^2} \right]\xi(t) \right\} + 
\mathcal{O}(c^{-2}) , \\
T &= \sqrt{9 \chi_1^2 - 2 \chi_1 \chi_2 + 9 \chi_2^2}.
\end{align}

\section{Precession phases}
\label{app:precphases}

In this section, we give the exact expressions for the precession phases that
we used in our implementation. $\phi_{\PP,\pm}$ is calculated from 
$\phi_{\PP,\pm} = \phi_{\PP,m} \pm \delta \phi_\PP$.
We give them in terms of 
the $a_i$ given in appendix~\ref{app:freqevol}, and $\delta m = (m_1 -
m_2)/M$.

The mean precession phase we used in our implementation is
\begin{align}
 \phi_{\PP,m} &= \frac{1}{2} \int \left( \omega_{\PP,+} + \omega_{\PP,-} \right)
\frac{dt}{d\xi}
d\xi = \frac{1}{2} \int \frac{ b+c+d+e+f+g}{a\xi} d\xi \nonumber\\
&= \phi_{\PP,m,0} + \phi_{\PP,m}^{(-3)} \xi^{-3} + \phi_{\PP,m}^{(-2)} \xi^{-2} 
+
\phi_{\PP,m}^{(-1)} \xi^{-1} + \phi_{\PP,m}^{(0)} \log\xi + \phi_{\PP,m}^{(1)} 
\xi +
\phi_{\PP,m}^{(2)} \xi^{2} + \phi_{\PP,m}^{(3)} \xi^{3}, \\
\phi_{\PP,m}^{(-3)} &= - \frac{7 - \delta m^2}{8a_0}, \\
\phi_{\PP,m}^{(-2)} &= -\frac{3}{32a_0} [(5 + 4 \delta m - \delta m^2) \chi_1
+ (5 - 4 \delta m - \delta m^2) \chi_2 ], \\
\phi_{\PP,m}^{(-1)} &= \frac{3}{8 a_0}   [ a_2 (7 - \delta m^2) + 3 (1 -
\delta m^2) \chi_1 \chi_2 ] , \\
\phi_{\PP,m}^{(0)} &= - \frac{3}{16 a_0} \{ 2 a_3 (7 - \delta m^2) + a_2 [(5 + 4
\delta m - \delta m^2) \chi_1
+ (5 - 4 \delta m - \delta m^2) \chi_2 ] \} , \\
\phi_{\PP,m}^{(1)} &= \frac{3}{16 a_0} \{ 2 (a_2^2 - a_4) (7 - \delta m^2) - a_3
[(5 + 4 \delta m - \delta m^2) \chi_1
+ (5 - 4 \delta m - \delta m^2) \chi_2 ] +
6 a_2 (1 - \delta m^2) \chi_1 \chi_2 \} , \\
\phi_{\PP,m}^{(2)} &= \frac{3}{32a_0} \{ (4 a_2 a_3 - 2 a_5) (7 - \delta m^2) +
(a_2^2 - a_4) [(5 + 4
\delta m - \delta m^2) \chi_1
+ (5 - 4 \delta m - \delta m^2) \chi_2 ] + 6 a_3 (1 -
\delta m^2) \chi_1 \chi_2 \} , \\
\phi_{\PP,m}^{(3)} &= -\frac{1}{16 a_0} \{ (2 a_2^3 -
2 a_3^2 - 4 a_2 a_4 ) (7 - \delta m^2) \nonumber\\
&- (2 a_2 a_3 - a_5) [(5 +
4 \delta m - \delta m^2) \chi_1
+ (5 - 4 \delta m - \delta m^2) \chi_2 ] + 6 (a_2^2 - a_4) (1 -
\delta m^2) \chi_1 \chi_2 \} .
\end{align}

$\delta\phi_\PP$ for $\delta m \gg \mathcal{O}(c^{-1})$ is given by
\begin{align}
 \delta \phi_{\PP,1} &= \frac{1}{2} \int \left( \omega_{\PP,+} - \omega_{\PP,-}
\right)
\frac{dt}{d\xi}
d\xi = \frac{1}{2} \int \frac{\sqrt{(b-c+d-e+f-g)^2 + 4(c-f)(b-g)}}{a\xi} d\xi
\nonumber\\
&= \delta \phi_{\PP,1,0} + \delta\phi_{\PP,1}^{(-3)} \xi^{-3} +
\delta\phi_{\PP,1}^{(-2)} \xi^{-2} + \delta\phi_{\PP,1}^{(-1)}
\xi^{-1} + \delta\phi_{\PP,1}^{(0)} \log\xi + \delta\phi_{\PP,1}^{(1)} \xi +
\delta\phi_{\PP,1}^{(2)}
\xi^2 + \delta\phi_{\PP,1}^{(3)}
\xi^3 , \\
\delta \phi_{\PP,1}^{(-3)} &= - \frac{3 \delta m}{4a_0} , \\
\delta \phi_{\PP,1}^{(-2)} &= \frac{9}{32 a_0} [(3 + 4 \delta m +
\delta m^2) \chi_1 - (3 - 4 \delta m +
\delta m^2) \chi_2]
, \\
\delta \phi_{\PP,1}^{(-1)} &= \frac{9}{8 a_0 \delta m} [2 a_2 \delta m^2
- (1 -
\delta m^2) \chi_1 \chi_2 ], \\
\delta \phi_{\PP,1}^{(0)} &= -\frac{9}{32 a_0 \delta m^2} \{ 8 a_3 \delta m^3
- 2 a_2 \delta m^2 [(3 + 4 \delta m +
\delta m^2) \chi_1 - (3 - 4 \delta m +
\delta m^2) \chi_2] \nonumber\\
&- (1 -
\delta m^2) [(3 + 2 \delta m +
\delta m^2) \chi_1 - (3 - 2 \delta m +
\delta m^2) \chi_2] \chi_1 \chi_2 \} , \\
\delta \phi_{\PP,1}^{(1)} &= \frac{9}{128 a_0\delta m^3} \{ 8 \delta
m^5 a_3 ( \chi_1 - \chi_2) + 32 \delta
m^4 [ a_2^2 - a_4 + a_3 (\chi_1 + \chi_2)] + 24 \delta m^3 a_3(
\chi_1 - \chi_2) \nonumber\\
& - [ 16 \delta m^2 a_2 - (1 + \delta m)^2 ( 9 - \delta m^2) \chi_1^2
+ 2 (\delta m^4 - 12 \delta m^2 + 11) \chi_1 \chi_2
- (1 -
\delta m)^2 ( 9 - \delta m^2) \chi_2^2 ] (1 - \delta m^2) \chi_1 \chi_2
\} , \\
\delta \phi_{\PP,1}^{(2)} &= -\frac{9}{1024 a_0 \delta m^4} \bm{[} 32
\delta m^6 (a_2^2 - a_4 ) ( \chi_1 - \chi_2) + 128 \delta m^5 [ a_5 - 2 a_2 a_3
+ (a_2^2 - a_4) (\chi_1 + \chi_2) ] \nonumber \\
&+ 96 \delta m ^4 ( a_2^2 - a_4 ) ( \chi_1
- \chi_2 ) + \bm{(} \delta m^6 ( \chi_1^2 - \chi_2^2) ( \chi_1 + \chi_2) + 2
\delta m^5 (3 \chi_1^2 + 2\chi_1 \chi_2 + 3\chi_2^2 ) (\chi_1 + \chi_2 )
\nonumber\\
& - \delta m^4 [ 16 a_2 - 3 ( \chi_1 + \chi_2 )^2 ] ( \chi_1 - \chi_2 ) + 4
\delta m^3 \{ 16 a_3 + [8 a_2 - 11 ( \chi_1 + \chi_2 )^2 ] ( \chi_1 + \chi_2 )\}
\nonumber\\
&+ \delta m^2 (48 a_2 - 105 \chi_1^2 - 226 \chi_1 \chi_2 - 105
\chi_2^2) ( \chi_1 - \chi_2) - 2 \delta m (45 \chi_1^2 - 106 \chi_1 \chi_2 + 45
\chi_2^2 ) (\chi_1 + \chi_2 ) \nonumber\\
&- 9 ( 3 \chi_1^2 - 10 \chi_1 \chi_2 + 3
\chi_2^2) ( \chi_1 - \chi_2)
\bm{)} (1 - \delta m^2 ) \chi_1 \chi_2
\bm{]}, \\
\delta \phi_{\PP,1}^{(3)} &= \frac{3}{128 a_0 \delta m^3} \bm{(} 8 \delta m^3 
(a_5
- 2 a_2 a_3) [(3 + 4 \delta m +
\delta m^2) \chi_1 - (3 - 4 \delta m +
\delta m^2) \chi_2] + \{ \delta m^4 a_2 ( \chi_1 + \chi_2 )^2 \nonumber\\
&+ 2 \delta m^3 [2 a_3 + a_2(\chi_1 + \chi_2 )] ( \chi_1 - \chi_2)
- 8 \delta m^2 [ 2(a_4 - a_2^2 ) + a_3 (\chi_1 + \chi_2 ) + a_2 ( \chi_1^2 + 3
\chi_1 \chi_2 + \chi_2^2 )] \nonumber \\
& - 6 \delta m [ 2 a_3 + 3 a_2 ( \chi_1 + \chi_2 ) ] ( \chi_1 - \chi_2 )
- a_2 ( 9 \chi_1^2 - 22 \chi_1 \chi_2 + 9 \chi_2^2 )
\} (1 - \delta m^2 ) \chi_1 \chi_2
\bm{)}.
\end{align}

When $\delta m \sim \mathcal{O} ( c^{-1} )$, $\delta \phi_\PP$ is given by
\begin{align}
 \delta \phi_{\PP,2} &= \frac{1}{2} \int \left( \omega_{\PP,+} - \omega_{\PP,-}
\right)
\frac{dt}{d\xi}
d\xi = \frac{1}{2} \int \frac{\sqrt{(b-c+d-e+f-g)^2 + 4(c-f)(b-g)}}{a\xi} d\xi
\nonumber\\
&= \delta \phi_{\PP,2,0} + \frac{1}{T \xi^3} \left( \delta \phi_{\PP,2}^{(-3)} +
\delta \phi_{\PP,2}^{(-2)} \xi + \delta \phi_{\PP,2}^{(-1)} \xi^2 + \delta
\phi_{\PP,2}^{(0)} \xi^3 + \delta \phi_{\PP,2}^{(1)} \xi^4 + \delta
\phi_{\PP,2}^{(2)} \xi^5 \right) \nonumber\\
& + \delta \phi_{\PP,2}^{(l,1)} [ \log\xi - \log | 4 \delta m - 3 ( \chi_1 -
\chi_2) \xi + T | ] \nonumber\\
&+ \delta
\phi_{\PP,2}^{(l,2)} \log | 12 \delta m ( \chi_1 - \chi_2 ) - 
( 9 \chi_1^2 - 2
\chi_1 \chi_2 + 9 \chi_2^2 ) \xi - (9 \chi_1^2 - 2
\chi_1 \chi_2 + 9 \chi_2^2)^{1/2} T |, \\
T &= [ 16 \delta m^2 - 24 \delta m ( \chi_1 - \chi_2 ) \xi + ( 9 \chi_1^2 - 2
\chi_1 \chi_2 + 9 \chi_2^2 ) \xi^2 ]^{1/2}, \\
\delta \phi_{\PP,2}^{(-3)} &= - \frac{3}{16 a_0} , \\
\delta \phi_{\PP,2}^{(-2)} &= \frac{9}{128 a_0 \delta m} [ (1 + 4 \delta m +
\delta m^2) \chi_1 - ( 1 - 4 \delta m^2 + \delta m^2 ) \chi_2 ] , \\
\delta \phi_{\PP,2}^{(-1)} &= \frac{3}{512 a_0 \delta m^2} [ 96 a_2 \delta m^2 
+ 
9
(1 + 4 \delta m + \delta
m^2) \chi_1^2 - 10 ( 5 - 3 \delta m^2) \chi_1 \chi_2 + 9
(1 - 4 \delta m + \delta
m^2) \chi_2^2]  , \\
\delta \phi_{\PP,2}^{(0)} &= \frac{3}{16 a_0 ( 9 \chi_1^2 - 2 \chi_1 \chi_2 + 9
\chi_2^2)^4 } \{ 3456 \delta m^4 a_3 (81 \chi_1^6 - 567 \chi_1^5 \chi_2 +
1123 \chi_1^4 \chi_2^2 - 
  1530 \chi_1^3 \chi_2^3 + 1123 \chi_1^2 \chi_2^4 - 
  567 \chi_1 \chi_2^5 \nonumber \\
  &+ 81 \chi_2^6) (\chi_1 + \chi_2 )^2 + 864 \delta m^3 a_2 (27 \chi_1^4 - 63
\chi_1^3 \chi_2 + 88 \chi_1^2 \chi_2^2 - 
  63 \chi_1 \chi_2^3 + 27 \chi_2^4 )  (9 \chi_1^2 - 
   2 \chi_1 \chi_2 \nonumber \\
&+ 9 \chi_2^2) (\chi_1 +
\chi_2)^2  (\chi_1 - \chi_2) - \delta m^2 a_3 (6561 \chi_1^8 + 17496
\chi_1^7
\chi_2 + 
  89676 \chi_1^6 \chi_2^2 - 229976 \chi_1^5 \chi_2^3 + 
  355366 \chi_1^4 \chi_2^4 \nonumber\\
  &- 229976 \chi_1^3 \chi_2^5 + 
  89676 \chi_1^2 \chi_2^6 + 17496 \chi_1 \chi_2^7 + 
  6561 \chi_2^8) - 6 \delta m [ 6 a_3 (9 \chi_1^2 - 
   8 \chi_1 \chi_2 + 9 \chi_2^2)(9 \chi_1^2 - 
   2 \chi_1 \chi_2 \nonumber \\
   &+ 9 \chi_2^2)(\chi_1 + \chi_2) + 2 a_2
(27 \chi_1^4 + 225 \chi_1^3 \chi_2 - 184 \chi_1^2 \chi_2^2 + 
  225 \chi_1 \chi_2^3 + 27 \chi_2^4) \chi_1 \chi_2
+ 9 (9 \chi_1^2 - 
   2 \chi_1 \chi_2 + 9 \chi_2^2)^2 ( 3
\chi_1^2 \nonumber \\
& - \chi_1 \chi_2 + 3 \chi_2^2)  (\chi_1 + \chi_2)^2
  ] (9 \chi_1^2 - 
   2 \chi_1 \chi_2 + 9 \chi_2^2) ( \chi_1 - \chi_2 )
   - 3 [ a_3 (9 \chi_1^2 - 
   2 \chi_1 \chi_2 + 9 \chi_2^2) \nonumber \\
   &- 4 a_2 ( \chi_1 + \chi_2 ) \chi_1
\chi_2 ](9 \chi_1^2 - 
   2 \chi_1 \chi_2 + 9 \chi_2^2)^3
   \} , \\
\delta \phi_{\PP,2}^{(1)} &= -\frac{9}{32 a_0 ( 9 \chi_1^2 - 2 \chi_1 \chi_2 + 9
\chi_2^2)^4 } \{ 
576 \delta m^3 a_3 (243 \chi_1^6 - 2781 \chi_1^5 \chi_2 + 5577
\chi_1^4 \chi_2^2 - 
  7870 \chi_1^3 \chi_2^3 + 5577 \chi_1^2 \chi_2^4 \nonumber \\
&-  2781 \chi_1 \chi_2^5 + 243 \chi_2^6) ( \chi_1 + \chi_2 )^2 (\chi_1 -
\chi_2 ) + 144 \delta m^2 a_2 (81 \chi_1^6 - 567 \chi_1^5 \chi_2 + 1123
\chi_1^4 \chi_2^2 - 
  1530 \chi_1^3 \chi_2^3 + 1123 \chi_1^2 \chi_2^4 \nonumber \\
&- 567 \chi_1 \chi_2^5 + 81 \chi_2^6)(9 \chi_1^2 - 2 \chi_1 \chi_2 + 9
\chi_2^2) (\chi_1 + \chi_2)^2 + 4 \delta m a_3  (27 \chi_1^4 + 230 \chi_1^3
\chi_2 - 194 \chi_1^2 \chi_2^2 + 
  230 \chi_1 \chi_2^3 \nonumber \\
  &+ 27 \chi_2^4) (9 \chi_1^2 - 2 \chi_1 \chi_2 + 9
\chi_2^2) (\chi_1 - \chi_2) \chi_1 \chi_2 - 4  a_3 (9 \chi_1^2 - 2 \chi_1 \chi_2
+ 9 \chi_2^2)^3 (\chi_1 + \chi_2 ) \chi_1 \chi_2 + 10 a_2 ( 9
\chi_1^2 - 2 \chi_1 \chi_2 \nonumber \\
&+ 9 \chi_2^2)^2 (\chi_1 - \chi_2 )^2 \chi_1^2
\chi_2^2 - 9 (9 \chi_1^4 - 13 \chi_1^3 \chi_2 + 24 \chi_1^2 \chi_2^2 - 
  13 \chi_1 \chi_2^3 + 9 \chi_2^4) ( 9 \chi_1^2 - 2 \chi_1
\chi_2 + 9 \chi_2^2)^3 (\chi_1 + \chi_2 )^2
\} , \\
\delta \phi_{\PP,2}^{(2)} &= -\frac{15 a_3 ( \chi_1 - \chi_2 )^2 \chi_1^2
\chi_2^2}{8 a_0 ( 9 \chi_1^2 - 2 \chi_1 \chi_2 + 9 \chi_2^2)^2 } , \\
\delta \phi_{\PP,2}^{(l,1)} &= -\frac{9}{32 a_0 \delta m^2} \{ 8 \delta m^3 a_3
-2 \delta m^2 a_2 [ ( 3 + 4 \delta m + \delta m^2 ) \chi_1 - ( 3 - 4 \delta m +
\delta m^2 ) \chi_2 ] \nonumber \\
&- [ (3 + 2 \delta m - 4 \delta m^2) \chi_1 -
(3 - 2 \delta m - 4 \delta m^2 ) \chi_2 ]
\}, \\
\delta \phi_{\PP,2}^{(l,2)} &= - \frac{9}{16 a_0 ( 9 \chi_1^2 - 2 \chi_1
\chi_2 + 9 \chi_2^2)^{9/2}} \{ 
32 \delta m^3 a_3 (12393 \chi_1^6 + 5238 \chi_1^5 \chi_2 - 
  25065 \chi_1^4 \chi_2^2 + 53780 \chi_1^3 \chi_2^3 - 
  25065 \chi_1^2 \chi_2^4 \nonumber\\
  &+ 5238 \chi_1 \chi_2^5 + 
  12393 \chi_2^6) (\chi_1 - \chi_2 ) \chi_1 \chi_2 - \delta m^2 [ 64 a_3 (45
\chi_1^2 - 34 \chi_1 \chi_2 + 45 \chi_2^2) (9 \chi_1^2 - 2 \chi_1 \chi_2
\nonumber\\
&+ 9 \chi_2^2)^2 (\chi_1 + \chi_2) \chi_1 \chi_2 - a_2 (2187 \chi_1^8 + 5184
\chi_1^7 \chi_2 +   25668 \chi_1^6 \chi_2^2 - 61184 \chi_1^5 \chi_2^3 + 
  97250 \chi_1^4 \chi_2^4 - 61184 \chi_1^3 \chi_2^5 \nonumber\\
  &+  25668 \chi_1^2 \chi_2^6 + 5184 \chi_1 \chi_2^7 + 2187 \chi_2^8) (9
\chi_1^2 - 2 \chi_1 \chi_2 + 9 \chi_2^2) ]
- 12 \delta m [a_3 (9 \chi_1^2 - 2 \chi_1 \chi_2 + 9 \chi_2^2) \nonumber\\
&- 3 a_2  (3
\chi_1^2 - 2 \chi_1 \chi_2 + 3 \chi_2^2)  (\chi_1 + \chi_2) ]  (9
\chi_1^2 - 2 \chi_1 \chi_2 + 9 \chi_2^2)^3 ( \chi_1 - \chi_2 ) + [ a_2 (9
\chi_1^2 - 2 \chi_1 \chi_2 + 9 \chi_2^2)^2 \nonumber\\
&- 10 ( \chi_1^2 - \chi_2^2
) \chi_1^2 \chi_2^2 ] (9 \chi_1^2 - 2 \chi_1 \chi_2 + 9 \chi_2^2)^3
\}.
\end{align}

When $\delta m \ll \mathcal{O} ( c^{-1} )$, $\delta \phi_\PP$ is given by
\begin{align}
 \delta \phi_{\PP,3} &= \frac{1}{2} \int \left( \omega_{\PP,+} - \omega_{\PP,-}
\right)
\frac{dt}{d\xi}
d\xi = \frac{1}{2} \int \frac{\sqrt{(b-c+d-e+f-g)^2 + 4(c-f)(b-g)}}{a\xi} d\xi
\nonumber\\
&= \delta \phi_{\PP,3,0} + \delta
\phi_{\PP,3}^{(0)} + \frac{T}{\xi^3} \left( \delta \phi_{\PP,3}^{(-3)} +
\delta \phi_{\PP,3}^{(-2)} \xi + \delta \phi_{\PP,3}^{(-1)} \xi^2  + \delta
\phi_{\PP,3}^{(1)} \xi^4 \right) \nonumber\\
&+ \delta
\phi_{\PP,3}^{(l,1)}  \log \left| \frac{2 (\chi_1 + \chi_2 ) - 2 \chi_1 \chi_2 
\xi
- T}{2 (\chi_1 + \chi_2 ) - (9 \chi_1^2 - 2 \chi_1
\chi_2 + 9 \chi_2^2)^{1/2}} \right| \nonumber\\
&+
\delta
\phi_{\PP,3}^{(l,2)}  [ \log \xi - \log | 9 \chi_1^2 - 2 \chi_1 \chi_2 + 9
\chi_2^2 - 4 ( \chi_1 + \chi_2 ) \chi_1 \chi_2 \xi + (9 \chi_1^2 - 2 \chi_1
\chi_2 + 9 \chi_2^2)^{1/2} T | ], \\
T &= [ (9 \chi_1^2 - 2 \chi_1 \chi_2 + 9
\chi_2^2) - 8 (
\chi_1 + \chi_2 ) \chi_1 \chi_2 \xi + 4 \chi_1^2 \chi_2^2 \xi^2]^{1/2}, \\
\delta \phi_{\PP,3}^{(-3)} &= \frac{9 \delta m (\chi_1 - \chi_2)}{4 a_0 (9
\chi_1^2 - 2 \chi_1
\chi_2 + 9 \chi_2^2)} , \\
\delta \phi_{\PP,3}^{(-2)} &= - \frac{9}{32 a_0} - \frac{9 \delta m (27 
\chi_1^2 
-
26 \chi_1 \chi_2 + 27 \chi_2^2) ( \chi_1^2 - \chi_2^2) }{8 a_0 (9 \chi_1^2
- 2 \chi_1
\chi_2 + 9 \chi_2^2)^2} , \\
\delta \phi_{\PP,3}^{(-1)} &= \frac{9 ( \chi_1 + \chi_2 ) \chi_1 \chi_2}{8 a_0 
(9
\chi_1^2
- 2 \chi_1
\chi_2 + 9 \chi_2^2)} - \frac{27 a_2 \delta m (\chi_1 - \chi_2)}{4 a_0 (9
\chi_1^2
- 2 \chi_1
\chi_2 + 9 \chi_2^2)} \nonumber\\
&- \frac{9 \delta m (81 \chi_1^4 + 276 \chi_1^3
\chi_2 - 170 \chi_1^2 \chi_2^2 + 
  276 \chi_1 \chi_2^3 + 81 \chi_2^4) ( \chi_1 - \chi_2) \chi_1 \chi_2 }{4
a_0 (9
\chi_1^2
- 2 \chi_1
\chi_2 + 9 \chi_2^2)^3}
,\\
\delta \phi_{\PP,3}^{(0)} &= - \frac{9 a_2 T}{16 a_0} + \frac{9 a_3 [ \chi_1 +
\chi_2 + 2 \delta m (\chi_1 - \chi_2 )]}{32 a_0 \chi_1 \chi_2} \left( T -
\sqrt{9
\chi_1^2
- 2 \chi_1
\chi_2 + 9 \chi_2^2} \right) ,
\\
\delta \phi_{\PP,3}^{(1)} &= - \frac{9 a_3}{32 a_0} , \\
\delta \phi_{\PP,3}^{(l,1)} &= \frac{9 \{ a_3 [(5 + 16 \delta m) \chi_1 - (5 -
16\delta m) \chi_2 ] ( \chi_1 - \chi_2 ) - 8 a_2 [ \chi_1 + \chi_2 + \delta m (
\chi_1 - \chi_2 )] \chi_1 \chi_2 \} }{64 a_0 \chi_1 \chi_2} , \\
\delta \phi_{\PP,3}^{(l,2)} &= - \frac{9}{16 a_0 (9 \chi_1^2 - 2 \chi_1
\chi_2 + 9 \chi_2^2)^{7/2}} \{(9 \chi_1^2 - 2 \chi_1
\chi_2 + 9 \chi_2^2)^2 [
a_2 (9 \chi_1^2 - 2 \chi_1
\chi_2 + 9 \chi_2^2)^2 - 10 ( \chi_1 - \chi_2)^2 \chi_1^2 \chi_2^2 ] \nonumber\\
&
- \delta m
[ 12 a_3  (9 \chi_1^2 - 2 \chi_1
\chi_2 + 9 \chi_2^2)^3 ( \chi_1 - \chi_2)
-36 a_2 (9 \chi_1^2 - 2 \chi_1
\chi_2 + 9 \chi_2^2)^2 (3 \chi_1^2 - 2 \chi_1
\chi_2 + 3 \chi_2^2) ( \chi_1^2 - \chi_2^2 ) \nonumber \\
&-8 (81 \chi_1^4 - 840 \chi_1^3 \chi_2 + 878 \chi_1^2 \chi_2^2 - 
  840 \chi_1 \chi_2^3 + 81 \chi_2^4)  ( \chi_1^2 - \chi_2^2 ) \chi_1^2
\chi_2^2
]\}.
\end{align}

Note that despite appearances, $\delta \phi_{\PP,3}$ is regular in the
limits $\chi_1 \to 0$ and $\chi_2 \to 0$.

From $\omega_{\PP,m}(\xi)$, $\delta \omega_{\PP}(\xi)$ and $\dot{\xi}$, we 
can compute $\ddot{\phi}_{\PP,\pm} = \dot{\omega}_{\PP,\pm} = 
\dot{\omega}_{\PP,m} \pm \delta \dot{\omega}_{\PP}$. We get
\begin{align}
 \dot{\omega}_{\PP,m} &= \dot{\omega} \left\{ \frac{5}{24} \left( 7 - \delta 
m^2\right) \xi^2 + \frac{1}{8} \left[ \left( 5 + 4 \delta m - \delta m^2\right) 
\chi_1 + \left(5 - 4 \delta m - \delta m^2 \right) \chi_2 \right] \xi^3 - 
\frac{7}{8} \left( 1 - \delta m^2 \right) \chi_1 \chi_2 \xi^4 \right\}, \\
 \delta \dot{\omega}_{\PP} &= \frac{\dot{\omega}}{\delta \omega_\PP} \bigg\{ 
\frac{15 \delta m^2}{16} \xi^7 + \frac{33 \delta m}{64} \left[ \left( 3 + 4 
\delta m  + \delta m^2 \right) \chi_1 - \left( 3 - 4 \delta m + \delta 
m^2\right) \chi_2 \right] \xi^8 \nonumber\\
&+ \frac{9}{128} \left[ \left( 3 + 4 \delta m  + 
\delta m^2 \right)^2 \chi_1^2 - 2 \left(1 - \delta m^2\right)^2 \chi_1 \chi_2  
+ \left( 3 - 4 \delta m + \delta m^2\right)^2 \chi_2^2 \right] \xi^9 \nonumber\\
& + \frac{39}{64} \left(1 - \delta m^2\right) \left[\left(1+ \delta m\right) 
\chi_1 + \left(1 - \delta m\right)\chi_2\right] \chi_1 \chi_2 \xi^{10} + 
\frac{21}{64} \left( 1 + \delta m\right)^2 \left(1 - \delta m\right)^2 \chi_1^2 
\chi_2^2 \xi^{11}
\bigg\}.
\end{align}

\section{Amplitude and Phase Modulations}
\label{app:amplitudeandphasemodulations}

Using $\bm{V}_1 = \uvec{z} \times \uvec{N}$ and $\bm{V}_2 = \bm{z} - \uvec{N} 
(\uvec{z} \cdot \uvec{N})$,
the amplitudes and phases of Eq.~\eqref{conversion} are given by
\begin{align}
A_{0,n,k,m} &= k \arccos \hat{N}_z + m \arctan \frac{V_{1,z}}{V_{2,z}} \\
 A_{\delta \phi,\pm} &=  \mbox{sign}\left( \hat{N}_x 
B_{\pm,2}^{(1,0)} - \hat{N}_y
B_{\pm,1}^{(1,0)} \right) \left[ 
\frac{\left.B_{\pm,1}^{(1,0)}\right.^2 +
\left.B_{\pm,2}^{(1,0)}\right.^2}{\hat{N}_x^2 + \hat{N}_y^2} \right]^{1/2} 
\hat{N}_z \frac{\xi}{\nu}, \\
\phi_{\delta\phi,\pm}^{(0)} &= - \arctan \left( \frac{\hat{N}_x 
B_{\pm,1}^{(1,0)} + \hat{N}_y 
B_{\pm,2}^{(1,0)}}{\hat{N}_x 
B_{\pm,2}^{(1,0)} - \hat{N}_y 
B_{\pm,1}^{(1,0)}} \right), \\
 A_{i_L,\pm} &= -\mbox{sign}\left( \hat{N}_x 
B_{\pm,1}^{(1,0)} + \hat{N}_y
B_{\pm,2}^{(1,0)} \right) \left[ 
\left.B_{\pm,1}^{(1,0)}\right.^2 +
\left.B_{\pm,2}^{(1,0)}\right.^2 \right]^{1/2} \frac{\xi}{\nu}, \\
\phi_{i_L,\pm}^{(0)} &= \arctan \left( \frac{\hat{N}_x 
B_{\pm,2}^{(1,0)} - \hat{N}_y 
B_{\pm,1}^{(1,0)}}{\hat{N}_x 
B_{\pm,1}^{(1,0)} + \hat{N}_y 
B_{\pm,2}^{(1,0)}} \right), \\
A_{\psi_N,\pm} &= \mbox{sign} \Big[ V_{1,z} \Big( B_{\pm,1}^{(1,0)} 
V_{2,x} +
B_{\pm,2}^{(1,0)} V_{2,y} \Big) - V_{2,z} \Big( 
B_{\pm,1}^{(1,0)} V_{1,x} + 
B_{\pm,2}^{(1,0)} V_{1,y} \Big) \Big] \nonumber\\
&\times
\bigg\{ \Big[ V_{1,z} \Big( B_{\pm,1}^{(1,0)} 
V_{2,x} +
B_{\pm,2}^{(1,0)} V_{2,y} \Big) - V_{2,z} \Big( 
B_{\pm,1}^{(1,0)} V_{1,x} + 
B_{\pm,2}^{(1,0)} V_{1,y} \Big) \Big]^2 \nonumber\\
&+ \Big[ V_{1,z} \Big( 
B_{\pm,1}^{(1,0)} V_{2,y} - 
B_{\pm,2}^{(1,0)} V_{2,x} \Big) - V_{2,z} \Big( B_{\pm,1}^{(1,0)} V_{1,y} - 
B_{\pm,2}^{(1,0)} V_{1,x}\Big) \Big]^2 \bigg\}^{1/2}  \left( V_{1,z}^2 
+ V_{2,z}^2   \right)^{-1} \frac{\xi}{\nu} , \\
\phi_{\psi_N,\pm}^{(0)} &= - \arctan \bigg\{ \Big[ V_{1,z} \Big( 
B_{\pm,1}^{(1,0)} V_{2,y} - 
B_{\pm,2}^{(1,0)} V_{2,x} \Big) - V_{2,z} \Big( B_{\pm,1}^{(1,0)} V_{1,y} - 
B_{\pm,2}^{(1,0)} V_{1,x}\Big) \Big] \nonumber\\
&\times\Big[ V_{1,z} \Big( B_{\pm,1}^{(1,0)} 
V_{2,x} +
B_{\pm,2}^{(1,0)} V_{2,y} \Big) - V_{2,z} \Big( 
B_{\pm,1}^{(1,0)} V_{1,x} + 
B_{\pm,2}^{(1,0)} V_{1,y} \Big) \Big]^{-1} \bigg\}.
\end{align}

Using this, we can express
\begin{align}
 A_{\pm,n,k,m} &= \mbox{sign}(A_{c,\pm}) \sqrt{A_{c,\pm}^2 + A_{s,\pm}^2}, \\
 \phi_{\pm,n,k,m} &= \arctan \left( \frac{A_{s,\pm}}{A_{c,\pm}} \right), \\
 A_{c,\pm} &= n A_{\delta\phi,\pm} \cos (\phi_{\delta\phi,\pm}^{(0)}) + k 
A_{i_L,\pm} \cos (\phi_{i_L,\pm}^{(0)})
+ m 
A_{\psi_N,\pm} \cos ( \phi_{\psi_N,\pm}^{(0)}), \\
A_{s,\pm} &= n A_{\delta\phi,\pm} \sin 
(\phi_{\delta\phi,\pm}^{(0)}) + k 
A_{i_L,\pm} \sin (\phi_{i_L,\pm}^{(0)}) 
+ m 
A_{\psi_N,\pm} \sin ( \phi_{\psi_N,\pm}^{(0)}) ,
\end{align}

\section{Mode-Decomposed, Time-Domain Amplitudes}
\label{app:amplitudes}

The waveforms are expressed as
\begin{align}
 h(t) = \frac{G\mu \xi^2}{D_L c^2} \sum_{n \geq 0} \ \sum_{k \in \mathbb{Z}} \ 
\sum_{m = -2,2}
 \mathcal{A}_{n,k,m}(\theta_N, \phi_N) e^{i [ n\psi_{\C} + n\delta\phi + k
i_L + m \psi_N]} + \rm{c.c.} \, ,
\end{align}
where $\psi_\C = \phi_\C - (6 - 3 \nu \xi^2) \xi^3 \log\xi$.

Defining
\begin{align}
\mathcal{A}_F &= \frac{1}{2} \left( 1 + \cos^2 \theta_N
\right) \cos 2\phi_N, \\
\mathcal{B}_F &= \cos \theta_N \sin 2\phi_N,
\end{align}
and using $\delta m = (m_1 - m_2)/M$ the dimensionless mass difference,
we can express the $\mathcal{A}_{n,k,m}$ at 2PN for $n \neq 0$ as 
(see~\cite{abiq})
\begin{align}
 \mathcal{A}_{1,1,2} &= - \mathcal{A}_{1,-1,2} = - \delta m ( \mathcal{B}_F - i 
\mathcal{A}_F ) \left\{
\frac{21}{128}\xi - \left( \frac{575}{6144} - \frac{367 \nu}{3072}\right) \xi^3
+ \left[ \frac{21\pi}{128} + i \left(\frac{21 \log 2}{64} + \frac{51}{640}
\right) \right] \xi^4 \right\}, \\
 \mathcal{A}_{1,2,2} &= - \mathcal{A}_{1,-2,2} = \delta m ( \mathcal{B}_F - i 
\mathcal{A}_F ) \left\{
\frac{3}{32}\xi - \left( \frac{121}{1536} - \frac{41 \nu}{768}\right) \xi^3
+ \left[ \frac{3\pi}{32} + i \left(\frac{3 \log 2}{16} + \frac{9}{160}
\right) \right] \xi^4 \right\}, \\
 \mathcal{A}_{1,3,2} &= - \mathcal{A}_{1,-3,2} = - \delta m ( \mathcal{B}_F - i 
\mathcal{A}_F ) \left\{
\frac{1}{128}\xi - \left( \frac{79}{4096} + \frac{17 \nu}{2048}\right) \xi^3
+ \left[ \frac{\pi}{128} + i \left(\frac{\log 2}{64} + \frac{7}{640}
\right) \right] \xi^4 \right\}, \\
 \mathcal{A}_{1,4,2} &= - \mathcal{A}_{1,-4,2} = \delta m ( \mathcal{B}_F - i 
\mathcal{A}_F ) \left( \frac{5}{3072}
- \frac{5 \nu}{1536}\right) \xi^3, \\
 \mathcal{A}_{1,5,2} &= - \mathcal{A}_{1,-5,2} = - \delta m ( \mathcal{B}_F - i 
\mathcal{A}_F ) \left( \frac{1}{12288}
- \frac{\nu}{6144}\right) \xi^3, \\
 \mathcal{A}_{1,1,-2} &= - \mathcal{A}_{1,-1,-2} = \delta m ( \mathcal{B}_F + i 
\mathcal{A}_F ) \left\{
\frac{21}{128}\xi - \left( \frac{575}{6144} - \frac{367 \nu}{3072}\right) \xi^3
+ \left[ \frac{21\pi}{128} + i \left(\frac{21 \log 2}{64} + \frac{51}{640}
\right) \right] \xi^4 \right\}, \\
 \mathcal{A}_{1,2,-2} &= - \mathcal{A}_{1,-2,-2} = \delta m ( \mathcal{B}_F + i 
\mathcal{A}_F ) \left\{
\frac{3}{32}\xi - \left( \frac{121}{1536} - \frac{41 \nu}{768}\right) \xi^3
+ \left[ \frac{3\pi}{32} + i \left(\frac{3 \log 2}{16} + \frac{9}{160}
\right) \right] \xi^4 \right\}, \\
 \mathcal{A}_{1,3,-2} &= - \mathcal{A}_{1,-3,-2} = \delta m ( \mathcal{B}_F + i 
\mathcal{A}_F ) \left\{
\frac{1}{128}\xi - \left( \frac{79}{4096} + \frac{17 \nu}{2048}\right) \xi^3
+ \left[ \frac{\pi}{128} + i \left(\frac{\log 2}{64} + \frac{7}{640}
\right) \right] \xi^4 \right\}, \\
 \mathcal{A}_{1,4,-2} &= - \mathcal{A}_{1,-4,-2} = \delta m ( \mathcal{B}_F + i 
\mathcal{A}_F ) \left( \frac{5}{3072}
- \frac{5 \nu}{1536}\right) \xi^3, \\
 \mathcal{A}_{1,5,-2} &= - \mathcal{A}_{1,-5,-2} = \delta m ( \mathcal{B}_F + i 
\mathcal{A}_F ) \left( \frac{1}{12288}
- \frac{\nu}{6144}\right) \xi^3, \\
 \mathcal{A}_{2,0,2} &= - (\mathcal{A}_F + i \mathcal{B}_F) \left[ \frac{3}{4} 
- 
\left( \frac{91}{48} -
\frac{15\nu}{16} \right) \xi^2 + \frac{3\pi}{2} \xi^3 - \left( \frac{291}{256}
+ \frac{8137\nu}{2304} - \frac{257\nu^2}{256} \right) \xi^4 \right], \\
 \mathcal{A}_{2,1,2} &= \mathcal{A}_{2,-1,2} = (\mathcal{A}_F + i 
\mathcal{B}_F) 
\left[ \frac{1}{2} - \left(
\frac{7}{6} -
\frac{\nu}{3} \right) \xi^2 + \pi \xi^3 - \left( \frac{365}{384}
+ \frac{1861\nu}{1152} - \frac{103\nu^2}{384} \right) \xi^4 \right], \\
 \mathcal{A}_{2,2,2} &= \mathcal{A}_{2,-2,2} = -(\mathcal{A}_F + i 
\mathcal{B}_F) \left[ \frac{1}{8} - \left(
\frac{7}{48} +
\frac{17\nu}{48} \right) \xi^2 + \frac{\pi}{4} \xi^3 - \left( \frac{8581}{15360}
+ \frac{7247\nu}{9216} - \frac{1541\nu^2}{3072} \right) \xi^4 \right], \\
 \mathcal{A}_{2,3,2} &= \mathcal{A}_{2,-3,2} = (\mathcal{A}_F + i 
\mathcal{B}_F) 
\left[ \left(
\frac{1}{12} -
\frac{\nu}{4} \right) \xi^2 - \left( \frac{829}{3840}
- \frac{1735\nu}{2304} + \frac{205\nu^2}{768} \right) \xi^4 \right], \\
 \mathcal{A}_{2,4,2} &= \mathcal{A}_{2,-4,2} = -(\mathcal{A}_F + i 
\mathcal{B}_F) \left[ \left(
\frac{1}{96} -
\frac{\nu}{32} \right) \xi^2 - \left( \frac{55}{1536}
- \frac{457\nu}{4608} - \frac{29\nu^2}{1536} \right) \xi^4 \right], \\
 \mathcal{A}_{2,5,2} &= \mathcal{A}_{2,-5,2} = (\mathcal{A}_F + i 
\mathcal{B}_F) 
 \left( \frac{1}{256}
- \frac{5\nu}{256} + \frac{5\nu^2}{256} \right) \xi^4, \\
 \mathcal{A}_{2,6,2} &= \mathcal{A}_{2,-6,2} = -(\mathcal{A}_F + i 
\mathcal{B}_F)  \left( \frac{1}{3072}
- \frac{5\nu}{3072} + \frac{5\nu^2}{3072} \right) \xi^4, \\
 \mathcal{A}_{2,0,-2} &= - (\mathcal{A}_F - i \mathcal{B}_F) \left[ \frac{3}{4} 
- \left( \frac{91}{48} -
\frac{15\nu}{16} \right) \xi^2 + \frac{3\pi}{2} \xi^3 - \left( \frac{291}{256}
+ \frac{8137\nu}{2304} - \frac{257\nu^2}{256} \right) \xi^4 \right], \\
 \mathcal{A}_{2,1,-2} &= \mathcal{A}_{2,-1,-2} = -(\mathcal{A}_F - i 
\mathcal{B}_F) \left[ \frac{1}{2} - \left(
\frac{7}{6} -
\frac{\nu}{3} \right) \xi^2 + \pi \xi^3 - \left( \frac{365}{384}
+ \frac{1861\nu}{1152} - \frac{103\nu^2}{384} \right) \xi^4 \right], \\
 \mathcal{A}_{2,2,-2} &= \mathcal{A}_{2,-2,-2} = -(\mathcal{A}_F - i 
\mathcal{B}_F) \left[ \frac{1}{8} - \left(
\frac{7}{48} +
\frac{17\nu}{48} \right) \xi^2 + \frac{\pi}{4} \xi^3 - \left( \frac{8581}{15360}
+ \frac{7247\nu}{9216} - \frac{1541\nu^2}{3072} \right) \xi^4 \right], \\
 \mathcal{A}_{2,3,-2} &= \mathcal{A}_{2,-3,-2} = -(\mathcal{A}_F - i 
\mathcal{B}_F) \left[ \left(
\frac{1}{12} -
\frac{\nu}{4} \right) \xi^2 - \left( \frac{829}{3840}
- \frac{1735\nu}{2304} + \frac{205\nu^2}{768} \right) \xi^4 \right], \\
 \mathcal{A}_{2,4,-2} &= \mathcal{A}_{2,-4,-2} = -(\mathcal{A}_F - i 
\mathcal{B}_F) \left[ \left(
\frac{1}{96} -
\frac{\nu}{32} \right) \xi^2 - \left( \frac{55}{1536}
- \frac{457\nu}{4608} - \frac{29\nu^2}{1536} \right) \xi^4 \right], \\
 \mathcal{A}_{2,5,-2} &= \mathcal{A}_{2,-5,-2} = -(\mathcal{A}_F - i 
\mathcal{B}_F)  \left( \frac{1}{256}
- \frac{5\nu}{256} + \frac{5\nu^2}{256} \right) \xi^4, \\
 \mathcal{A}_{2,6,-2} &= \mathcal{A}_{2,-6,-2} = -(\mathcal{A}_F - i 
\mathcal{B}_F)  \left( \frac{1}{3072}
- \frac{5\nu}{3072} + \frac{5\nu^2}{3072} \right) \xi^4, \\
 \mathcal{A}_{3,1,2} &= - \mathcal{A}_{3,-1,2} = \delta m ( \mathcal{B}_F - i 
\mathcal{A}_F ) \left\{
\frac{45}{128}\xi - \left( \frac{5895}{4096} - \frac{1575 \nu}{2048}\right)
\xi^3
+ \left[ \frac{135\pi}{128} - i \left(\frac{135 \log(3/2)}{64} -
\frac{189}{128}
\right) \right] \xi^4 \right\}, \\
 \mathcal{A}_{3,2,2} &= - \mathcal{A}_{3,-2,2} = -\delta m ( \mathcal{B}_F - i 
\mathcal{A}_F ) \left\{
\frac{9}{32}\xi - \left( \frac{1071}{1024} - \frac{207 \nu}{512}\right)
\xi^3
+ \left[ \frac{27\pi}{32} + i \left(\frac{27 \log (3/2)}{16} -
\frac{189}{160}
\right) \right] \xi^4 \right\}, \\
 \mathcal{A}_{3,3,2} &= - \mathcal{A}_{3,-3,2} = \delta m ( \mathcal{B}_F - i 
\mathcal{A}_F ) \left\{
\frac{9}{128}\xi - \left( \frac{1197}{8192} + \frac{531 \nu}{4096}\right)
\xi^3
+ \left[ \frac{27\pi}{128} - i \left(\frac{27 \log(3/2)}{64} -
\frac{189}{640}
\right) \right] \xi^4 \right\}, \\
 \mathcal{A}_{3,4,2} &= - \mathcal{A}_{3,-4,2} = -\delta m ( \mathcal{B}_F - i 
\mathcal{A}_F ) \left( \frac{135}{2048} -
\frac{135 \nu}{1024}\right)
\xi^3, \\
 \mathcal{A}_{3,5,2} &= - \mathcal{A}_{3,-5,2} = \delta m ( \mathcal{B}_F - i 
\mathcal{A}_F ) \left( \frac{81}{8192} -
\frac{81 \nu}{4096}\right)
\xi^3, \\
 \mathcal{A}_{3,1,-2} &= - \mathcal{A}_{3,-1,-2} = - \delta m ( \mathcal{B}_F + 
i \mathcal{A}_F ) \left\{
\frac{45}{128}\xi - \left( \frac{5895}{4096} - \frac{1575 \nu}{2048}\right)
\xi^3
+ \left[ \frac{135\pi}{128} - i \left(\frac{135 \log(3/2)}{64} -
\frac{189}{128}
\right) \right] \xi^4 \right\}, \\
 \mathcal{A}_{3,2,-2} &= - \mathcal{A}_{3,-2,-2} = -\delta m ( \mathcal{B}_F + 
i \mathcal{A}_F ) \left\{
\frac{9}{32}\xi - \left( \frac{1071}{1024} - \frac{207 \nu}{512}\right)
\xi^3
+ \left[ \frac{27\pi}{32} - i \left(\frac{27 \log (3/2)}{16} -
\frac{189}{160}
\right) \right] \xi^4 \right\}, \\
 \mathcal{A}_{3,3,-2} &= - \mathcal{A}_{3,-3,-2} = -\delta m ( \mathcal{B}_F + 
i \mathcal{A}_F ) \left\{
\frac{9}{128}\xi - \left( \frac{1197}{8192} + \frac{531 \nu}{4096}\right)
\xi^3
+ \left[ \frac{27\pi}{128} - i \left(\frac{27 \log(3/2)}{64} -
\frac{189}{640}
\right) \right] \xi^4 \right\}, \\
 \mathcal{A}_{3,4,-2} &= - \mathcal{A}_{3,-4,-2} = -\delta m ( \mathcal{B}_F + 
i \mathcal{A}_F ) \left( \frac{135}{2048}
-
\frac{135 \nu}{1024}\right)
\xi^3, \\
 \mathcal{A}_{3,5,-2} &= - \mathcal{A}_{3,-5,-2} = -\delta m ( \mathcal{B}_F + 
i \mathcal{A}_F ) \left( \frac{81}{8192}
-
\frac{81 \nu}{4096}\right)
\xi^3, \\
 \mathcal{A}_{4,0,2} &= - (\mathcal{A}_F + i \mathcal{B}_F) \left[ \left( 
\frac{5}{12} -
\frac{5 \nu}{4} \right) \xi^2  - \left( \frac{89}{40}
- \frac{571 \nu}{72} + \frac{77\nu^2}{24} \right) \xi^4 \right], \\
 \mathcal{A}_{4,1,2} &= \mathcal{A}_{4,-1,2} = (\mathcal{A}_F + i 
\mathcal{B}_F) 
\left[ \left(
\frac{1}{6} -
\frac{\nu}{2} \right) \xi^2 - \left( \frac{49}{60}
- \frac{101\nu}{36} + \frac{11 \nu^2}{12} \right) \xi^4 \right], \\
 \mathcal{A}_{4,2,2} &= \mathcal{A}_{4,-2,2} = (\mathcal{A}_F + i 
\mathcal{B}_F) 
\left[ \left(
\frac{1}{6} -
\frac{\nu}{2} \right) \xi^2 - \left( \frac{119}{120}
- \frac{265 \nu}{72} + \frac{43 \nu^2}{24} \right) \xi^4 \right], \\
 \mathcal{A}_{4,3,2} &= \mathcal{A}_{4,-3,2} = - (\mathcal{A}_F + i 
\mathcal{B}_F) \left[ \left(
\frac{1}{6} -
\frac{\nu}{2} \right) \xi^2 - \left( \frac{13}{15}
- \frac{55 \nu}{18} + \frac{7 \nu^2}{6} \right) \xi^4 \right], \\
 \mathcal{A}_{4,4,2} &= \mathcal{A}_{4,-4,2} = (\mathcal{A}_F + i 
\mathcal{B}_F) 
\left[ \left(
\frac{1}{24} -
\frac{\nu}{8} \right) \xi^2 - \left( \frac{31}{240}
- \frac{47 \nu}{144} - \frac{7 \nu^2}{48} \right) \xi^4 \right], \\
 \mathcal{A}_{4,5,2} &= \mathcal{A}_{4,-5,2} = -(\mathcal{A}_F + i 
\mathcal{B}_F)  \left( \frac{1}{20}
- \frac{\nu}{4} + \frac{\nu^2}{4} \right) \xi^4, \\
 \mathcal{A}_{4,6,2} &= \mathcal{A}_{4,-6,2} = (\mathcal{A}_F + i 
\mathcal{B}_F) 
 \left( \frac{1}{120}
- \frac{\nu}{24} + \frac{\nu^2}{24} \right) \xi^4, \\
 \mathcal{A}_{4,0,-2} &= - (\mathcal{A}_F - i \mathcal{B}_F) \left[ \left( 
\frac{5}{12} -
\frac{5 \nu}{4} \right) \xi^2  - \left( \frac{89}{40}
- \frac{571 \nu}{72} + \frac{77\nu^2}{24} \right) \xi^4 \right], \\
 \mathcal{A}_{4,1,-2} &= \mathcal{A}_{4,-1,-2} = -(\mathcal{A}_F - i 
\mathcal{B}_F) \left[ \left(
\frac{1}{6} -
\frac{\nu}{2} \right) \xi^2 - \left( \frac{49}{60}
- \frac{101\nu}{36} + \frac{11 \nu^2}{12} \right) \xi^4 \right], \\
 \mathcal{A}_{4,2,-2} &= \mathcal{A}_{4,-2,-2} = (\mathcal{A}_F - i 
\mathcal{B}_F) \left[ \left(
\frac{1}{6} -
\frac{\nu}{2} \right) \xi^2 - \left( \frac{119}{120}
- \frac{265 \nu}{72} + \frac{43 \nu^2}{24} \right) \xi^4 \right], \\
 \mathcal{A}_{4,3,-2} &= \mathcal{A}_{4,-3,-2} = (\mathcal{A}_F - i 
\mathcal{B}_F) \left[ \left(
\frac{1}{6} -
\frac{\nu}{2} \right) \xi^2 - \left( \frac{13}{15}
- \frac{55 \nu}{18} + \frac{7 \nu^2}{6} \right) \xi^4 \right], \\
 \mathcal{A}_{4,4,-2} &= \mathcal{A}_{4,-4,-2} = (\mathcal{A}_F - i 
\mathcal{B}_F) \left[ \left(
\frac{1}{24} -
\frac{\nu}{8} \right) \xi^2 - \left( \frac{31}{240}
- \frac{47 \nu}{144} - \frac{7 \nu^2}{48} \right) \xi^4 \right], \\
 \mathcal{A}_{4,5,-2} &= \mathcal{A}_{4,-5,-2} = (\mathcal{A}_F - i 
\mathcal{B}_F)  \left( \frac{1}{20}
- \frac{\nu}{4} + \frac{\nu^2}{4} \right) \xi^4, \\
 \mathcal{A}_{4,6,-2} &= \mathcal{A}_{4,-6,-2} = (\mathcal{A}_F - i 
\mathcal{B}_F)  \left( \frac{1}{120}
- \frac{\nu}{24} + \frac{\nu^2}{24} \right) \xi^4, \\
 \mathcal{A}_{5,1,2} &= - \mathcal{A}_{5,-1,2} = \delta m ( \mathcal{B}_F - i 
\mathcal{A}_F ) \left( \frac{4375}{12288}
- \frac{4375 \nu}{6144}\right)
\xi^3, \\
 \mathcal{A}_{5,2,2} &= - \mathcal{A}_{5,-2,2} = - \delta m ( \mathcal{B}_F - i 
\mathcal{A}_F ) \left( \frac{625}{3072} -
\frac{625 \nu}{1536}\right)
\xi^3, \\
 \mathcal{A}_{5,3,2} &= - \mathcal{A}_{5,-3,2} = - \delta m ( \mathcal{B}_F - i 
\mathcal{A}_F ) \left( \frac{625}{8192}
- \frac{625 \nu}{4096}\right)
\xi^3, \\
 \mathcal{A}_{5,4,2} &= - \mathcal{A}_{5,-4,2} = \delta m ( \mathcal{B}_F - i 
\mathcal{A}_F ) \left( \frac{625}{6144} -
\frac{625 \nu}{3072}\right)
\xi^3, \\
 \mathcal{A}_{5,5,2} &= - \mathcal{A}_{5,-5,2} = -\delta m ( \mathcal{B}_F - i 
\mathcal{A}_F ) \left( \frac{625}{24576}
-
\frac{625 \nu}{12288}\right)
\xi^3,  \\
 \mathcal{A}_{5,1,-2} &= - \mathcal{A}_{5,-1,-2} = -\delta m ( \mathcal{B}_F + 
i \mathcal{A}_F ) \left(
\frac{4375}{12288}
- \frac{4375 \nu}{6144}\right)
\xi^3, \\
 \mathcal{A}_{5,2,-2} &= - \mathcal{A}_{5,-2,-2} = -\delta m ( \mathcal{B}_F + 
i \mathcal{A}_F ) \left( \frac{625}{3072}
-
\frac{625 \nu}{1536}\right)
\xi^3, \\
 \mathcal{A}_{5,3,-2} &= - \mathcal{A}_{5,-3,-2} = \delta m ( \mathcal{B}_F + i 
\mathcal{A}_F ) \left(
\frac{625}{8192}
- \frac{625 \nu}{4096}\right)
\xi^3, \\
 \mathcal{A}_{5,4,-2} &= - \mathcal{A}_{5,-4,-2} = \delta m ( \mathcal{B}_F + i 
\mathcal{A}_F ) \left( \frac{625}{6144}
-
\frac{625 \nu}{3072}\right)
\xi^3, \\
 \mathcal{A}_{5,5,-2} &= - \mathcal{A}_{5,-5,-2} = \delta m ( \mathcal{B}_F + i 
\mathcal{A}_F ) \left(
\frac{625}{24576}
-
\frac{625 \nu}{12288}\right)
\xi^3,  \\
 \mathcal{A}_{6,0,2} &= - (\mathcal{A}_F + i \mathcal{B}_F) \left( 
\frac{567}{1280}
- \frac{567 \nu}{256} + \frac{567\nu^2}{256} \right) \xi^4 , \\
 \mathcal{A}_{6,1,2} &= \mathcal{A}_{6,-1,2} = (\mathcal{A}_F + i 
\mathcal{B}_F) 
\left( \frac{81}{640}
- \frac{81\nu}{128} + \frac{81 \nu^2}{128} \right) \xi^4 , \\
 \mathcal{A}_{6,2,2} &= \mathcal{A}_{6,-2,2} = (\mathcal{A}_F + i 
\mathcal{B}_F) 
\left( \frac{1377}{5120}
- \frac{1377 \nu}{1024} + \frac{1377 \nu^2}{1024} \right) \xi^4 , \\
 \mathcal{A}_{6,3,2} &= \mathcal{A}_{6,-3,2} = - (\mathcal{A}_F + i 
\mathcal{B}_F) \left( \frac{243}{1280}
- \frac{243 \nu}{256} + \frac{243 \nu^2}{256} \right) \xi^4 , \\
 \mathcal{A}_{6,4,2} &= \mathcal{A}_{6,-4,2} = -(\mathcal{A}_F + i 
\mathcal{B}_F) \left( \frac{81}{2560}
- \frac{81 \nu}{512} + \frac{81 \nu^2}{512} \right) \xi^4 , \\
 \mathcal{A}_{6,5,2} &= \mathcal{A}_{6,-5,2} = (\mathcal{A}_F + i 
\mathcal{B}_F) 
 \left( \frac{81}{1280}
- \frac{81\nu}{256} + \frac{81\nu^2}{256} \right) \xi^4, \\
 \mathcal{A}_{6,6,2} &= \mathcal{A}_{6,-6,2} = -(\mathcal{A}_F + i 
\mathcal{B}_F)  \left( \frac{81}{5120}
- \frac{81\nu}{1024} + \frac{81\nu^2}{1024} \right) \xi^4, \\
 \mathcal{A}_{6,0,-2} &= - (\mathcal{A}_F - i \mathcal{B}_F) \left( 
\frac{567}{1280}
- \frac{567 \nu}{256} + \frac{567\nu^2}{256} \right) \xi^4 , \\
 \mathcal{A}_{6,1,-2} &= \mathcal{A}_{6,-1,-2} = -(\mathcal{A}_F - i 
\mathcal{B}_F) \left( \frac{81}{640}
- \frac{81\nu}{128} + \frac{81 \nu^2}{128} \right) \xi^4 , \\
 \mathcal{A}_{6,2,-2} &= \mathcal{A}_{6,-2,-2} = (\mathcal{A}_F - i 
\mathcal{B}_F) \left( \frac{1377}{5120}
- \frac{1377 \nu}{1024} + \frac{1377 \nu^2}{1024} \right) \xi^4 , \\
 \mathcal{A}_{6,3,-2} &= \mathcal{A}_{6,-3,-2} = (\mathcal{A}_F - i 
\mathcal{B}_F) \left( \frac{243}{1280}
- \frac{243 \nu}{256} + \frac{243 \nu^2}{256} \right) \xi^4 , \\
 \mathcal{A}_{6,4,-2} &= \mathcal{A}_{6,-4,-2} = -(\mathcal{A}_F - i 
\mathcal{B}_F) \left( \frac{81}{2560}
- \frac{81 \nu}{512} + \frac{81 \nu^2}{512} \right) \xi^4 , \\
 \mathcal{A}_{6,5,-2} &= \mathcal{A}_{6,-5,-2} = -(\mathcal{A}_F - i 
\mathcal{B}_F)  \left( \frac{81}{1280}
- \frac{81\nu}{256} + \frac{81\nu^2}{256} \right) \xi^4, \\
 \mathcal{A}_{6,6,-2} &= \mathcal{A}_{6,-6,-2} = -(\mathcal{A}_F - i 
\mathcal{B}_F)  \left( \frac{81}{5120}
- \frac{81\nu}{1024} + \frac{81\nu^2}{1024} \right) \xi^4,
\end{align}

\end{widetext}


\end{document}